\documentclass[twocolumn,tighten]{aastex6} 
\usepackage{rotating,fp,amsmath}
\usepackage{color}

\newcommand{\toe}[1]{% code to automatically update the times of eruption given very truncated MJD-57000 as input
  \FPeval{\result}{round(#1-734.32,3)}
  \FPifneg{\result} \result \else \phs\result \fi 
}

\def\m31{{M\,31}}

\def\chandra{{\it Chandra~}}

\def\swift{{\it Swift~}}
\def\swiftk{{\it Swift}}
\def\xmm{{\it XMM-Newton~}}
\def\xmmk{{\it XMM-Newton}}
\def\hst{{\it HST~}}

\def\nova{{M31N\,2008-12a~}}
\def\novak{{M31N\,2008-12a}}

\newcommand{\nh}{\hbox{$N_{\rm H}$}~}

\newcommand{\hcm}[1]{$\times 10^{#1}$\,cm$^{-2}$}
\newcommand{\ohcm}[1]{$10^{#1}$\,cm$^{-2}$}

\newcommand{\cts}[1]{$\times 10^{#1}$ ct s$^{-1}$}

\newcommand{\power}[1]{$10^{#1}$}

\newcommand{\eton}{t_{\mbox{\small{on}}}}
\newcommand{\etoff}{t_{\mbox{\small{off}}}}

\defcitealias{2014A&A...563L...9D}{DWB14}
\defcitealias{2014A&A...563L...8H}{HND14}
\defcitealias{2014ApJ...786...61T}{TBW14}
\defcitealias{2015A&A...580A..45D}{DHS15}
\defcitealias{2015A&A...580A..46H}{HND15}
\defcitealias{2015A&A...582L...8H}{HDK15}
\defcitealias{2016ApJ...833..149D}{DHB16}
\defcitealias{2017ApJ...847...35D}{DHG17S}
\defcitealias{2017ApJ...849...96D}{DHG17P}

\newcommand{\oonek}{\citetalias{2014A&A...563L...9D}}
\newcommand{\xonek}{\citetalias{2014A&A...563L...8H}}
\newcommand{\ponek}{\citetalias{2014ApJ...786...61T}}
\newcommand{\otwok}{\citetalias{2015A&A...580A..45D}}
\newcommand{\xtwok}{\citetalias{2015A&A...580A..46H}}
\newcommand{\halfk}{\citetalias{2015A&A...582L...8H}}
\newcommand{\othreek}{\citetalias{2016ApJ...833..149D}}
\newcommand{\hstspec}{\citetalias{2017ApJ...847...35D}}
\newcommand{\hstphot}{\citetalias{2017ApJ...849...96D}}

\begin{document} 

\received{?}
\revised{?}
\accepted{?}
\slugcomment{The Astrophysical Journal, \today, Draft version}

\title{Breaking the habit --- the peculiar 2016 eruption of \\the unique recurrent nova \novak.}
\shorttitle{\novak: The peculiar 2016 eruption}
\shortauthors{Henze et al.\ 2017}

\author{
M. Henze,\altaffilmark{1,2}
M. J. Darnley,\altaffilmark{3}
S. C. Williams,\altaffilmark{3,4} 
M. Kato,\altaffilmark{5}
I. Hachisu,\altaffilmark{6} 
G. C. Anupama,\altaffilmark{7}
A. Arai,\altaffilmark{8}
D. Boyd,\altaffilmark{9,10}\\
D. Burke,\altaffilmark{11}
R. Ciardullo,\altaffilmark{12,13}
K. Chinetti,\altaffilmark{14}
L. M. Cook,\altaffilmark{10}
M. J. Cook,\altaffilmark{10}
P. Erdman,\altaffilmark{11}
X. Gao,\altaffilmark{15}
B. Harris,\altaffilmark{10}\\
D. H. Hartmann,\altaffilmark{16}
K. Hornoch,\altaffilmark{17}
J. Chuck Horst,\altaffilmark{1}
R. Hounsell,\altaffilmark{18,19}
D. Husar,\altaffilmark{10,20}
K. Itagaki,\altaffilmark{21}\\
F. Kabashima,\altaffilmark{22}
S. Kafka,\altaffilmark{10}
A. Kaur,\altaffilmark{16}
S. Kiyota,\altaffilmark{23}
N. Kojiguchi,\altaffilmark{24}
H. Ku\v{c}\'akov\'a,\altaffilmark{17,25,26}
K. Kuramoto,\altaffilmark{27}\\
H. Maehara,\altaffilmark{28}
A. Mantero,\altaffilmark{10}
F. J. Masci,\altaffilmark{29}
K. Matsumoto,\altaffilmark{23}
H. Naito,\altaffilmark{30}
J.-U. Ness,\altaffilmark{31}
K. Nishiyama,\altaffilmark{22}\\
A. Oksanen,\altaffilmark{10}
J. P. Osborne,\altaffilmark{32}
K. L. Page,\altaffilmark{32}
E. Paunzen,\altaffilmark{33}
M. Pavana,\altaffilmark{7}
R. Pickard,\altaffilmark{9,10}
J. Prieto-Arranz,\altaffilmark{34,35}\\
P. Rodr\'\i guez-Gil,\altaffilmark{34,35}
G. Sala,\altaffilmark{36,37}
Y. Sano,\altaffilmark{27}
A. W. Shafter,\altaffilmark{1}
Y. Sugiura,\altaffilmark{24}
H. Tan,\altaffilmark{38}
T. Tordai,\altaffilmark{39}
J. Vra\v{s}til,\altaffilmark{17,25}\\
R. M. Wagner,\altaffilmark{40}
F. Watanabe,\altaffilmark{30}
B. F. Williams,\altaffilmark{41}
M. F. Bode,\altaffilmark{3,42}
A. Bruno,\altaffilmark{10}
B. Buchheim,\altaffilmark{10}
T. Crawford,\altaffilmark{10}\\
B. Goff,\altaffilmark{10}
M. Hernanz,\altaffilmark{2}
A. S. Igarashi,\altaffilmark{1}
J. Jos\'{e},\altaffilmark{36,37}
M. Motta,\altaffilmark{10}
T. J. O'Brien,\altaffilmark{43}
T. Oswalt,\altaffilmark{11}
G. Poyner,\altaffilmark{9,10}\\
V. A. R. M. Ribeiro,\altaffilmark{44,45,46}
R. Sabo,\altaffilmark{10}
M. M. Shara,\altaffilmark{47}
J. Shears,\altaffilmark{9}
D. Starkey,\altaffilmark{10}
S. Starrfield,\altaffilmark{48}
C. E. Woodward\altaffilmark{49}
}

\altaffiltext{1}{Department of Astronomy, San Diego State University, San Diego, CA 92182, USA}  % Henze, Shafter, Horst, Quimby, Igarashi
\altaffiltext{2}{Institut de Ci\`encies de l'Espai (CSIC-IEEC), Campus UAB, C/Can Magrans s/n, 08193 Cerdanyola del Valles, Spain} % Henze, Hernanz
\altaffiltext{3}{Astrophysics Research Institute, Liverpool John Moores University, IC2 Liverpool Science Park, Liverpool, L3 5RF, UK} % Darnley, Steele, Bode, SWilliams, Harman
\altaffiltext{4}{Physics Department, Lancaster University, Lancaster, LA1 4YB, UK}  % SWilliams
\altaffiltext{5}{Department of Astronomy, Keio University, Hiyoshi, Yokohama 223-8521, Japan}  % Kato
\altaffiltext{6}{Department of Earth Science and Astronomy, College of Arts and Sciences, University of Tokyo, 3-8-1 Komaba, Meguro-ku, Tokyo 153-8902, Japan} % Hachisu
\altaffiltext{7}{Indian Institute of Astrophysics, Koramangala, Bangalore 560 034, India} % Anupama, pavana
\altaffiltext{8}{Koyama Astronomical Observatory, Kyoto Sangyo University, Motoyama, Kamigamo, Kita-ku, Kyoto, Kyoto 603-8555, Japan}  % Arai
\altaffiltext{9}{British Astronomical Association Variable Star Section, Burlington House, Piccadilly, London W1J 0DU, UK}  % Shears, Boyd, Poyner, Pickard
\altaffiltext{10}{American Association of Variable Star Observers, 49 Bay State Rd., Cambridge, MA 02138, USA} % Kafka, LCook, MCook, BHarris, Boyd, Pickard, Oksanen, Mantero
\altaffiltext{11}{Embry-Riddle Aeronautical University, Daytona Beach, FL, USA} % Burke, Oswalt, Erdman
\altaffiltext{12}{Department of Astronomy and Astrophysics, The Pennsylvania State University, 525 Davey Lab, University Park, PA 16802, USA} % Ciardullo
\altaffiltext{13}{Institute for Gravitation and the Cosmos, The Pennsylvania State University, University Park, PA 16802, USA} % Ciardullo
\altaffiltext{14}{Division of Physics, Mathematics and Astronomy, California Institute of Technology, Pasadena, CA 91125, USA} % Chinetti
\altaffiltext{15}{Xinjiang Astronomical Observatory, Urumqi, Xinjiang 830011, PR China}  % Gao
\altaffiltext{16}{Department of Physics and Astronomy, Clemson University, Clemson, SC 29634, USA}  % Hartmann, Kaur
\altaffiltext{17}{Astronomical Institute, Academy of Sciences, CZ-251 65 Ond\v{r}ejov, Czech Republic}  % Hornoch, Kucakova, Vrastil
\altaffiltext{18}{Department of Astronomy and Astrophysics, University of California, Santa Cruz, CA 95064, USA} %Hounsell
\altaffiltext{19}{Department of Astronomy, University of Illinois at Urbana-Champaign, 1002 W.\ Green Street, Urbana, IL 61801, USA} % Hounsell
\altaffiltext{20}{Stiftung Interaktive Astronomie und Astrophysik, Geschwister-Scholl-Platz, 72074 T\"ubingen, Germany} % Husar
\altaffiltext{21}{Itagaki Astronomical Observatory, Teppo, Yamagata 990-2492, Japan}  % Itagaki
\altaffiltext{22}{Miyaki-Argenteus Observatory, Miyaki, Saga 840-1102, Japan}  % Kabashima, Nishiyama
\altaffiltext{23}{Variable Stars Observers League in Japan (VSOLJ), 7-1 Kitahatsutomi, Kamagaya 273-0126, Japan}  % Kiyota
\altaffiltext{24}{Osaka Kyoiku University, 4-698-1 Asahigaoka, Kashiwara, Osaka 582-8582, Japan} % Kojiguchi, Matsumoto, Sugiura
\altaffiltext{25}{Astronomical Institute of the Charles University, Faculty of Mathemathics and Physics, V Hole\v{s}ovi\v{c}k\'ach 2, 180 00 Praha 8, Czech Republic}  % Kucakova, Vrastil
\altaffiltext{26}{Institute of Physics, Faculty of Philosophy and Science, Silesian University in Opava, Bezru\v{c}ovo n\'am.\ 13, CZ-746 01 Opava, Czech Republic}  % Kucakova
\altaffiltext{27}{Department of Cosmosciences, Hokkaido University, Kita 10, Nishi 8, Kita-ku, Sapporo, Hokkaido 060-0810, Japan} % Kuramoto, Sano
\altaffiltext{28}{Okayama Astrophysical Observatory, NAOJ, NINS, 3037-5 Honjo, Kamogata, Asakuchi, Okayama 719-0232, Japan}  % Maehara, Kuroda(?)
\altaffiltext{29}{Infrared Processing and Analysis Center, California Institute of Technology, Pasadena, CA 91125, USA} % Masci
\altaffiltext{30}{Nayoro Observatory, 157-1 Nisshin, Nayoro, Hokkaido 096-0066, Japan}  % Naito,  Nakajima, FWantanabe
\altaffiltext{31}{XMM-Newton Observatory SOC, European Space Astronomy Centre, Camino Bajo del Castillo s/n, Urb.\ Villafranca del Castillo, 28692 Villanueva de la Ca\~{n}ada, Madrid, Spain}  % Ness
\altaffiltext{32}{X-Ray and Observational Astronomy Group, Department of Physics \& Astronomy, University of Leicester, LE1 7RH, UK}  % Osborne, Page
\altaffiltext{33}{Department of Theoretical Physics and Astrophysics, Masaryk University, Kotl\'a\v{r}sk\'a 2, 611 37 Brno, Czech Republic} %Paunzen
\altaffiltext{34}{Instituto de Astrof\'\i sica de Canarias, V\'\i a L\'actea, s/n, La Laguna, E-38205, Santa Cruz de Tenerife, Spain}  % R-G, Prieto-Arranz
\altaffiltext{35}{Departamento de Astrof\'\i sica, Universidad de La Laguna, La Laguna, E-38206, Santa Cruz de Tenerife, Spain}  % R-G, Prieto-Arranz
\altaffiltext{36}{Departament de F\'{i}sica, EEBE, Universitat Polit\`{e}cnica de Catalunya, BarcelonaTech, Av/ d'Eduard Maristany, 10-14, E-08019 Barcelona, Spain} % Jose, Sala
\altaffiltext{37}{Institut d'Estudis Espacials de Catalunya, c/Gran Capit\`{a} 2-4, Ed.\ Nexus-201, E-08034, Barcelona, Spain} % Jose, Sala
\altaffiltext{38}{Department of Optoelectric Physics, Chinese Culture University, Taipei 11114, Taiwan}  % Tan
\altaffiltext{39}{Polaris Observatory, Hungarian Astronomical Association, Laborc u.\ 2/c, 1037 Budapest, Hungary} % Tordai
\altaffiltext{40}{Department of Astronomy, The Ohio State University, 140 West 18th Avenue, Columbus, OH 43210, USA} % Wagner
\altaffiltext{41}{Department of Astronomy, Box 351580, University of Washington, Seattle, WA 98195, USA}  % BWilliams
\altaffiltext{42}{Office of the Vice Chancellor, Botswana International University of Science \& Technology, Private Bag 16, Palapye, Botswana} % Bode
\altaffiltext{43}{Jodrell Bank Centre for Astrophysics, Alan Turing Building, University of Manchester, Manchester, M13 9PL, UK}  % O'Brien
\altaffiltext{44}{CIDMA, Departamento de F\'isica, Universidade de Aveiro, Campus de Santiago, 3810-193 Aveiro, Portugal} % Ribeiro
\altaffiltext{45}{Instituto de Telecomunica\c{c}\~oes, Campus de Santiago, 3810-193 Aveiro, Portugal} % Ribeiro
\altaffiltext{46}{Department of Physics and Astronomy, Botswana International University of Science \& Technology, Private Bag 16, Palapye, Botswana} % Ribeiro
\altaffiltext{47}{American Museum of Natural History, 79th Street and Central Park West, New York, NY 10024, USA}  % Shara
\altaffiltext{48}{School of Earth and Space Exploration, Arizona State University, Tempe, AZ 85287-1504, USA} % Starrfield
\altaffiltext{49}{University of Minnesota, Minnesota Institute for Astrophysics, 116 Church St. SE, Minneapolis, MN 55455, USA}  % Woodward

\begin{abstract}
Since its discovery in 2008, the Andromeda galaxy nova \novak\ has been observed in eruption every single year. This unprecedented frequency indicates an extreme object, with a massive white dwarf and a high accretion rate, which is the most promising candidate for the single-degenerate progenitor of a type-Ia supernova known to date. The previous three eruptions of \novak\ have displayed remarkably homogeneous multi-wavelength properties:\ (i) From a faint peak, the optical light curve declined rapidly by two magnitudes in less than two days; (ii) Early spectra showed initial high velocities that slowed down significantly within days and displayed clear He/N lines throughout; (iii) The supersoft X-ray source (SSS) phase of the nova began extremely early, six days after eruption, and only lasted for about two weeks. In contrast, the peculiar 2016 eruption was clearly different. Here we report (i) the considerable delay in the 2016 eruption date, (ii) the significantly shorter SSS phase, and (iii) the brighter optical peak magnitude (with a hitherto unobserved cusp shape). Early theoretical models suggest that these three different effects can be consistently understood as caused by a lower quiescence mass-accretion rate. The corresponding higher ignition mass caused a brighter peak in the free-free emission model. The less-massive accretion disk experienced greater disruption, consequently delaying re-establishment of effective accretion. Without the early refueling, the SSS phase was shortened. Observing the next few eruptions will determine whether the properties of the 2016 outburst make it a genuine outlier in the evolution of \novak.
\end{abstract}


\section{Introduction}
Recurrent novae with frequent eruptions are new and exciting objects at the interface between the parameter spaces of novae and type Ia supernovae (SNe\,Ia). Novae are periodic thermonuclear eruptions on the surfaces of white dwarfs (WDs) in mass-transfer binaries \citep[see][for comprehensive reviews on nova physics]{2008clno.book.....B, Jos16,2016PASP..128e1001S}. In SNe\,Ia, a carbon-oxygen (CO) WD approaches the \citet{1931ApJ....74...81C} mass limit to be destroyed in a thermonuclear explosion. Theoretical models show that a CO WD can indeed grow from a low initial mass through many nova cycles to eventually become a SN\,Ia \citep[e.g.,][]{2005ApJ...623..398Y,2014ASPC..490..287N, 2016ApJ...819..168H}.

Only for massive WDs with high accretion rates do the periods of the nova cycles become shorter than $\sim100$\,yr \citep{1985ApJ...291..136S,2005ApJ...623..398Y,2008NewAR..52..386H,2014ApJ...793..136K} --- the (current) empirical limit to observe a nova erupting more than once. These are called recurrent novae (RNe) and have been observed in the Galaxy and its closest neighbors \citep[see, for example,][]{1991ApJ...370..193S,2010ApJS..187..275S,2015ApJS..216...34S,2016ApJ...818..145B}.  The extreme physics necessary to power the high eruption frequency of the RNe with the shortest periods makes them the most promising (single-degenerate) SN\,Ia progenitor candidates known today \citep{2015ApJ...808...52K}.

Among the ten RNe in the Galaxy, U\,Scorpii has the shortest period with inter-eruption durations as short as eight years \citep{2010ApJS..187..275S}. Another nova with rapid eruptions has recently been found in the Large Magellanic Cloud \citep[LMCN\,1968-12a with 5~yr;][]{2016ATel.8578....1M,2016ATel.8587....1D,KuinPaper}. However, it is the nearby Andromeda galaxy (\m31) which hosts six RNe with eruption periods of less than 10\,yr. Due to its proximity and relatively high stellar mass (within the Local Group), \m31 has been a target of optical nova surveys for a century. Starting with the first discovery by \citet{1917PASP...29..210R}, exactly 100\,yr ago, and the first monitoring survey by \citet{1929ApJ....69..103H}, the community has gradually built a rich database of more than 1000 nova candidates in \m31 \citep[see][and their on-line database\footnote{\url{http://www.mpe.mpg.de/~m31novae/opt/m31/index.php}}]{2007A&A...465..375P,2010AN....331..187P}. Crucially, the low foreground extinction toward \m31 \citep[\nh = 0.7\hcm{21},][]{1992ApJS...79...77S} favours X-ray monitoring surveys for novae \citep{2007A&A...465..375P,2010A&A...523A..89H,2011A&A...533A..52H,2014A&A...563A...2H}.

The unparalleled \m31 nova sample contains 18 known RNe \citep{2015ApJS..216...34S,2015ATel.7116....1H,2017ATel10001....1S}. Among them there are five RNe with recurrence periods between four and nine years. Those objects are: M31N\,1990-10a \citep[9\,yr period;][]{2016ATel.9276....1H,2016ATel.9280....1H,2016ATel.9281....1E,2016ATel.9383....1F}, M31N\,2007-11f \citep[9\,yr period;][]{2017ATel10001....1S,2017ATel.9942....1F}, M31N\,1984-07a \citep[8\,yr period][]{2012ATel.4364....1H,2015ApJS..216...34S}, M31N\,1963-09c \citep[5\,yr period][]{1973A&AS....9..347R,2014A&A...563A...2H,2015ATel.8234....1W,2015ATel.8242....1W,2015ATel.8235....1H,2015ATel.8290....1H}, and M31N\,1997-11k \citep[4\,yr period][]{2009ATel.2286....1H,2015ApJS..216...34S}. 

The indisputable champion of all RNe, however, is \novak. Since its discovery in 2008 \citep[by][]{2008Nis}, this remarkable nova has been seen in eruption every single year \citep[][hereafter \othreek, see Table~\ref{eruption_history}]{2016ApJ...833..149D}. Beginning in 2013, our group has been studying the eruptions of \nova with detailed multi-wavelength observations. For the 2013 eruption we found a fast optical evolution \citep[hereafter \oonek]{2014A&A...563L...9D} and a supersoft X-ray source \citep[SSS;][]{2008ASPC..401..139K} phase of only two weeks (\citealt[hereafter \xonek]{2014A&A...563L...8H}, also see \citealt{2014ApJ...786...61T}). The SSS stage, powered by nuclear burning within the hydrogen-rich envelope remaining on the WD after the eruption, typically lasts years to decades in regular novae \citep{2011ApJS..197...31S,2014A&A...563A...2H,2015JHEAp...7..117O}. The SSS phase of the 2014 eruption was similarly short \citep[hereafter \xtwok]{2015A&A...580A..46H} and we collected high-cadence, multi-color optical photometry \citep[hereafter \otwok]{2015A&A...580A..45D}. In \citet[hereafter \halfk]{2015A&A...582L...8H} we predicted the date of the 2015 eruption with an accuracy of better than a month and followed it with a large multi-wavelength fleet of telescopes (\othreek).

\begin{table*}
\caption{All Known Eruption Dates of \novak.\label{eruption_history}}
\begin{center}
\begin{tabular}{lllll}
\hline\hline
Eruption date\tablenotemark{a} & SSS-on date\tablenotemark{b} & Days since & Detection wavelength & References\\
(UT) & (UT) & last eruption\tablenotemark{c} & (observatory) & \\
\hline
(1992 Jan 28) & 1992 Feb 03 & \nodata & X-ray ({\it ROSAT}) & 1, 2 \\
(1993 Jan 03) & 1993 Jan 09 & 341 & X-ray ({\it ROSAT}) & 1, 2 \\
(2001 Aug 27) & 2001 Sep 02 & \nodata & X-ray ({\it Chandra}) & 2, 3 \\
2008 Dec 25 & \nodata & \nodata & Visible (Miyaki-Argenteus) & 4 \\
2009 Dec 02 & \nodata & 342 & Visible (PTF) & 5 \\
2010 Nov 19 & \nodata & 352 & Visible (Miyaki-Argenteus) & 2 \\
2011 Oct 22.5 & \nodata & 337.5 & Visible (ISON-NM) & 5--8 \\
2012 Oct 18.7 & $<2012$ Nov 06.45 & 362.2 & Visible (Miyaki-Argenteus) & 8--11 \\
2013 Nov $26.95\pm0.25$ & $\le2013$ Dec 03.03 & 403.5 & Visible (iPTF); UV/X-ray (\swiftk) & 5, 8, 11--14 \\
2014 Oct $02.69\pm0.21$ & 2014 Oct $08.6\pm0.5$ & $309.8\pm0.7$ & Visible (LT); UV/X-ray (\swiftk) & 8, 15 \\
2015 Aug $28.28\pm0.12$ & 2015 Sep $02.9\pm0.7$ & $329.6\pm0.3$ & Visible (LCO); UV/X-ray (\swiftk) & 14, 16--18\\
2016 Dec $12.32\pm0.17$ & 2016 Dec $17.2\pm1.1$ & $471.7\pm0.2$ & Visible (Itagaki); UV/X-ray (\swiftk) & 19--23\\ 
\hline
\end{tabular}
\end{center}
\catcode`\&=12
\tablecomments{This is an updated version of Table~1 as it was published by \protect \citet{2014ApJ...786...61T}, \protect \citet{2015A&A...580A..45D}, \protect \citet{2015A&A...582L...8H}, and \protect\citet{2016ApJ...833..149D}. Here we add the 2016 eruption information.}
\tablenotetext{a}{Derived eruption time in the optical bands. The values in parentheses were estimated from the archival X-ray detections \protect \citep[cf.][]{2015A&A...582L...8H}.}
\tablenotetext{b}{Emergence of the SSS counterpart. There is sufficient ROSAT data to estimate the SSS turn-on time accurately. The \chandra detection comprises of only one data point, on September 8th, which we assume to be midpoint of a typical 12-day SSS light curve. Due to the very short SSS phase the associated uncertainties will be small ($\pm6$\,d).}
\tablenotetext{c}{The gaps between eruption dates is only given for the case of observed eruptions in consecutive years.}
\tablerefs{(1)~\citet{1995ApJ...445L.125W}, (2)~\citet{2015A&A...582L...8H}, (3)~\citet{2004ApJ...609..735W}, (4)~\citet{2008Nis}, (5)~\citet{2014ApJ...786...61T}, (6)~\citet{2011Kor}, (7)~\citet{2011ATel.3725....1B}, (8)~\citet{2015A&A...580A..45D}, (9)~\citet{2012Nis}, (10)~\citet{2012ATel.4503....1S}, (11)~\citet{2014A&A...563L...8H}, (12)~\citet{2013ATel.5607....1T}, (13)~\citet{2014A&A...563L...9D}, (14)~\citet{2016ApJ...833..149D}, (15)~\citet{2015A&A...580A..46H}, (16)~\citet{2015ATel.7964....1D}, (17)~\citet{2015ATel.7965....1D}, (18)~\citet{2015ATel.7984....1H}, (19)~this paper, (20)~\citet{2016Ita}, (21)~\citet{2016ATel.9848....1I}, (22)~\citet{2016ATel.9853....1D}, (23)~\citet{2016ATel.9872....1H}, (24)~\citet{2017ATel11116....1B}, (25)~\citet{2018ATel11121....1H}, (26)~\citet{2018ATel11130....1H}, (27)~\citet{2018ATel11149....1D}.}
\end{table*}

The overall picture of \nova that had been emerging through the recent campaigns indicated very regular properties (see \othreek\ for a detailed description): Successive eruptions occurred every year with a predictable observed period of almost one year ($347\pm10$\,d). The optical light curve rose within about a day to a maximum below 18th mag (faint for an \m31 nova) and then immediately declined rapidly by 2 mag in about 2\,d throughout the UV/optical bands. The SSS counterpart brightened at around day 6 after eruption and disappeared again into obscurity around day 19 ($\eton = 5.6\pm0.7$\,d and $\etoff = 18.6\pm0.7$\,d in 2015). Even the time evolution of the SSS effective temperatures in 2013--2015, albeit derived from low-count \swift spectra, closely resembled each other.

Far UV spectroscopy of the 2015 eruption uncovered no evidence for neon in the ejecta \citep[hereafter \hstspec]{2017ApJ...847...35D}. Therefore, these observations could not constrain the composition of the WD, since an ONe core might be shielded by a layer of He that grows with each eruption and H-burning episode.  Modeling of the accretion disk, based on late-time and quiescent {\it Hubble Space Telescope (HST)} photometry, indicated that the accretion disk survives the eruptions, and that the quiescent accretion rate was both extremely variable and remarkably high $\sim10^{-6}\,M_\odot\,\mathrm{yr}^{-1}$ \citep[hereafter \hstphot]{2017ApJ...849...96D}.  Theoretical simulations found the eruption properties to be consistent with an $1.38\,M_\sun$ WD accreting at a rate of $1.6 \times 10^{-7}\,M_\sun$\,yr$^{-1}$ \citep{2015ApJ...808...52K,2016ApJ...830...40K,2017ApJ...838..153K}. \hstphot\ also produced the first constraints on the mass donor a, possibly irradiated, red-clump star with $L_\mathrm{donor}=103^{+12}_{-11}\,L_\odot$, $R_\mathrm{donor}=14.14^{+0.46}_{-0.47}\,R_\odot$, and $T_\mathrm{eff, donor}=4890\pm110$\,K.  Finally, \hstphot\ utilized these updated system parameters to refine the time remaining for the WD to grow to the Chandrasekhar mass to be $<20$\,kyr. 

By all accounts, \nova appeared to have become remarkably predictable even for a RN \citep[see also][for a recent review]{2017ASPC..509..515D}.  Then everything changed. The 2016 eruption, predicted for mid September, did not occur until December 12th \citep{2016ATel.9848....1I}; leading to a frankly suspenseful monitoring campaign. Once detected, the optical light curve was observed to peak at a significantly brighter level than previously seen \citep{2016ATel.9857....1E,2016ATel.9861....1B}, before settling into the familiar rapid decline. When the SSS duly appeared around day 6 \citep{2016ATel.9872....1H} we believed the surprises were over. We were wrong \citep{2016ATel.9907....1H}. This paper studies the unexpected behavior of the 2016 eruption of \nova and discusses its impact on past and future observations.

\section{Observations and data analysis of the 2016 Eruption}\label{sec:observations}

In this section, we describe the multi-wavelength set of telescopes used in studying the 2016 eruption together with the corresponding analysis procedures. All errors are quoted to $1\sigma$ and all upper limits to $3\sigma$, unless specifically stated otherwise. The majority of the statistical analysis was carried out within the \texttt{R} software environment \citep{R_manual}.  Throughout, all photometry through Johnson--Cousins filters, and the {\it HST}, \xmmk, and {\it Swift} flight filters are computed in the Vega system, all photometry through Sloan filters are quoted in AB magnitudes. We assume an eruption date of 2016-12-12.32 UT; discussed in detail in Sect.~\ref{sec:time} and \ref{sec:disc_date}.

\subsection{Visible Photometry}\label{sec:optical_photometry}

Like the 2014 and 2015 eruptions before it (\otwok, \othreek), the 2016 eruption of \novak\ was observed by a large number of ground-based telescopes operating in the visible regime.  Unfortunately, due to poor weather conditions at many of the planned facilities, observations of the 2016 eruption are much sparser than in recent years.

A major achievement for the 2016 eruption campaign was the addition of extensive observations from the American Association of Variable Star Observers (AAVSO\footnote{\url{https://www.aavso.org}}), along with the continued support of the Variable Star Observers League in Japan (VSOLJ\footnote{\url{http://vsolj.cetus-net.org}}; see Section~\ref{sec:time} and Appendix~\ref{app:optical_photometry}).  Observations were also obtained from the Mount Laguna Observatory (MLO) 1.0\,m telescope in California, the Ond\v{r}ejov Observatory 0.65\,m telescope in the Czech Republic, the Danish 1.54\,m telescope at La Silla in Chile, the fully-robotic 2\,m Liverpool Telescope \citep[LT;][]{2004SPIE.5489..679S} in La Palma, the 2.54\,m Isaac Newton Telescope (INT) at La Palma, the Palomar 48$^{\prime\prime}$ telescope in California, the 0.6\,m and 1\,m telescopes operated by members of the Embry Riddle Aeronautical University (ERAU) in Florida, the $2\times8.4$\,m (11.8\,m eq.) Large Binocular Telescope (LBT) on Mount Graham, Arizona, the 2\,m Himalayan Chandra Telescope (HCT) located at Indian Astronomical Observatory (IAO), Hanle, India, and the 2.4\,m {\it Hubble Space Telescope}.

\subsubsection{{\textit Hubble Space Telescope} photometry}

The 2016 eruption, and pre-eruption interval, of \novak\ were observed serendipitously by {\it HST} as part of Program ID:\,14651.  The aim of this program was to observe the proposed ``Super-Remnant'' surrounding \novak\ \citep[see \otwok\ and][]{2017arXiv171204872D}. Five pairs of orbits were tasked to obtain narrow band F657N (H$\alpha$+[N\,{\sc ii}]) and F645N (continuum) observations using Wide Field Camera 3 (WFC3) in the UVIS mode.  Each orbit utilized a three-point dither to enable removal of detector defects.  A `post-flash' of 12 electrons was included to minimize charge transfer efficiency (CTE) losses.

The WFC3/UVIS observations were reduced using the STScI {\tt calwf3} pipeline \citep[v3.4;][]{2012wfci.book.....D}, which includes CTE correction.  Photometry of \novak\ was subsequently performed using DOLPHOT \citep[v2.0\footnote{\url{http://americano.dolphinsim.com/dolphot}};][]{2000PASP..112.1383D} employing the standard WFC3/UVIS parameters as quoted in the accompanying manual.  The resultant photometry is reported in Table~\ref{hst_photometry}, a full description of these {\it HST} data and their analysis will be reported in a follow-up paper.

\begin{table*}
\caption{{\it Hubble Space Telescope} Photometry of the 2016 Eruption of \novak.\label{hst_photometry}}
\begin{center}
\begin{tabular}{llllllll}
\hline
Date & $\Delta t$\tablenotemark{\dag} & \multicolumn{2}{c}{MJD 57,000+} & Exposure & Filter & S/N\tablenotemark{\ddag} & Photometry \\
(UT) & (days) & Start & End & time (s) \\
\hline
2016-12-08.014 & \toe{730.014} & 729.971 & 730.058 & $3\times898$ & F657N & \phn19.7 & $23.143\pm0.055$ \\
2016-12-09.312 & \toe{731.312} & 731.295 & 731.329 & $3\times898$ & F657N & \phn14.5 & $23.500\pm0.075$ \\
2016-12-10.305 & \toe{732.305} & 732.288 & 732.322 & $3\times898$ & F657N & \phn16.8 & $23.421\pm0.065$ \\
2016-12-11.060 & \toe{733.060} & 733.016 & 733.104 & $3\times898$ & F657N & \phn17.8 & $23.327\pm0.061$ \\
2016-12-17.081 & \toe{739.081} & 739.043 & 739.118 & $3\times898$ & F657N & 165.3 & $19.348\pm0.007$\tablenotemark{a} \\
\hline
2016-12-08.140 & \toe{730.140} & 730.102 & 730.179 & $3\times935$ & F645N & \phn13.4 & $23.591\pm0.081$ \\
2016-12-09.378 & \toe{731.378} & 731.360 & 731.396 & $3\times935$ & F645N & \phn11.3 & $23.806\pm0.096$ \\
2016-12-10.371 & \toe{732.371} & 732.353 & 732.389 & $3\times935$ & F645N & \phn12.5 & $23.589\pm0.087$ \\
2016-12-11.186 & \toe{733.186} & 733.148 & 733.225 & $3\times935$ & F645N & \phn15.5 & $23.413\pm0.070$ \\
2016-12-17.159 & \toe{739.159} & 739.120 & 739.197 & $3\times935$ & F645N & \phn85.0 & $20.488\pm0.013$\tablenotemark{a}\\
\hline
\end{tabular}
\end{center}
\tablenotetext{\dag}{The time since eruption assumes an eruption date of 2016 December 12.32\,UT.}
\tablenotetext{\ddag}{Signal-to-noise ratio.}
\tablerefs{(a)~\citet{2016ATel.9874....1D}.}
\end{table*}

\subsubsection{Ground-Based Photometry}

Data from each contributing telescope were reduced following the standard procedures for those facilities, full details for those previously employed in observations of \novak\ are presented in the Appendix of \othreek. For all the new facilities successfully taking data in this campaign we provide detailed information in Appendix~\ref{app:optical_photometry}. Photometry was also carried out in a similar manner to that reported in \othreek, using the identified secondary standards as presented in \othreek\ (see their Table\,10).

Preliminary photometry from several instruments was first published by the following authors as the optical light curve was evolving: \citet{2016ATel.9848....1I}, \citet{2016ATel.9857....1E},  \citet{2016ATel.9861....1B}, \citet{2016ATel.9864....1S}, \citet{2016ATel.9874....1D}, \citet{2016ATel.9881....1K}, \citet{2016ATel.9883....1H}, \citet{2016ATel.9885....1T}, \citet{2016ATel.9891....1N}, \citet{2016ATel.9906....1D}, and \citet{2016ATel.9910....1D}.  All photometry from the 2016 eruption of \novak\ is provided in Table~\ref{optical_photometry_table}.

\subsection{Visible Spectroscopy}\label{optical_spectroscopy} 

The spectroscopic confirmation of the 2016 eruption of \novak\ was announced by \citet{2016ATel.9852....1D}, with  additional spectroscopic follow-up reported in \citet{2016ATel.9865....1P}. A summary of all optical spectra of the 2016 eruption of \novak\ is shown in Table~{\ref{tab:spec}}, all the spectra are reproduced in Figure~\ref{specall}.

We obtained several spectra of the 2016 eruption with SPRAT \citep{2014SPIE.9147E..8HP}, the low-resolution, high-throughput spectrograph on the LT. SPRAT covers the wavelength range of $4000-8000$\,\AA\ and uses a $1^{\prime\prime}\!\!.8$ slit, giving a resolution of $\sim$18\,\AA. We obtained our spectra using the blue-optimized mode. The data were reduced using a combination of the LT SPRAT reduction pipeline and standard routines in IRAF\footnote{IRAF is distributed by the National Optical Astronomy Observatory, which is operated by the Association of Universities for Research in Astronomy (AURA) under a cooperative agreement with the National Science Foundation.} \citep{1993ASPC...52..173T}. The spectra were calibrated using previous observations of the standard star G191-B2B against data from \citet{1990AJ.....99.1621O} obtained via ESO. Conditions on La Palma were poor during the time frame the nova was accessible with SPRAT during the 2016 eruption, so the absolute flux levels are possibly unreliable.

We obtained an early spectrum of the nova, 0.54\,days after eruption, using the Andaluc\'\i a Faint Object Spectrograph and Camera (ALFOSC) on the 2.5\,m Nordic Optical Telescope (NOT) at the Roque de los Muchachos Observatory on La Palma. Grism \#7 and a slit width of $1^{\prime\prime}\!\!.3$ yielded a spectral resolution of 8.5\,\AA\ at the centre of the useful wavelength range $4000-7070$\,\AA\ ($R \sim 650$). The 1500\,s spectrum was imaged on the $2048 \times 2048$ pixel CCD \#14 with binning $2 \times 2$. We performed the observation under poor seeing conditions ($\sim 2^{\prime\prime}\!\!.5$). We reduced the raw images using standard IRAF procedures, and then did an optical extraction of the target spectrum with {\sc starlink}/{\sc pamela} \citep{1989PASP..101.1032M}. The pixel-to-wavelength solution was computed by comparison with 25 emission lines of the spectrum of a HeNe arc lamp. We used a 4th-order polynomial that provided residuals with an rms more than 10 times smaller than the spectral dispersion.

In addition, 1.87 days after eruption, we obtained a spectrum of \novak\ using the blue channel of the 10\,m Hobby Eberly Telescope's (HET) new integral-field Low Resolution Spectrograph \citep[LRS2-B;][]{2014SPIE.9147E..0AC,2016SPIE.9908E..4CC}.  This dual-beam instrument uses 280 fibers and a lenslet array to produce spectra with a resolution of $R \sim 1910$ between the wavelengths 3700 and 4700\,\AA,  and $R \sim 1140$ between 4600 and 7000\,\AA\ over a $12^{\prime\prime} \times 6^{\prime\prime}$ region of sky. The seeing for our observations was relatively poor ($1\farcs 8$), and the total exposure time was 30 minutes, split into 3 ten-minute exposures.  

Reduction of the LRS2-B data was accomplished using Panacea\footnote{\url{https://github.com/grzeimann/Panacea}}, a general-purpose IFU reduction package built for HET. After performing the initial CCD reductions (overscan removal and bias subtraction), we derived the wavelength solution, trace model, and spatial profile of each fiber using data from twilight sky exposures taken at the beginning of the night.  From these models, we extracted each fiber's spectrum and rectified the wavelength to a common grid.  Finally, at each wavelength in the grid, we fit a second order polynomial to the M31's background starlight and subtracted that from the gaussian-shaped point-source assumed for the nova.  

Two epochs of spectra were obtained using the Himalayan Faint Object Spectrograph and Camera (HFOSC) mounted on the 2\,m Himalayan Chandra Telescope (HCT) located at Indian Astronomical Observatory (IAO), Hanle, India. HFOSC is equipped with a 2k$\times$4k E2V CCD with pixel size of $15\times15$\,$\mu$m. Spectra were obtained in the wavelength range $3800-8000$\,\AA\ on 2016 December 13.61 and 14.55\,UT. The spectroscopic data were bias subtracted and flat field corrected and extracted using the optimal extraction method. An FeAr arc lamp spectrum was used for wavelength calibration. The spectrophotometric standard star Feige\,34 was used to obtain the instrumental response for flux calibration.

Three spectra were obtained with the 3.5\,m Astrophysical Research Consortium (ARC) telescope at the Apache Point Observatory (APO), during the first half of the night on 2016 December 12, 13, and 17 (UT December 13, 14, and 18). We observed with the Dual Imaging Spectrograph (DIS):\ a medium dispersion long slit spectrograph with separate collimators for the red and blue part of the spectrum and two 2048$\times$1028 E2V CCD cameras, with the transition wavelength around 5350\,\AA.  For the blue branch, a 400 lines mm$^{-1}$ grating was used, while the red branch was equipped with a 300 lines mm$^{-1}$ grating. The nominal dispersions were 1.83 and 2.31\,\AA\,pixel$^{-1}$, respectively, with central wavelengths at 4500 and 7500\,\AA. The wavelength regions actually used were 3500--5400\,\AA\ and 5300--9900\,\AA\ for blue and red, respectively. A $1^{\prime\prime}\!\!.5$ slit was employed. Exposure times were 2700\,s.  At least three exposures were obtained per night. Each on-target series of exposures was followed by a comparison lamp exposure (HeNeAr) for wavelength calibration. A spectrum of a spectrophotometric flux standard (BD+28\,4211) was also acquired during each night, along with bias and flat field calibration exposures. The spectra were reduced using Python scripts to perform standard flat field and bias corrections to the 2-D spectral images. Extraction traces and sky regions were then defined interactively on the standard star and object spectral images.  Wavelength calibration was determined using lines identified on the extracted HeNeAr spectra. We then determined the solution by fitting a 3rd order polynomial to these measured wavelengths.  Flux calibration was determined by measuring the ratio of the star fluxes to the known fluxes as a function of wavelength.  We performed these calibrations independently for the red and blue spectra, so that the clear agreement in the overlapping regions of the wavelength ranges confirms that our calibration and reduction procedure was successful.

\begin{table}
\caption{Summary of the Optical Spectra of the 2016 Eruption of \novak.\label{tab:spec}}
\begin{center}
\begin{tabular}{llll}
\hline
Date (UT)&$\Delta t$ &Instrument & Exposure\\
2017 Dec & (days) & \& Telescope & time (s)\\
\hline
12.86 &0.54$\pm$0.01 &ALFOSC/NOT & $1\times1500$\\
12.93 &0.61$\pm$0.06 &SPRAT/LT & $6\times\phn900$ \\
13.14 &0.82$\pm$0.11 &DIS/ARC &\\
13.61 &1.29$\pm$0.02 &HFOSC/HCT & $1\times3600$\\
13.98 &1.66$\pm$0.07 &SPRAT/LT & $6\times\phn900$ \\
14.12 &1.80$\pm$0.08 &DIS/ARC &\\
14.19 &1.87$\pm$0.02 &LRS2-B/HET &$3\times\phn600$\\
14.55 &2.23$\pm$0.02 &HFOSC/HCT & $1\times2700$\\
14.90 &2.58$\pm$0.05 &SPRAT/LT & $6\times\phn900$ \\
15.91 &3.59$\pm$0.02 &SPRAT/LT & $3\times\phn900$ \\
16.85 &4.53$\pm$0.02 &SPRAT/LT & $3\times\phn900$ \\
18.15 &5.83$\pm$0.05 &DIS/ARC &\\
\hline
\end{tabular}
\end{center}
\tablecomments{The time since eruption assumes an eruption date of 2016 December 12.32\,UT. The error bars do not include the systematic error in this eruption date, but represent the total exposure time/time between combined exposures of a given epoch.}
\end{table}

\subsection{X-ray and UV observations}\label{swift_data}

A Neil Gehrels \swift Observatory \citep{2004ApJ...611.1005G} target of opportunity (ToO) request was submitted immediately after confirming the eruption and the satellite began observing the nova on 2016-12-12.65 UT \citep[cf.][]{2016ATel.9853....1H}, only four hours after the optical discovery. All \swift observations are summarized in Table~\ref{tab:swift}. The \swift target ID of \nova\ is always 32613. Because of the low-Earth orbit of the satellite, a \swift observation is normally split into several snapshots, which we list separately in Table~\ref{tab:swift_split}.

\begin{table*}
\caption{\swift Observations of \novak\ for the 2016 Eruption.}
\label{tab:swift}
\begin{center}
\begin{tabular}{rrrrrrrrrr}\hline\hline \noalign{\smallskip}
ObsID & Exp$^a$ & Date$^b$ & MJD$^b$ & $\Delta t^c$ & uvw2$^d$ & XRT Rate$^e$ \\
& (ks) & (UT) & (d) & (d) & (mag) & (\power{-2}\,ct\,s$^{-1}$)\\ \hline \noalign{\smallskip}
00032613183 & 3.97 & 2016-12-12.65 & 57734.65 & 0.33 & $16.7\pm0.1$ & $<0.3$ \\
00032613184 & 4.13 & 2016-12-13.19 & 57735.19 & 0.87 & $17.3\pm0.1$ & $<0.2$ \\
00032613185 & 3.70 & 2016-12-14.25 & 57736.26 & 1.94 & $17.9\pm0.1$ & $<0.3$ \\
00032613186 & 3.23 & 2016-12-15.65 & 57737.65 & 3.33 & $18.6\pm0.1$ & $<0.4$ \\
00032613188 & 1.10 & 2016-12-16.38 & 57738.38 & 4.06 & $18.7\pm0.1$ & $<0.7$ \\
00032613189 & 3.86 & 2016-12-18.10 & 57740.10 & 5.78 & $19.3\pm0.1$ & $0.6\pm0.1$ \\
00032613190 & 4.03 & 2016-12-19.49 & 57741.50 & 7.18 & $20.0\pm0.2$ & $0.4\pm0.1$ \\
00032613191 & 2.02 & 2016-12-20.88 & 57742.89 & 8.57 & $20.6\pm0.3$ & $1.9\pm0.3$ \\
00032613192 & 3.95 & 2016-12-21.49 & 57743.49 & 9.17 & $20.9\pm0.3$ & $1.5\pm0.2$ \\
00032613193 & 2.53 & 2016-12-22.68 & 57744.69 & 10.37 & $20.4\pm0.2$ & $1.7\pm0.3$ \\
00032613194 & 2.95 & 2016-12-23.67 & 57745.68 & 11.36 & $20.8\pm0.3$ & $1.4\pm0.2$ \\
00032613195 & 2.90 & 2016-12-24.00 & 57746.01 & 11.69 & $20.5\pm0.2$ & $0.7\pm0.2$ \\
00032613196 & 2.73 & 2016-12-25.00 & 57747.01 & 12.69 & $>21.1$ & $0.6\pm0.2$ \\
00032613197 & 2.71 & 2016-12-26.20 & 57748.20 & 13.88 & $>21.1$ & $0.3\pm0.2$ \\
00032613198 & 2.84 & 2016-12-27.72 & 57749.73 & 15.41 & $>21.1$ & $<0.5$ \\
00032613199 & 3.23 & 2016-12-28.19 & 57750.19 & 15.87 & $>21.2$ & $<0.4$ \\
00032613200 & 2.65 & 2016-12-29.45 & 57751.46 & 17.14 & $>21.1$ & $<0.5$ \\
00032613201 & 3.05 & 2016-12-30.05 & 57752.05 & 17.73 & $>20.9$ & $<0.4$ \\
00032613202 & 2.88 & 2016-12-31.58 & 57753.58 & 19.26 & $>21.1$ & $<0.3$ \\\hline
\end{tabular}
\end{center}
\noindent
\tablenotetext{a}{Exposure time includes dead-time corrections.}
\tablenotetext{b}{Observation start date.}
\tablenotetext{c}{Time in days after the eruption date on 2016-12-12.32 UT (MJD 57734.32)}
\tablenotetext{d}{The \swift UVOT uvw2 filter has a central wavelength of 1930\,\AA\ with a FWHM of about 660\,\AA.}
\tablenotetext{e}{Count rates are measured in the 0.3--1.5 keV range.}
\end{table*}

\begin{table*}
\caption{Stacked \swift UVOT Observations and Photometry as Plotted in Figure~\ref{fig:uvot_lc}.\label{tab:uvot_merge}}
\begin{center}
\begin{tabular}{lllllll}
\hline
{ObsIDs$^a$} & {Exp$^b$} & {Date$^c$} & {MJD$^c$} & {$\Delta t^c$} & {Length$^d$} & {uvw2} \\
& {(ks)} & {(UT)} & {(d)} & {(d)} & {(d)} & {(mag)} \\
\hline
00032613196/198 & 8.3 & 2016-12-26.37 & 57748.37 & 14.05 & 2.72 & $21.7\pm0.4$ \\
00032613199/200 & 5.9 & 2016-12-28.83 & 57750.83 & 16.51 & 1.27 & $<21.5$\\
\hline
\end{tabular}
\end{center}
\tablenotetext{a}{Start/End observation for each stack (cf.\ Table~\ref{tab:swift})}
\tablenotetext{b}{Summed up exposure.}
\tablenotetext{c}{Time between the eruption date (MJD 57734.32; cf.\ Section~\ref{sec:time}) and the stack midpoint.}
\tablenotetext{d}{Time in days from the first observation of the stack to the last one.}
\end{table*}

In addition, we triggered a 100\,ks \xmm \citep{2001A&A...365L...1J} ToO that was originally aimed at obtaining a high-resolution X-ray spectrum of the SSS variability phase. Due to the inconvenient eruption date, 14 days before the \xmm window opened, and the surprisingly fast light curve evolution, discussed in detail below, only low resolution spectra and light curves could be obtained. The \xmm object ID is 078400. The ToO was split into two observations which are summarized in Table~\ref{tab:xmm}. Since 2008, no eruption of \nova\ had occurred within one of the relatively narrow \xmm visibility windows from late December to mid February and July to mid August (cf.\ Table~\ref{eruption_history}).

The \swift UV/optical telescope \citep[UVOT,][]{2005SSRv..120...95R} magnitudes were obtained via the HEASoft (v6.18) tool \texttt{uvotsource}; based on aperture photometry of carefully selected source and background regions. We stacked individual images using \texttt{uvotimsum}. In contrast to previous years, our 2016 coverage exclusively used the uvw2 filter which has a central wavelength of 1930\,\AA. The photometric calibration assumes the UVOT photometric (Vega) system \citep{2008MNRAS.383..627P,2011AIPC.1358..373B} and has not been corrected for extinction. 

In the case of the \swift X-ray telescope \citep[XRT;][]{2005SSRv..120..165B} data we used the on-line software\footnote{\url{http://www.swift.ac.uk/user\_objects}} of \citet{2009MNRAS.397.1177E} to extract count rates and upper limits for each observation and snapshot, respectively. Following the recommendation for SSSs, we extracted only grade-zero events. The on-line software uses the Bayesian formalism of \citet{1991ApJ...374..344K} to estimate upper limits for low numbers of counts. All XRT observations were taken in the photon counting (PC) mode.

The \xmm X-ray data were obtained with the thin filter for the pn and MOS detectors of the European Photon Imaging Camera \citep[EPIC;][]{2001A&A...365L..18S,2001A&A...365L..27T}. They were processed with XMM-SAS (v15.0.0) starting from the observation data files (ODF) and using the most recent current calibration files (CCF). We used \texttt{evselect} to extract spectral counts and light curves from source and background regions that were defined by eye on the event files from the individual detectors. We filtered the event list by extracting a background light curve in the 0.2--0.7\,keV range (optimized after extracting the first spectra, see Section~\ref{sec:xmm_spec}) and removing the episodes of flaring activity.

In addition, we obtained UV data using the \xmm optical/UV monitor telescope \citep[OM;][]{2001A&A...365L..36M}. All OM exposures were taken with the uvw1 filter, which has a slightly different but comparable throughput as the \swift UVOT filter of the same name \citep[cf.][]{2005SSRv..120...95R}. The central wavelength of the OM uvw1 filter is 2910\,\AA\ with a width of 830\,\AA\ \citep[cf.\ UVOT uvw1:\ central wavelength 2600\,\AA, width 693\,\AA; see][]{2008MNRAS.383..627P}. We estimated the magnitude of \nova\ in both observations via carefully selected source and background regions, which were based on the \swift UVOT apertures. Our estimates include (small) coincidence corrections and a PSF curve-of-growth correction. The latter became necessary because the size of the source region needed to be restricted to avoid contamination by neighboring sources. The count rate and uncertainties were converted to magnitudes using the CCF zero points.

As in previous papers on this object (\xonek, \xtwok, \othreek), the X-ray spectral fitting was performed in \texttt{XSPEC} \citep[v12.8.2;][]{1996ASPC..101...17A} using the T\"ubingen-Boulder ISM absorption model (\texttt{TBabs} in \texttt{XSPEC}) and the photoelectric absorption cross-sections from \citet{1992ApJ...400..699B}. We assumed the ISM abundances from \citet{2000ApJ...542..914W} and applied Poisson likelihood ratio statistics \citep{1979ApJ...228..939C}.

\begin{table*}
\caption{\xmm Observations of \novak\ in 2016.}
\label{tab:xmm}
\begin{center}
\begin{tabular}{lrrrrrrrr}\hline\hline \noalign{\smallskip}
 ObsID & Exp$^a$ & GTI$^b$ & MJD$^c$ & $\Delta t^d$ & uvw1$^e$ & EPIC Rate & Equivalent XRT Rate$^f$\\
 & (ks) & (ks) & (UT) & (d) & (mag) & (\power{-2}\,ct\,s$^{-1}$) & (\power{-4}\,ct\,s$^{-1}$)\\ \hline \noalign{\smallskip}
 0784000101 & 33.5 & 16.1 & 57748.533 & 14.21 & $21.6^{+0.3}_{-0.2}$ & $1.9\pm0.2$ & $7.3\pm0.6$ \\
 0784000201 & 63.0 & 40.0 & 57750.117 & 15.80 & $21.6\pm0.2$ & $1.0\pm0.1$ & $3.3\pm0.2$ \\
\hline
\end{tabular}
\end{center}
\noindent
\tablenotetext{a}{Dead-time corrected exposure time for \xmm EPIC pn prior to GTI filtering for high background.}
\tablenotetext{b}{Exposure time for \xmm EPIC pn after GTI filtering for high background.}
\tablenotetext{c}{Start date of the observation.}
\tablenotetext{d}{Time in days after the eruption of nova \nova in the optical on 2016-12-12.32 UT \citep[MJD = 57734.32; see][]{2016ATel.9848....1I}}
\tablenotetext{e}{The OM filter was uvw1 (central wavelength 2910\,\AA\ with a width of 830\,\AA.)}
\tablenotetext{f}{Theoretical \swift XRT count rate (0.3--10.0 keV) extrapolated based on the 0.2--1.0\,keV EPIC pn count rates, in the previous column, and assuming the best-fit blackbody spectrum and foreground absorption.}
\end{table*}

\section{Panchromatic eruption light curve (visible to soft X-ray)}\label{sec:vis_lc}

\subsection{Detection and time of the eruption}\label{sec:time}

With a nova that evolves as rapidly as \novak, early detection of each eruption is crucial.  Following the successful eruption detection campaigns for the 2014 and 2015 outbursts, in 2016 we grew our large, multi-facility monitoring campaign into a global collaboration. The professional telescopes at the LT, Las Cumbres \citep[LCO;][the 2\,m at Haleakala, Hawai'i, the 1\,m at McDonald, Texas]{2013PASP..125.1031B}, and Ond\v{r}ejov Observatory, were joined by a network of highly motivated and experienced amateur observers in Canada, China, Finland, Germany, Italy, Japan, the United Kingdom, and the United States. A large part of their effort was coordinated through the AAVSO and VSOLJ, respectively (see Appendix~\ref{app:optical_photometry} for details). The persistence of the amateur observers in our team, during 6 suspenseful months of monitoring, allowed us to discover the eruption at an earlier stage than in previous years.

The 2016 eruption of \novak\ was first detected on 2016 December 12.4874 (UT) by the 0.5\,m f/6 telescope at the Itagaki Astronomical Observatory in Japan at an unfiltered magnitude of 18.2 \citep{2016Ita}. The previous non-detection took place at the LCO 1\,m (McDonald) just 0.337\,days earlier, providing an upper limit of $r'>19.1$.  A deeper upper limit of $u'>22.2$ was provided by the LT and its automated real-time alert system \citep[see][]{2007ApJ...661L..45D} 0.584\,days pre-detection.  The 2016 eruption was spectroscopically confirmed almost simultaneously by the NOT and LT, 0.37 and 0.39\,days post-detection, respectively \citep{2016ATel.9852....1D}.

All subsequent analysis assumes that the 2016 eruption of nova \nova\ ($\Delta t=0$) occurred on 2016-12-12.32 UT ($\mathrm{MJD}= 57734.32$). This date is defined as the midpoint between the last upper limit (2016-12-12.15 UT; LCO) and the discovery observation (2016-12-12.49 UT; Itagaki observatory), as first reported by \citet{2016ATel.9848....1I}. The corresponding uncertainty on the eruption date is $\pm0.17$\,d. The corresponding dates of the 2013, 2014, and 2015 eruptions, to which we will compare our new results, are listed in Table~\ref{eruption_history}.

\subsection{Pre-eruption evolution?}

\begin{figure}
\includegraphics[width=\columnwidth]{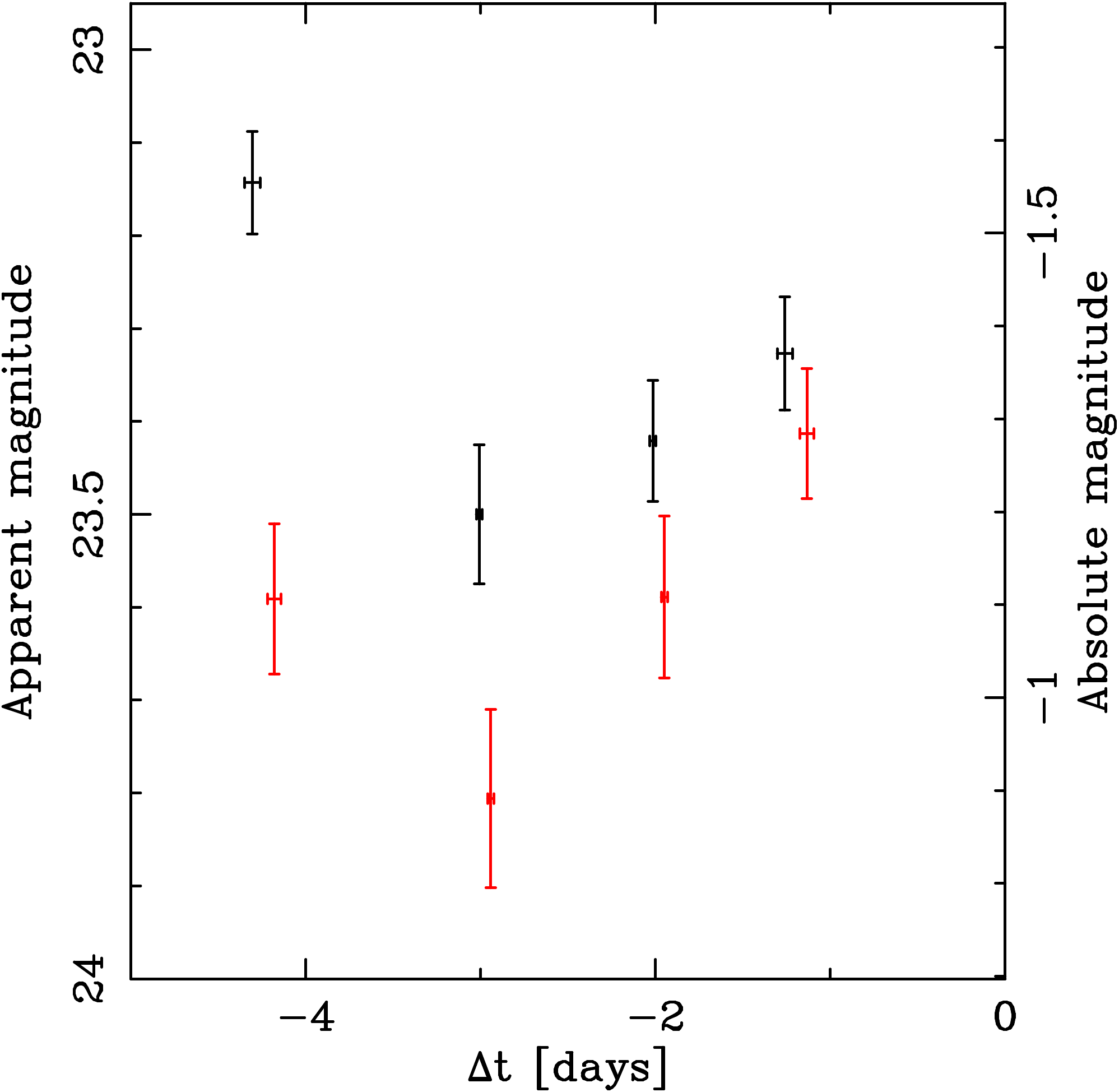}
\caption{{\it Hubble Space Telescope} WFC3/UVIS narrow-band photometry of \novak\ over the five days before the onset of the 2016 eruption.  Red points: F645N ``continuum'' photometry; black points: F657N ``H$\alpha$+[N\,{\sc ii}]'' photometry.  The absolute magnitude assumes a distance to M\,31 of 770\,kpc \citep{1990ApJ...365..186F} and reddening toward \novak\ of $E_{B-V}=0.1$ (\hstspec). \label{hst_pre}}
\end{figure}

The {\it HST} photometry serendipitously obtained over the five day pre-eruption period is shown in Figure~\ref{hst_pre}.  The H$\alpha$ photometry is shown by the black points and the narrow-band continuum by the red.  Clear variability is seen during this pre-eruption phase.  As this variability appears in both H$\alpha$ and the continuum it is possible that it is continuum driven.  The system has a clear H$\alpha$ excess immediately before eruption, but the H$\alpha$ excess appears to diminish as the continuum rises.  Following the discussion presented in \hstphot, it is possible that such H$\alpha$ emission arrises from the \novak\ accretion disk, which may be generating a significant disk wind.

The continuum flux during this period is broadly consistent with the quiescent luminosity of the system (see \hstphot).  Therefore, it is unclear whether this behavior is a genuine pre-eruption phenomenon, or related to variability at quiescence with a characteristic time scale of up to a few days, with possible causes being accretion disk flickering, or even orbital modulation.  Through constraining the mass donor, \hstphot\ indicated that the orbital period for the \novak\ binary should be $\gtrsim5$\,days.  Such variation, as shown in Figure~\ref{hst_pre} would not be inconsistent with that constraint.

\subsection{Visible and ultraviolet light curve}
\label{sec:vis_lc_vis}

\begin{figure*}
\includegraphics[width=\textwidth]{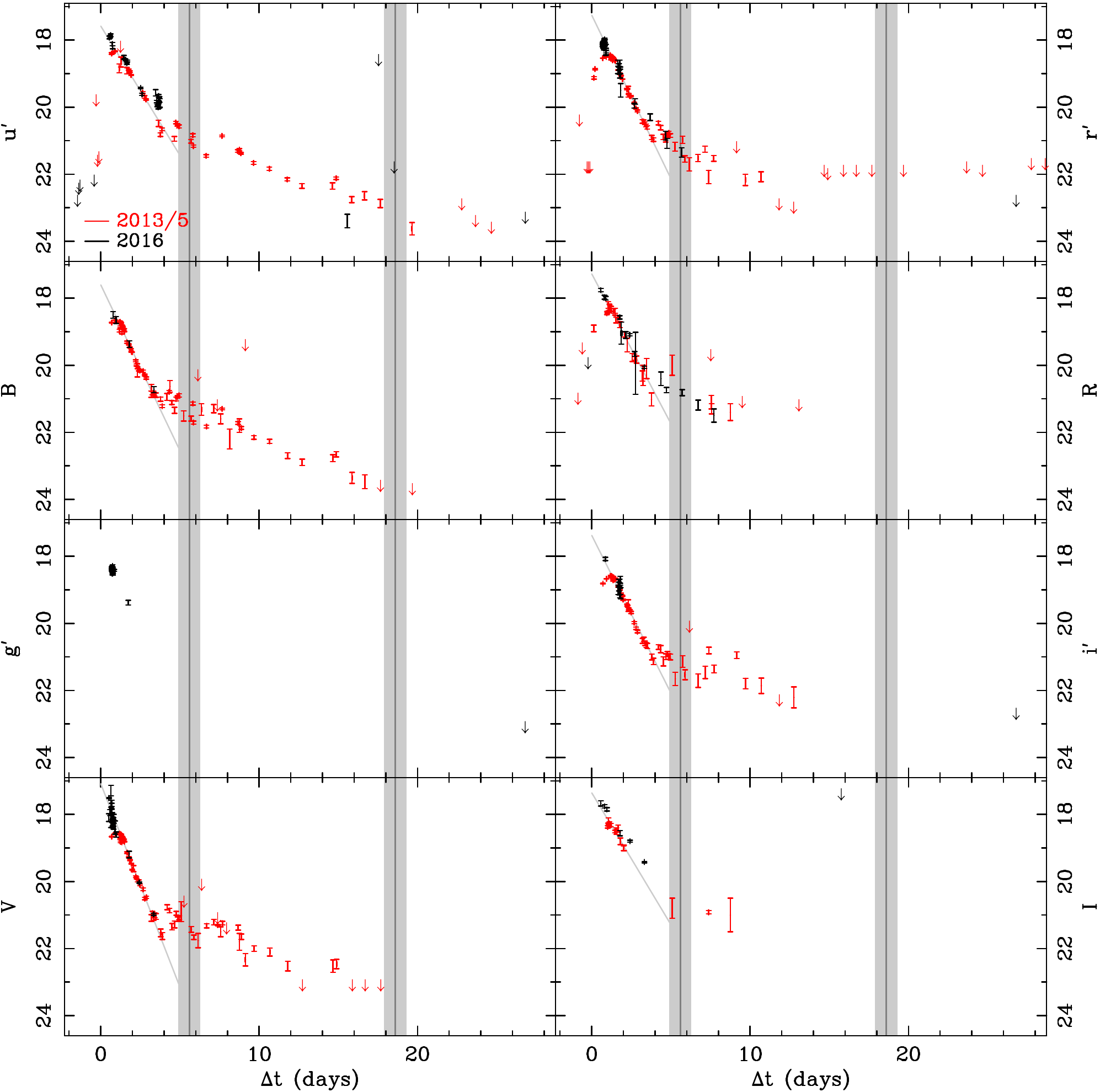}
\caption{Visible photometry of the past four eruptions of \novak. Black points show the 2016 data (see Table~\ref{optical_photometry_table}). The red points indicate combined data from the 2013--2015 eruptions (\oonek, \otwok, \othreek, and \ponek). We show the SSS turn-on/off times of the \textit{2015} eruption as vertical gray lines, with their uncertainties marked by the shaded areas. For the 2013--2015 light curves combined, the inclined gray lines indicate an exponential decay in luminosity during the range of $1\leq\Delta t\leq4$ days (\othreek).\label{optical_lc}}
\end{figure*}

\begin{figure*}
\includegraphics[width=\textwidth]{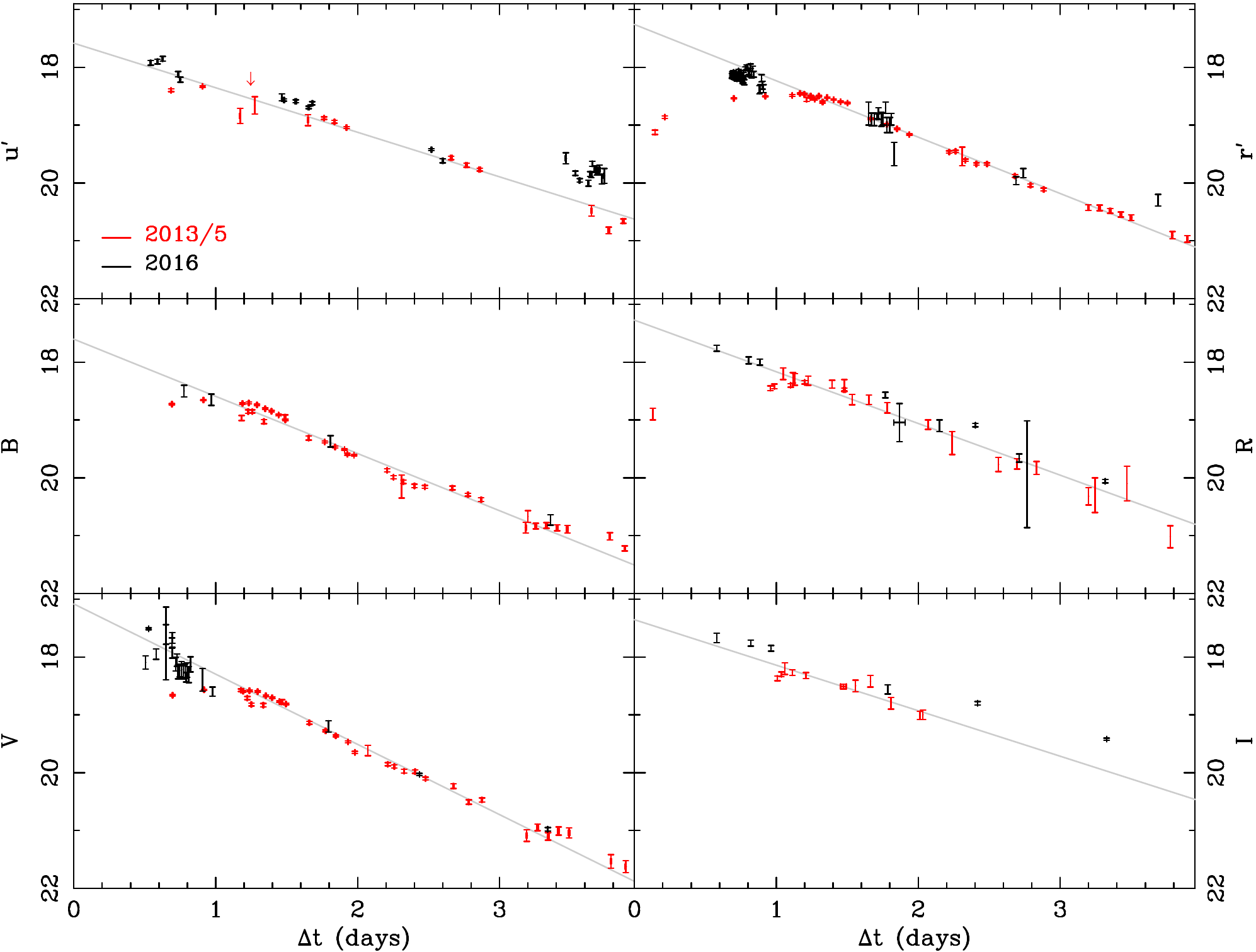}
\caption{As Figure~\ref{optical_lc}, but focusing on $0\leq t\leq4$\,days. The $i'$-band data are excluded as there were no discrepancies between the very limited 2016 $i'$ dataset and the extensive dataset from 2013--2015, $g'$-band data were excluded as no pre-2016 data exist.\label{optical_zoom}}
\end{figure*}

Following the 2015 eruption, \othreek\ noted that the 2013, 2014, and 2015 eruption light curves were remarkably similar spanning from the $I$-band to the near-UV (redder pass-bands only have data from 2015), see red data points in Figure~\ref{optical_lc}.  Based on those observations, \othreek\ defined four phases of the light curve: {\it the final rise (Day 0--1)} is a regime sparsely populated with data due to the rapid increase to maximum light; {\it the initial decline (Day 1--4)} where a exponential decline in flux (linear in magnitude) is observed from the NUV to the near-infrared (see, in particular, the red data points in Figure~\ref{optical_zoom}; {\it the plateau (Day 4--8)} a relatively flat, but jittery, region of the light curve which is time coincident with the SSS onset; and {\it the final decline (Day $>8$)} where a power-law (in flux) decline may be present.

The combined 2013--2015 light curve defined these four phases, the individual light curves from each of those eruptions were also consistent with those patterns (see Figures~\ref{optical_lc} and \ref{optical_zoom}). A time-resolved SED of the well-covered 2015 eruption was presented by \othreek. Unfortunately, due to severe weather constraints our 2016 campaign did not obtain sufficient simultaneous multi-filter data to compare the SED evolution. However, we find that the 2015 and 2016 light curves are largely consistent (Figure~\ref{optical_lc}) except for the surprising features we will present in the following text.

First, we look at the initial decline phase for the 2016 eruption.  We examine this region of the light curve first as, in previous eruptions, it has shown the simplest evolution -- a linear decline -- which was used by \othreek\ to tie together the epochs of the 2013, 2014, and 2015 eruptions.  But, due to the poor conditions at many of the planned sites, the data here are admittedly sparse, but are generally consistent with the linear behavior seen in the past three eruptions.  There may however, be evidence for a deviation, approximately one magnitude upward, toward the end of this phase in the $u'$ and $r'$-band data at $t\gtrsim3.6$\,days post-eruption.

\begin{figure*}
\includegraphics[width=\columnwidth]{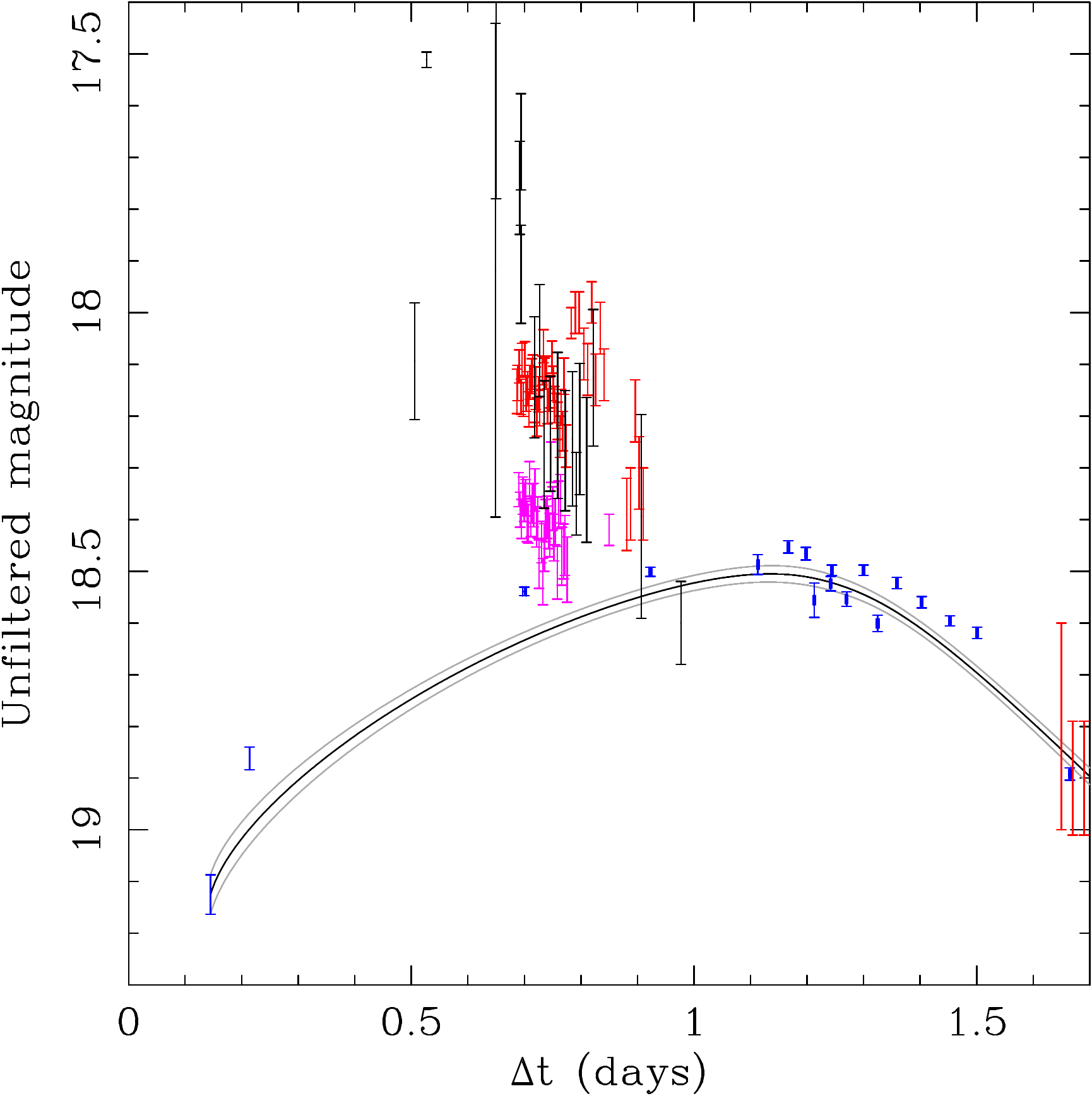}\hfill
\includegraphics[width=\columnwidth]{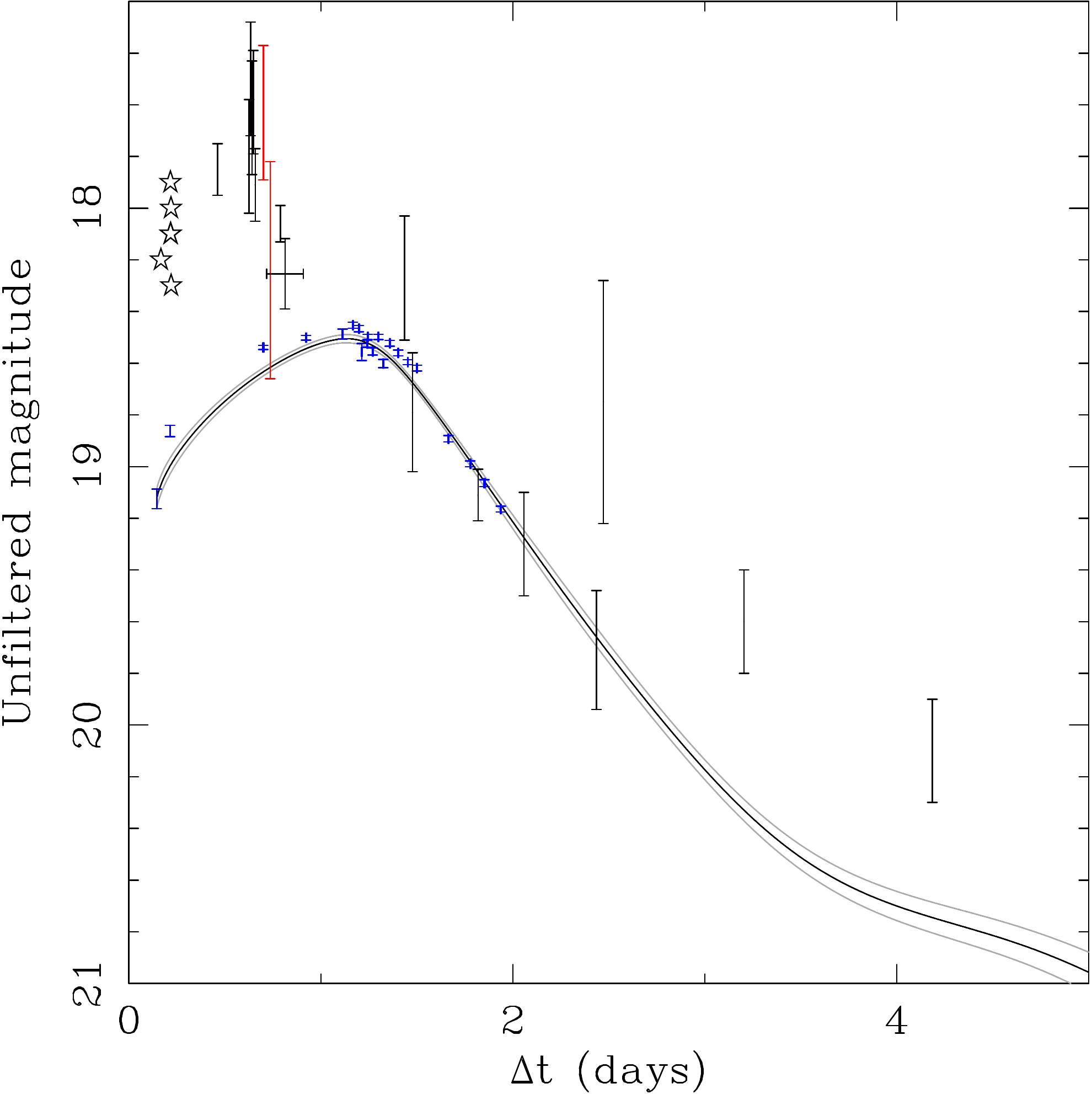}
\caption{Broad-band and unfiltered photometry of the \novak\ `cusp'. In both sub-plots, the blue points note the combined $r'$-band photometry from the 2013, 2014, and 2015 eruptions, with the solid line showing the template 2013--2015 $r'$-band light curve and associated uncertainties (see \othreek).  {\bf Left:} Broad-band photometry of the `cusp', of the 2016 eruption of \novak.  Red points $r'$-band, magenta points $g'$-band, and the black points are $V$-band.\label{fastphot} {\bf Right:} Here we show a comparison between the {\it unfiltered} photometry of the 2010 (red) and 2016 (black) eruptions of \novak, the black stars indicate photometry of the 2016 eruption with no computed uncertainties.\label{2010comp}}
\end{figure*}

However, the largest deviation from the 2013--2015 behavior occurs during the final rise phase, between $0\leq t\leq1$\,days.  There appears to be a short-lived, `cuspy' feature in the light curves seen through all filters (except the $B$-band where there was limited coverage) and the unfiltered observations (see Figures~\ref{optical_lc}, \ref{optical_zoom}, and \ref{fastphot}, which progressively focus on the `cusp').  The variation between the peak luminosity of the 2013--2015 eruptions and the 2016 eruption is shown in Table~\ref{max_deviation}, in all useful bands the deviation was significant.  The average (across all bands) increase in maximum magnitude was 0.64\,mag, or almost twice as luminous as the 2013--2015 eruptions at peak.  Notably, this over-luminous peak occurred much earlier than the 2013--2015 peaks.  The mean time of peak in 2013--2015 was $t\simeq1.0$\,days (across the $u'$, $B$, $R$, $r'$, and $I$ filters), whereas the bright cusp in 2016 occurred at $t\simeq0.65$\,days.  

\begin{table}
\caption{Comparison Between the Maximum Observed Magnitudes from the 2013--2015 and 2016 Eruptions of \novak.\label{max_deviation}}
\begin{center}
\begin{tabular}{llll}
\hline
Filter & \multicolumn{2}{c}{$m_\mathrm{max}$ (mag)} & `$\Delta m_\mathrm{max}$'\\
& 2013--2015\tablenotemark{a} & 2016\tablenotemark{b} & (mag)\\
\hline
$u'$ & $18.35\pm0.03$ & $17.85\pm0.04$ & $0.50\pm0.05$\\
$B$\tablenotemark{c} & $18.67\pm0.02$ & $18.50\pm0.10$ & $0.17\pm0.10$\\
$V$ & $18.55\pm0.01$ & $17.6$ & 1.0\\
$R$ & $18.38\pm0.02$ & $17.76\pm0.05$ & $0.62\pm0.05$ \\
$r'$ & $18.45\pm0.01$ & $17.98\pm0.04$ & $0.47\pm0.04$ \\
$I$ & $18.31\pm0.03$ & $17.68\pm0.08$ & $0.63\pm0.09$\\
\hline
\end{tabular}
\end{center}
\tablenotetext{a}{As calculated by \othreek, based on a fit to the combined 2013--2015 light curves.}
\tablenotetext{b}{The most luminous observation of the 2016 eruption, those without error bars are estimated maxima from multiple observations and observers.}
\tablenotetext{c}{The $B$-band coverage during the 2016 peak was limited.}
\end{table}

The INT and ERAU obtained a series of fast photometry of the 2016 eruption through $g'$, $i'$ (ERAU only), and $r'$-band filters during the final rise phase.  Figure~\ref{fastphot} (left) compares this photometry with the 2013--2015 $r'$-band eruption photometry.  This figure clearly illustrates the short-lived, bright, optical `cusp', but also its highly variable nature over a short time-scale with variation of up to 0.4\,mag occurring over just 90 minutes.  The $(g'-r')$ color during this period is consistent with the cusp light curve being achromatic.  We derive $(g'-r')_0=0.15\pm0.03$ for the cusp period, which is roughly consistent with the \novak\ color during the peak of the 2013--2015 eruptions \othreek.

The 2013--2015 eruptions exhibited a very smooth light curve evolution from, essentially, $t=0$ until $t\simeq4$\,days (see in particular the red $r'$-band light curve in Figure~\ref{optical_zoom}.  As well as never being seen before, the bright cusp appears to break this smooth evolution.  The 2016 eruption does not just appear more luminous than the observations of 2013--2015, there is evidence of a fundamental change, possibly in the emission mechanism, obscuration, or within the lines. 

There are sparse data covering both the plateau and final decline phases.   The $R$-band data from 2016 covers the entire plateau phase and is broadly consistent with the slow-jittery decline seen during this phase in the 2013--2015 eruptions.  The $u'$ and $r'$-band data show a departure from the linear early decline around day 3.6, this could indicate an early entry into the plateau, i.e.\ different behavior in 2016, or simply that the variation seen during the plateau always begins slightly earlier than the assumed 4\,day phase transition.

In essence, the 2016 light curves of \novak\ show a never before seen (but see Section~\ref{similar}), short-lived, bright cusp at all wavelengths during the final rise phase.  There is no further strong evidence of any deviation from previous eruptions -- however we again note the sparsity of the later-time data.  Possible explanations for the early bright light curve cusp are discussed in Section~\ref{sec:disc_peak} and ~\ref{cusp}, and Section~\ref{sec:disc_arch} re-examines earlier eruptions for possible indications of similar features.

\subsection{\swift and \xmm ultraviolet light curve}\label{sec:uvot_lc}

During the 2015 eruption we obtained a detailed \swift UVOT light curve through the uvw1 filter (\othreek). For the 2016 eruption our aim was to measure the uvw2 filter magnitudes instead to accumulate additional information on the broad-band SED evolution. With a central wavelength of 1930\,\AA\ the uvw2 band is the ``bluest'' UVOT filter (uvw1 central wavelength is 2600\,\AA). Therefore, the uvw1 range is more affected by spectral lines, for instance the prominent Mg\,{\sc ii} (2800\,\AA) resonance doublet, than the uvw2 magnitudes (see \hstspec\ for details). Due to the peculiar properties of the 2016 eruption, a direct comparison between both light curves is now more complex than initially expected.

\begin{figure}
\includegraphics[width=\columnwidth]{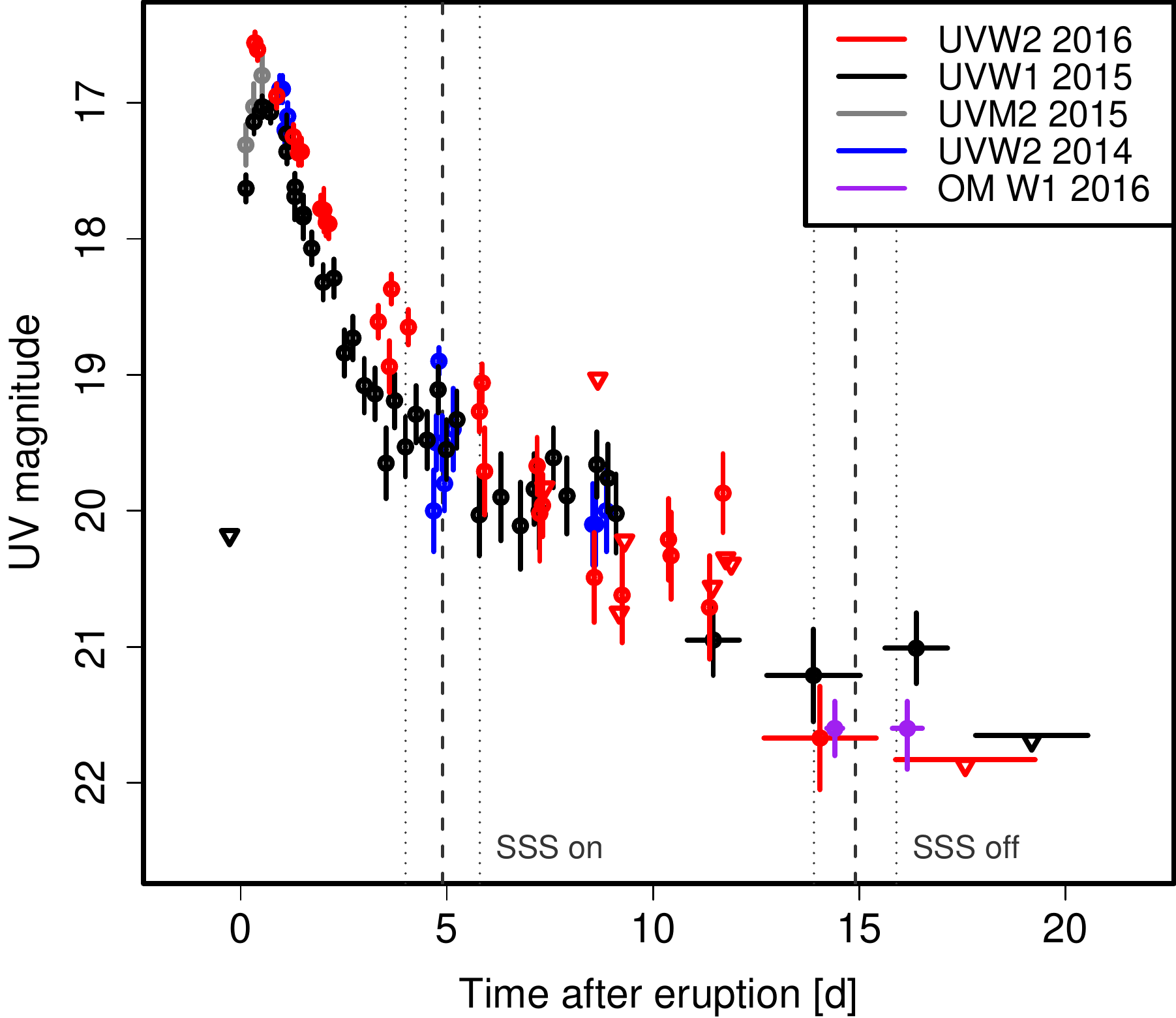}
\caption{\swift UVOT uvw2 light curve for the 2016 eruption of \nova\ (red) compared to (i) the detailed uvw1 coverage of the 2015 eruption (black; \othreek), (ii) a few uvm2 measurements around the 2015 peak (gray), (iii) the uvw2 magnitudes from the 2014 eruption (blue; \otwok, \xtwok), and (iv) the 2016 \xmm OM uvw1 magnitudes (cf.\ Table~\ref{tab:xmm}). The last two red data points were derived from stacking multiple images (see Table~\ref{tab:uvot_merge}). For better readability we only plot upper limits from individual observations until day 12 are plotted (cf. Tables~\ref{tab:swift} and \ref{tab:swift_split} for those). Uncertainties are the combined $1\sigma$ systematic and statistical values. Open triangles mark $3\sigma$ upper limits. Day zero is MJD = 57734.32 (see Section~\ref{sec:time}). The dark gray vertical lines indicate the SSS time scales (dashed) and their corresponding uncertainties (dotted) according to Section~\ref{sec:xrt_lc}.}
\label{fig:uvot_lc}
\end{figure}

In Figure~\ref{fig:uvot_lc} we show the 2016 uvw2 light curve compared to the 2015 uvw1 (plus a few uvm2) measurements (\othreek) as well as a few uvw2 magnitudes from the 2014 eruption (\xtwok, \otwok). The 2016 values are based on individual \swift snapshots (see Table~\ref{tab:swift_split}) except for the last two data points where we used stacked images (see Table~\ref{tab:uvot_merge}). Similarly to the uvw1 light curve in 2015, the uvw2 brightness initially declined linearly with a $t_2 = 2.8 \pm 0.2$\,d. This is comparable to the 2015 uvw1 value of $t_2 = 2.6\pm0.2$\,d.

From day three onward, the decline slowed down and became less monotonic. Viewed on its own, the UV light curve from this point onward would be consistent with a power-law decline (in flux) with an index of $-1.5\pm0.2$. However, in light of the well-covered 2015 eruption the 2016 light curve would also be consistent with the presence of three plateaus between (approximately) the days 3--5, 6--8, and 9--12; and with relatively sharp drops of about 1 mag connecting those. Around day 12, when the X-ray flux started to drop (cf.\ Figure~\ref{fig:xrt_xmm_lc}) there might even have been a brief rebrightening in the UV before it declined rapidly. The UV source had disappeared by day 16, which is noticeably earlier than in 2015 (in the uvw1 filter).  \hstphot\ presented evidence that the UV--optical flux is dominated by the surviving accretion disk from at least day 13 onward.  Therefore, a lower UV luminosity at this stage would imply a lower disk mass accretion rate. It is noteworthy that during the times where the 2014 and 2016 uvw2 measurements overlap they appear to be consistent.

The \xmm OM uvw1 magnitudes are given in Table~\ref{tab:xmm} and included in Figure~\ref{fig:uvot_lc}. The two OM measurements appear to be consistently fainter than the \swift UVOT uvw1 data at similar times during the 2015 eruption (cf.\ \othreek). However, the uncertainties are large and the filter response curves (and instruments) are not perfectly identical. Therefore, we do not consider this apparent difference to have any physical importance. In addition, there is a hint at variability in the uvw1 flux during the first \xmm observation. Of the seven individual OM exposures, the first five can be combined to a uvw1 = $21.3^{+0.3}_{-0.2}$\,mag whereas the last two give a $2\sigma$ upper limit of uvw1 $> 21.5$\,mag. The potential drop in UV flux corresponds to the drop in X-ray flux after the peak in Figure~\ref{fig:epic_lc}. Also here the significance of this fluctuation is low and we only mention it for completeness, in case similar effects will be observed in future eruption.

\subsection{\swift XRT light curve}
\label{sec:xrt_lc}
\smallskip
X-ray emission from \novak\ was first detected at a level of $0.6\pm0.1$\,\cts{-2} on 2016-12-18.101 UT, 5.8 days after the eruption \citep[see Table~\ref{tab:swift} and also][]{2016ATel.9872....1H}. Nothing was detected in the previous observation on 2016-12-16.38 UT (day 4.1) with an upper limit of $<0.7$ \cts{-2}. Although these numbers are comparable, there is a clear increase of counts at the nova position from the pre-detection observation (zero counts in 1.1 ks) to the detection (more than 30 counts in 3.9\,ks). Therefore, we conclude that the SSS phase had started by day 5.8.

\begin{figure}
\includegraphics[width=\columnwidth]{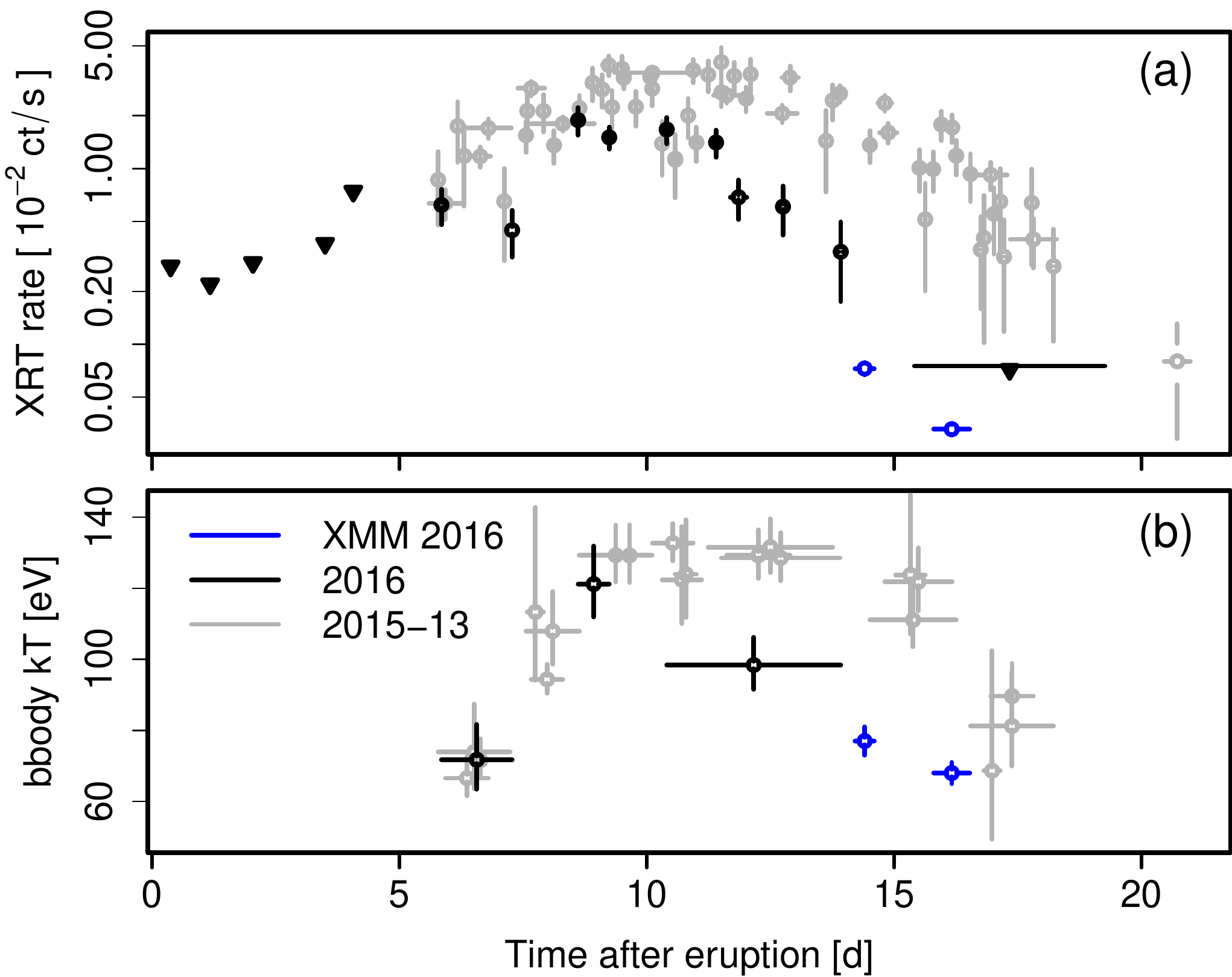}
\caption{\swift XRT (black) and \xmm EPIC pn (blue) (a) count rates (0.3--1.5\,keV) and (b) effective black body temperatures of \nova\ during the 2016 eruption compared to the XRT data of the 2013--15 eruptions (gray). {\it Panel a:} Triangles indicate upper limits (only shown for 2016 data). {\it Panel b:} Sets of observations with similar spectra have been fitted simultaneously assuming a fixed \nh = $0.7$ \hcm{21}. The error bars in time represent either the duration of a single observation or the time covering the sets of observations (for panel b and for the last 2016 XRT upper limit in panel a). The deviation of the 2016 eruption from the evolution of past events is clearly visible.}
\label{fig:xrt_xmm_lc}
\end{figure}

For a conservative estimate of the SSS turn-on time (and its accuracy) we use the midpoint between days 4.1 and 5.8 as $\eton = 4.9\pm1.1$\,d, which includes the uncertainty of the eruption date. This is consistent with the 2013--2015 X-ray light curves (see Figure~\ref{fig:xrt_xmm_lc}) for which we estimated turn-on times of $6\pm1$\,d (2013), $5.9\pm0.5$\,d (2014), and $5.6\pm0.7$\,d (2015) using the same method (see \xonek, \xtwok, \othreek). There is no evidence that the emergence of the SSS emission occurred at a different time than in the previous three eruptions.

The duration of the SSS phase, however, was significantly shorter than previously observed \citep[see Figure~\ref{fig:xrt_xmm_lc} and][]{2016ATel.9907....1H}. The last significant detection of X-ray emission in the XRT monitoring was on day 13.9 (Table~\ref{tab:swift}). However, the subsequent 2.9\,ks observation on day 15.4 still shows about 4 counts at the nova position which amount to a $2\sigma$ detection (Table~\ref{tab:swift} gives the $3\sigma$ upper limit). Nothing is visible on day 15.9. Again being conservative we estimate the SSS turn-off time as $\etoff = 14.9\pm1.2$\,d (including the uncertainty of the eruption date), which is the midpoint between observations 197 and 201 (see Table~\ref{tab:swift}).

In comparison, the SSS turn-off in previous eruptions happened on days $19\pm1$ (2013), $18.4\pm0.5$ (2014), and $18.6\pm0.7$ (2015); all significantly longer than in 2016. The upper limits in Figure~\ref{fig:xrt_xmm_lc} and Table~\ref{tab:swift} demonstrate that we would have detected each of the 2013, 2014, or 2015 light curves during the 2016 monitoring observations, which had similar exposure times (cf.\ \xonek, \xtwok, and \othreek). Therefore, the short duration of the 2016 SSS phase is real and not caused by an observational bias.

The full X-ray light curve, shown in Figure~\ref{fig:xrt_xmm_lc}a, is consistent with a shorter SSS phase which had already started to decline before day 12, instead of around day 16 as during the last three years. In a consistent way, the blackbody parametrization in Figure~\ref{fig:xrt_xmm_lc}b shows a significantly cooler effective temperature ($kT = 86\pm6$~eV) than in 2013--2015 ($kT\sim115\pm10$~eV) during days 10--14 (cf.\ \othreek). As previously, for this plot we fitted the XRT spectra in groups with similar effective temperature.

In contrast to our previous studies of \novak, here our blackbody parameterizations assume a fixed absorption of \nh = $0.7$ \hcm{21} throughout. (The X-ray analysis in \othreek\ had explored multiple \nh values). This value corresponds to the Galactic foreground. The extinction is based on {\it HST} extinction measurements during the 2015 eruption, which are consistent in indicating no significant additional absorption toward the binary system, e.g.\ from the \m31 disk \hstspec\ (also see \othreek). These {\it HST} spectra were taken about three days before the 2015 SSS phase onset, making it unlikely that the extinction varies significantly during the SSS phase. The new $N_\mathrm{H}$, also applied to the 2013--2015 data in Figure~\ref{fig:xrt_xmm_lc}, affects primarily the absolute blackbody temperature, now reaching almost 140\,eV, but not the relative evolution of the four eruptions.

\begin{figure}
\includegraphics[width=\columnwidth]{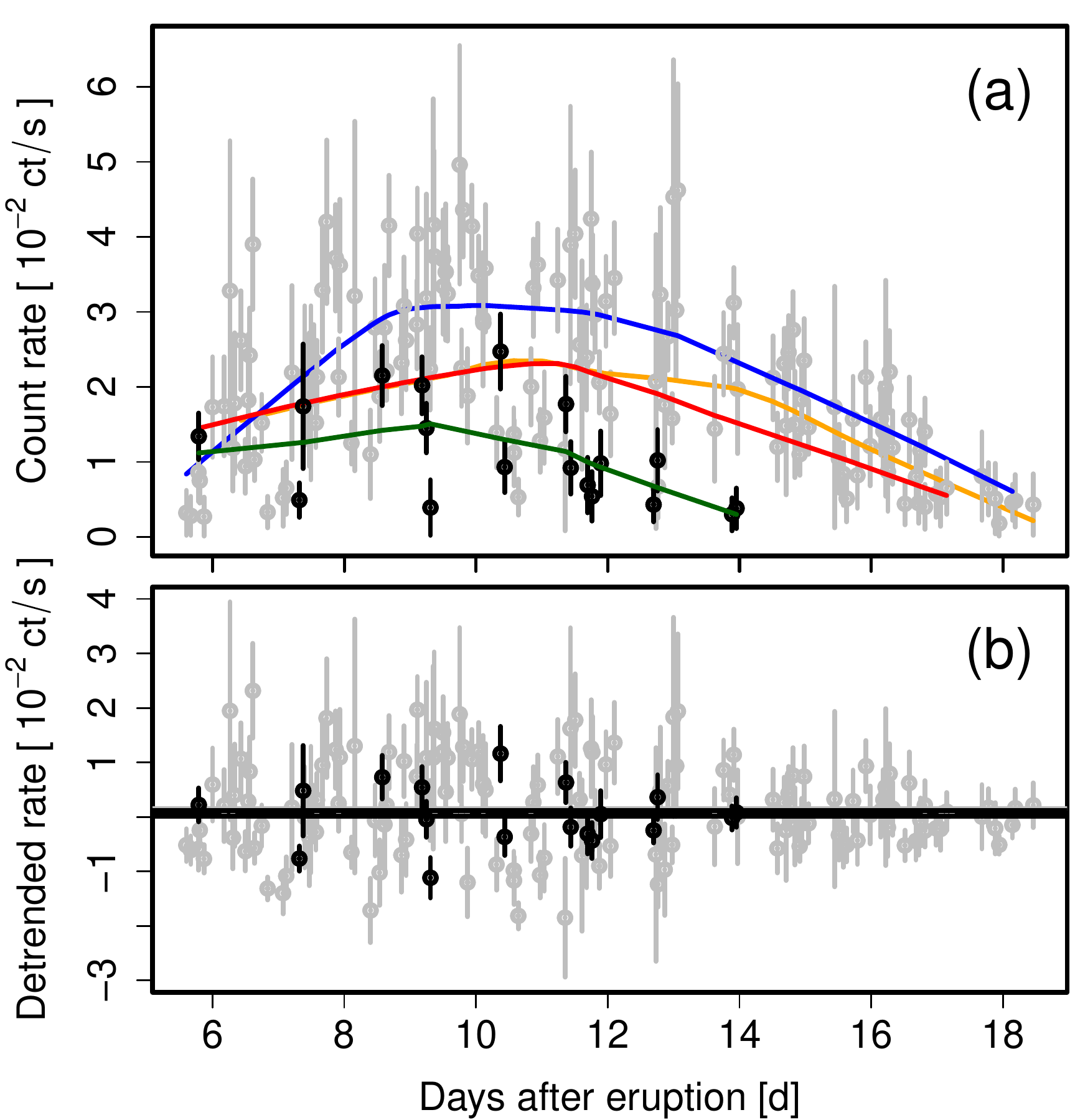}
\caption{\textit{Panel a}: The short-term SSS light curve of \nova\ derived from all XRT snapshots. The 2016 eruption data is shown in black in contrast to the gray 2013--2015 light curves. Instead of the logarithmic count rate scale in Figure~\ref{fig:xrt_xmm_lc} here we use a linear axis. The overlayed green (2016), red (2015), blue (2014), and orange (2013) curves show smoothing fits using local regression. The 2016 light curve is clearly shorter and appears to be less luminous than in 2013--2015. \textit{Panel b}: Detrended light curves after removing the smoothed trend. The 2016 light curve (black) suggests a drop in variability after day 11, whereas for the 2013--2015 light curves (gray) this drop happened around day 13.}
\label{fig:xrt_split}
\end{figure}

Figure~\ref{fig:xrt_xmm_lc}a also suggests that the SSS phase in 2016 was somewhat less luminous than in previous eruptions. The early SSS phase of this nova has shown significant flux variability, nevertheless a lower average luminosity is consistent with the XRT light curve binned per \swift snapshot, as shown in Figure~\ref{fig:xrt_split}. A lower XRT count rate would be consistent with the lower effective temperature suggested in Figure\,\ref{fig:xrt_xmm_lc}b. Note, that this refers to the observed characteristics of the SSS; not the theoretically possible maximum photospheric temperature if the hydrogen burning had not extinguished early.

We show the XRT light curve binned per \swift snapshot in Figure~\ref{fig:xrt_split}. As found in previous eruptions (\xonek, \xtwok, \othreek) the early SSS flux is clearly variable. However, here the variability level had already dropped by day $\sim11$ instead of after day 13 as in previous years. After day 11, the scatter (rms) decreased by a factor of two, which is significant on the 95\% confidence level (F-test, $p = 0.03$). This change in behavior can be seen better in the detrended \swift XRT count rate light curve in Figure~\ref{fig:xrt_split}b. The faster evolution is consistent with the overall shortening of the SSS duration.

\subsection{\xmm EPIC light curves}\label{sec:xmm_lc}

\begin{figure*}
\includegraphics[width=\columnwidth]{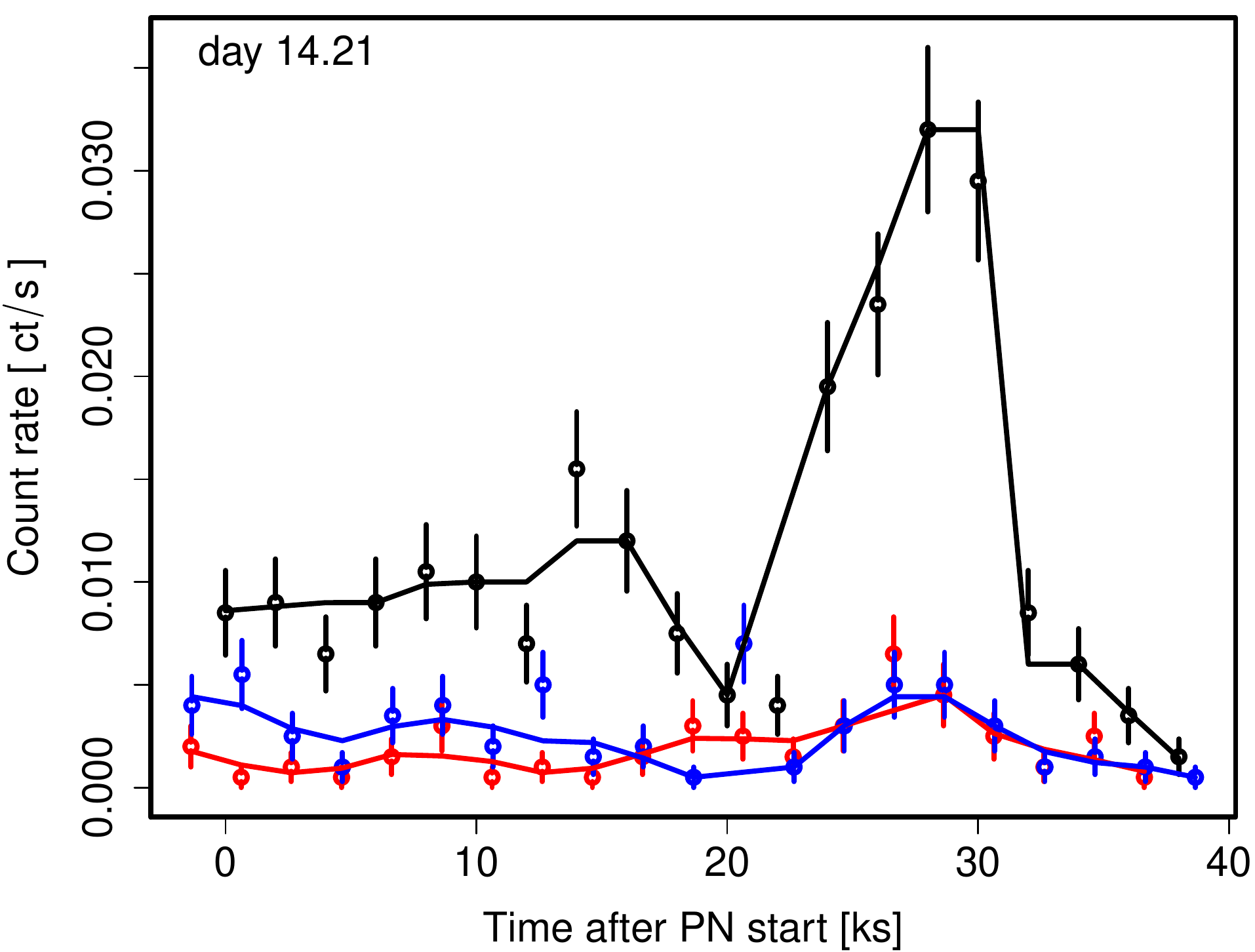}\hfill
\includegraphics[width=\columnwidth]{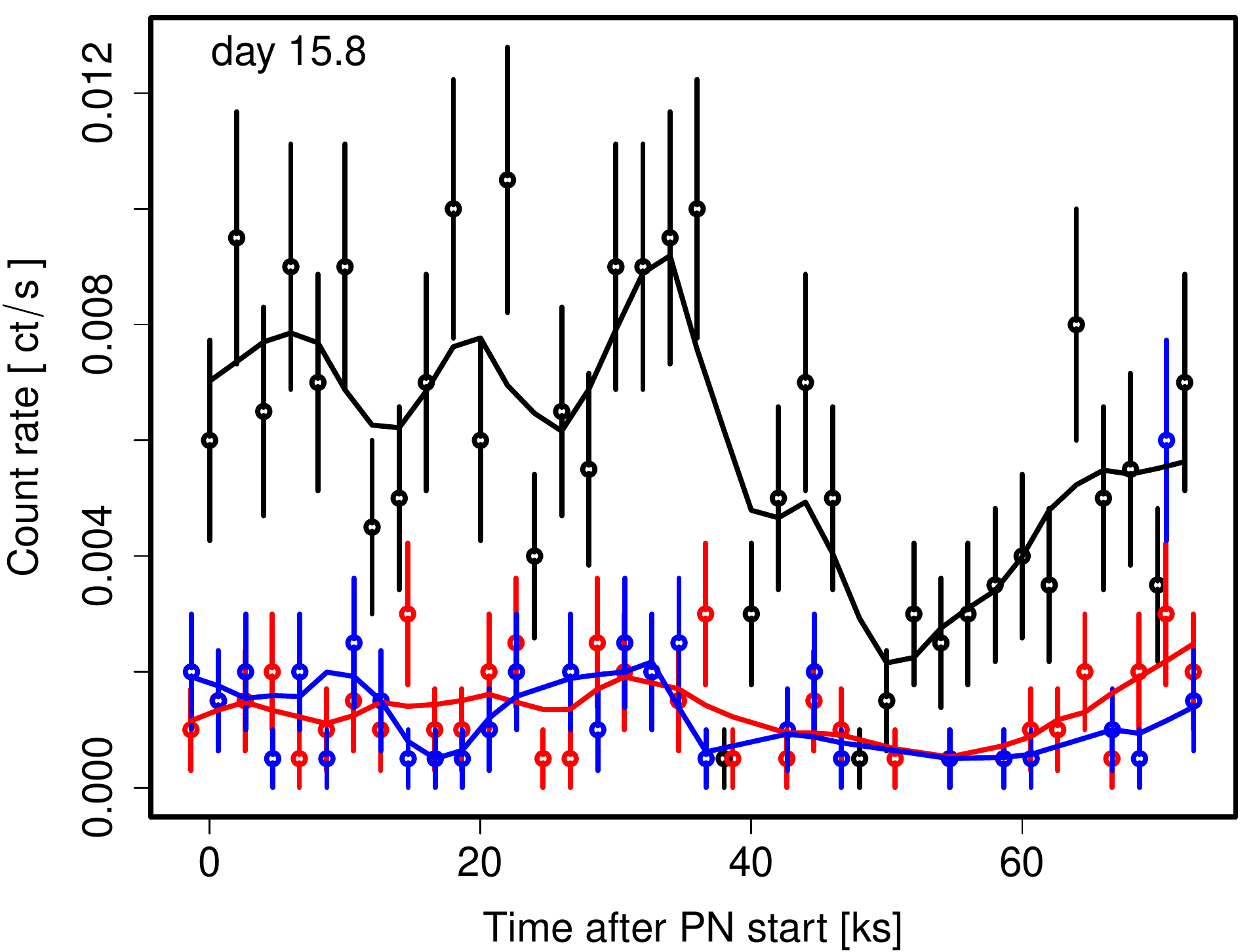}
\caption{\xmm EPIC light curves for observations 0784000101 (day 14.21; left) and 0784000201 (day 15.80; right) with 2\,ks binning. The EPIC pn (black), MOS1 (red), and MOS2 (blue) count rates and corresponding uncertainties are color-coded. The solid lines with the same colours are smoothed fits via locally-weighted polynomial regression (LOWESS).}
\label{fig:epic_lc}
\end{figure*}

The \xmm light curves from both pointings show clear variability over time scales of a few 1000\,s (Fig.\,\ref{fig:epic_lc}). This is an unexpected finding, since the variability in the \swift XRT light curve appeared to have ceased after day 11 (in general agreement with the 2013--15 light curve where this drop in variability occurred slightly later). Instead, we find that the late X-ray light curve around days 14--16 (corresponding to days 18--20 for the ``normal'' 2013--15 evolution) are still variable by factors of $\sim5$. The variability is consistent in the EPIC pn and MOS light curves (plotted without scaling in Figure~\ref{fig:epic_lc}).

Even with the lower XRT count rates during the late SSS phase, we would still be able to detect large variations similar to the high-amplitude spike and the sudden drop seen in the first and second EPIC light curve, respectively.

\section{Panchromatic eruption spectroscopy}
\label{sec:spec}

\subsection{Optical spectra}
\label{sec:vis_spec}

\begin{figure*}
\includegraphics[width=\textwidth]{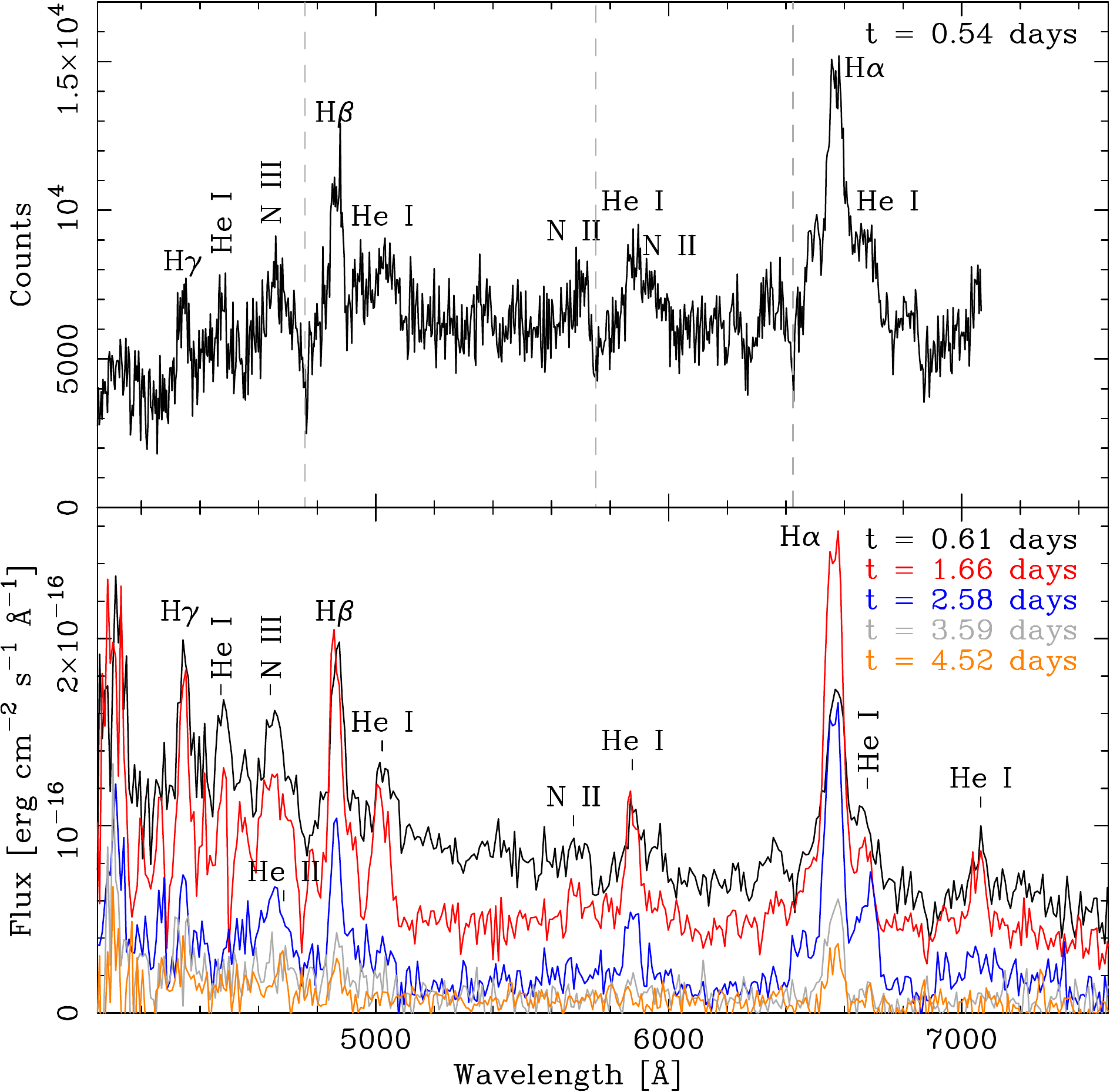}
\caption{{\bf Top:} NOT ALFOSC spectrum of \novak, taken 0.54\,days after the 2016 eruption, one of  the earliest spectra taken of any of the \novak\ eruptions. The gray dashed lines represent a velocity of $-6250$\,km\,s$^{-1}$ with respect the H$\beta$, He\,{\sc i} 5876\,\AA\ and H$\alpha$. Narrow absorption can be seen at this velocity accompanying the H$\alpha$ and H$\beta$ emission lines, and there is evidence for a similar absorption feature with He\,{\sc i} 5876\,\AA.\label{fig:not} {\bf Bottom:} LT spectra of the 2016 eruption, taken between 0.61 and 4.52\,days after eruption.\label{fig:optspec}}
\end{figure*}

The LT eruption spectra of 2016 are broadly similar to the 2015 (and prior) eruption (see \othreek),  with the hydrogen Balmer series being the strongest emission lines (Fig.\,\ref{fig:optspec}). He\,{\sc i} lines are detected at 4471, 5876, 6678 and 7065\,\AA, along with He\,{\sc ii} (4686\,\AA) blended with N\,{\sc iii} (4638\,\AA).  The broad N\,{\sc ii} (3)  multiplet around 5680\,\AA\ is also weakly detected. These emission lines are all typically associated with the He/N spectroscopic class of novae \citep{1992AJ....104..725W}. The five LT spectra are shown in Figure~\ref{fig:optspec} (bottom) and cover a similar time frame as those obtained during the 2015 eruption. These spectra are also displayed along with all of the other 2016 spectra at the end of this work in Figure~\ref{specall}.

The first 2016 spectrum, taken with NOT/ALFOSC 0.54\,days after eruption, shows P\,Cygni absorption profiles on the H$\alpha$ and H$\beta$ lines. We measure the velocity of the minima of these absorption lines to be at $-6320\pm160$ and $-6140\pm200$\,km\,s$^{-1}$ for H$\alpha$ and H$\beta$, respectively. This spectrum can be seen in Figure~\ref{fig:not} (top), which also shows evidence of a possible weak P\,Cygni absorption accompanying the He\,{\sc i} (5876\,\AA) line.  The first LT spectrum, taken 0.61\,days after eruption, also shows evidence of a P\,Cygni absorption profile on H$\alpha$ (and possibly H$\beta$) at $\sim-6000$\,km\,s$^{-1}$. 

This is the first time absorption lines have been detected in the optical spectra of \novak. We note that the HST FUV spectra of the 2015 eruption revealed strong, and possibly saturated, P\,Cygni absorptions still present on the resonance lines of N\,{\sc v}, Si\,{\sc iv}, and C\,{\sc iv} at $t=3.3$\,days with terminal velocities in the range 6500--9400\,km\,s$^{-1}$, the NUV spectra taken $\sim1.5$\,days later showed only emission lines (\hstspec). 

\begin{figure*}
\begin{center}
\includegraphics[width=0.48\textwidth]{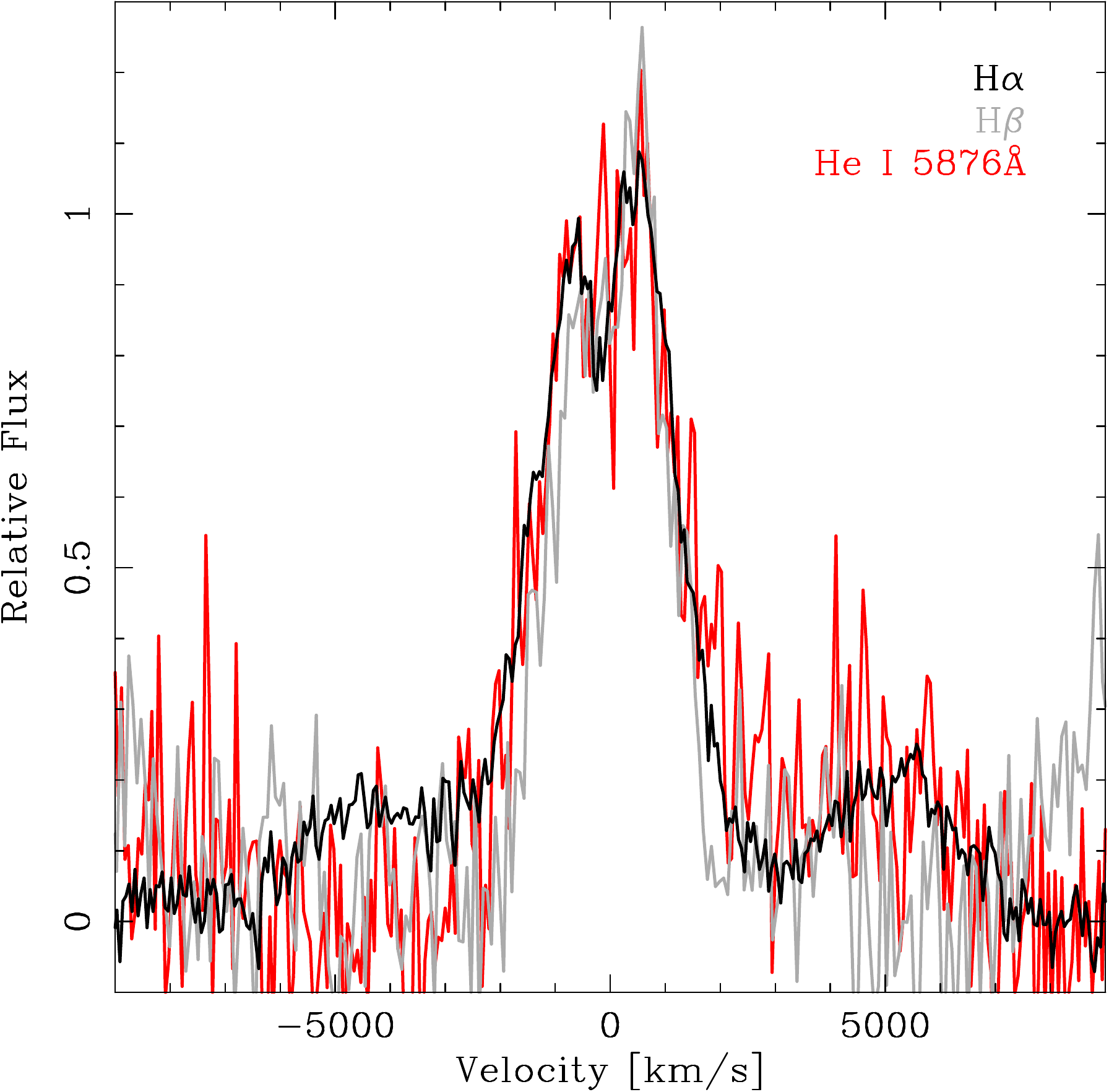}\hfill
\includegraphics[width=0.48\textwidth]{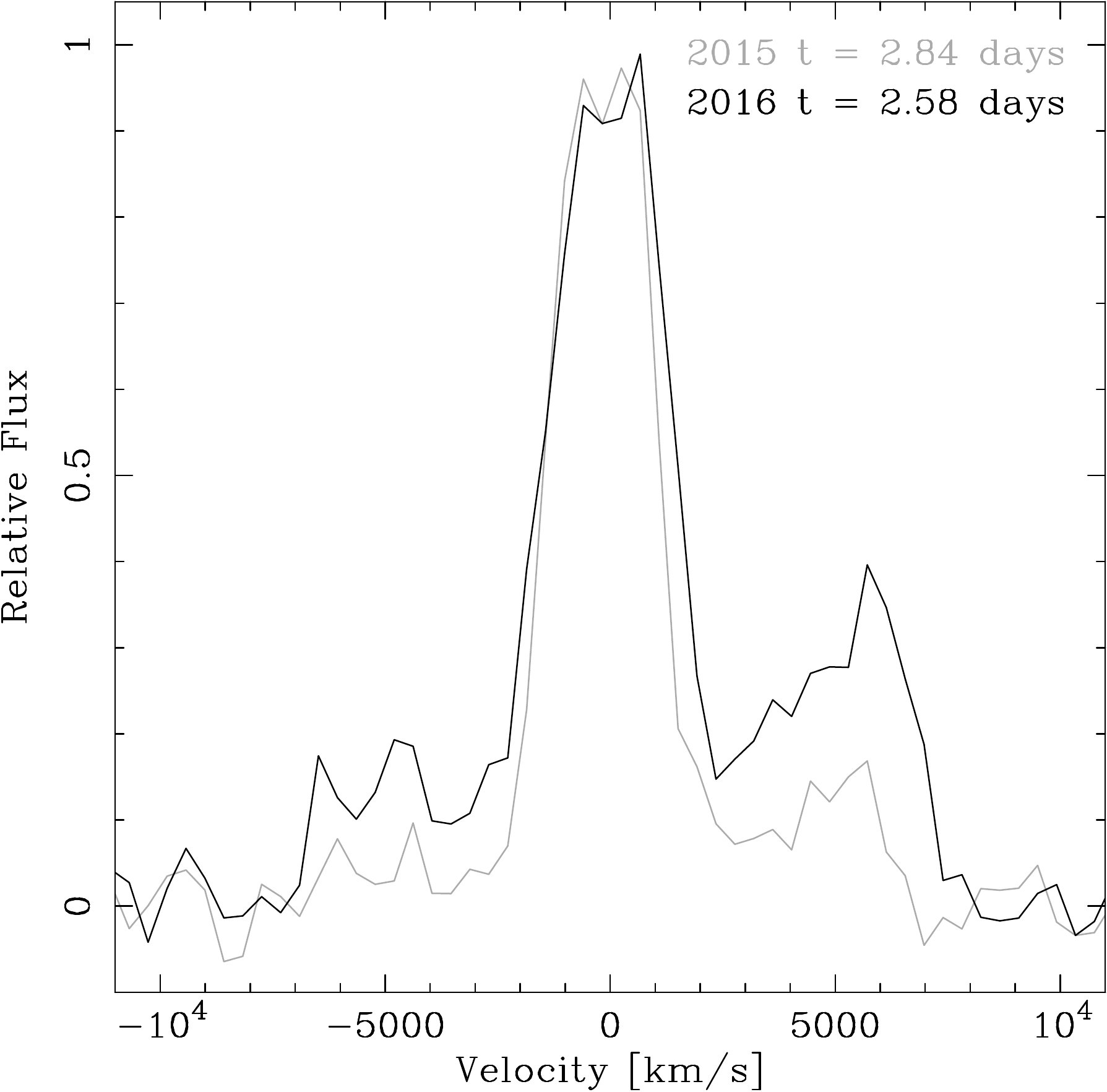}\\
\includegraphics[width=0.48\textwidth]{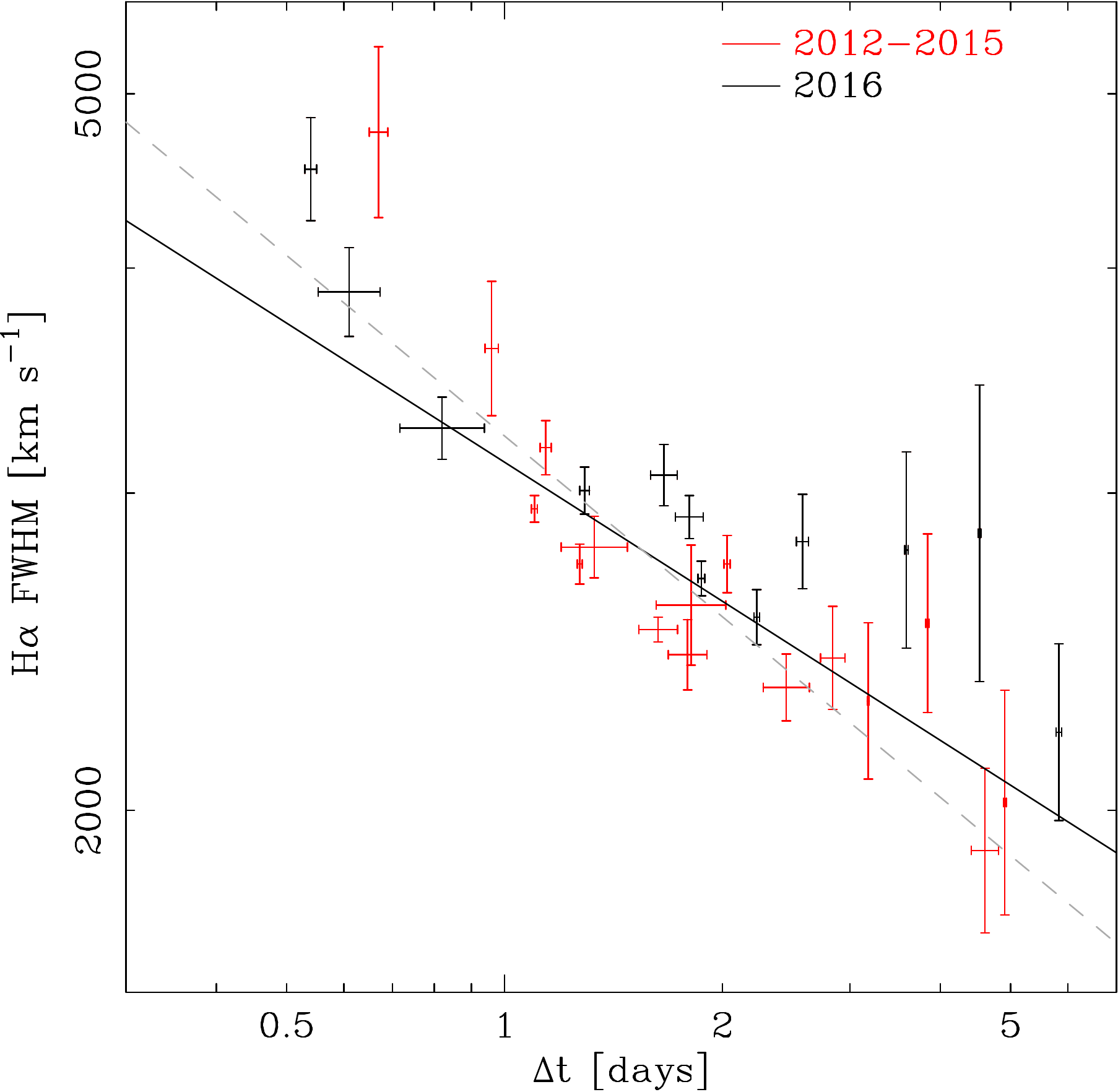}\hfill
\includegraphics[width=0.48\textwidth]{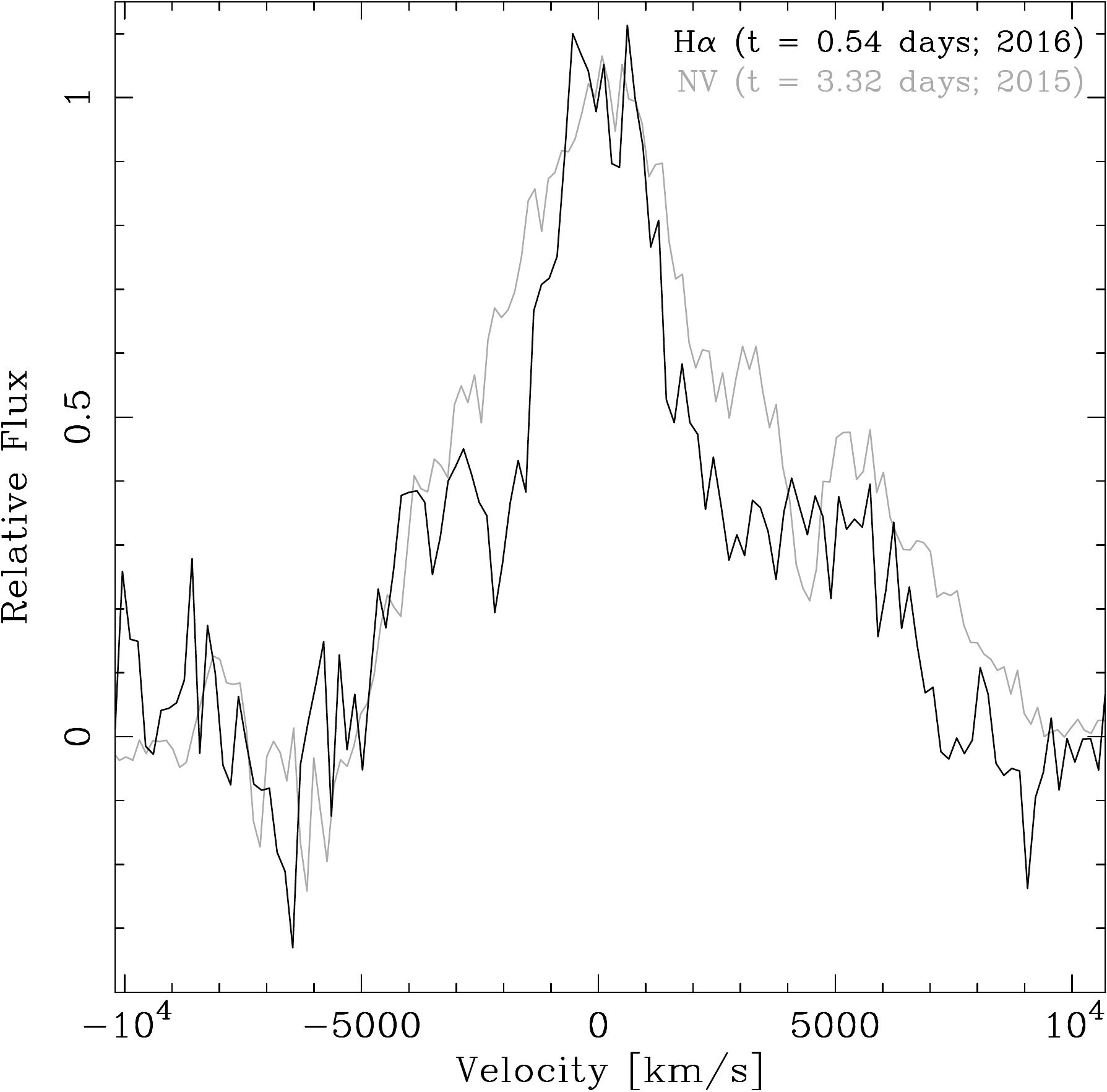}
\end{center}
\caption{\textbf{Top left}: HET spectrum at day 1.87, showing the similar line structures of H$\alpha$, H$\beta$ and He\,{\sc i} (5876\,\AA). {\textbf{Top right}}: LT spectra comparing the high-velocity material at day-2.84 of the 2015 eruption to day-2.58 of the 2016 eruption. These are normalized to the lower velocity component peak. \textbf{Bottom left}: FWHM velocity evolution of the H$\alpha$ profile during the 2016 eruption (black), compared to previous eruptions (red).  The gray dashed line is a power law of an index of $-1/3$ ($\chi^2_{/\mathrm{dof}}=3.7$; Phase II of shocked remnant development) and the solid black line is the best-fit power law with an index of $-0.26\pm$0.04 ($\chi^2_{/\mathrm{dof}}= 3.6$). \textbf{Bottom right:} comparison between the H$\alpha$ line profile 0.54\,days after the 2016 eruption (black) and the N\,{\sc v} (1240\,\AA) profile 3.32\,days after the 2015 eruption (gray; see \hstspec). Note that the N\,{\sc v} profile has been shifted 500\,km\,s$^{-1}$ blueward with respect to H$\alpha$.\label{fig:spec3}}
\end{figure*}

The HET spectrum taken 1.87\,d after eruption can be seen in Figure~\ref{fig:spec3}, showing that the central emission profiles of the Balmer lines and He\,{\sc i} are broadly consistent. Note that the emission around +5000\,km\,s$^{-1}$ from the H$\alpha$ rest velocity probably contains a significant contribution from He\,{\sc i} (6678\,\AA). By this time the P\,Cygni profiles appear to have dissipated. 

Figure~\ref{fig:optspec} clearly shows the existence of high velocity material around the central H$\alpha$ line at day 2.58 of the 2016 eruption. This can be seen in more detail, compared to the 2015 eruption, in Figure~\ref{fig:spec3}. Note that, as stated above, the redshifted part of the (2016) profile could be affected by He\,{\sc i} (6678\,\AA), although the weakness of the (isolated) He\,{\sc i} line at 7065\,\AA\ (see Figure~\ref{fig:optspec}) suggests this cannot explain all of the excess flux on this side of the profile. Also note the extremes of the profile indicate a similar velocity (HWZI $\sim$ 6500 to 7000 km\,s$^{-1}$).

\begin{figure*}
\includegraphics[width=0.96\textwidth]{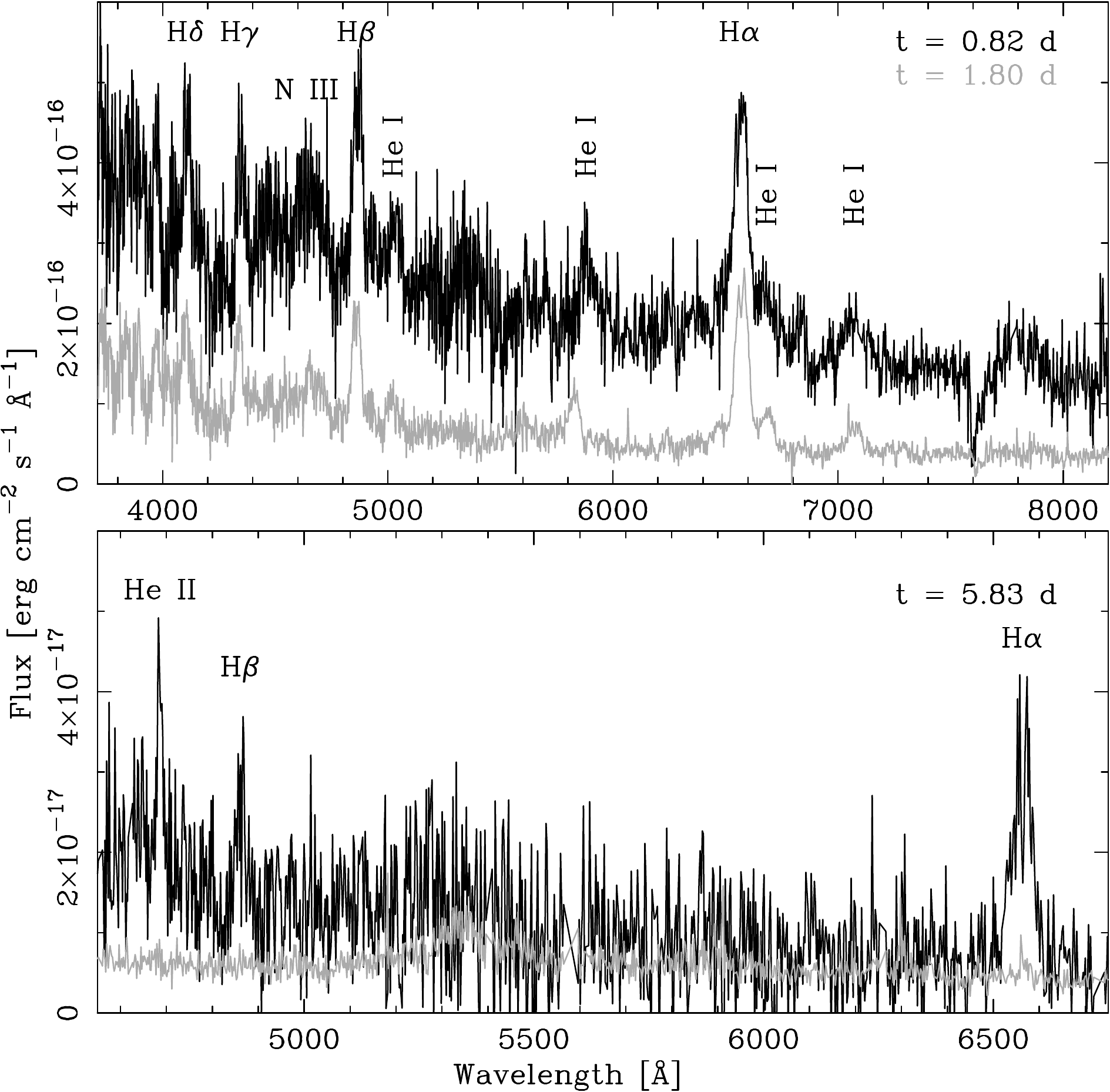}
\caption{ARC spectra of the 2016 eruption of \novak\ taken 0.82 and 1.80\,d post-eruption (top) and 5.83\,d post-eruption (bottom). The bottom panel shows a smaller wavelength range than the top panel, and here the gray line represents the errors for the $t=5.83$\,d spectrum.\label{fig:apo}}
\end{figure*}

\begin{figure}
\includegraphics[width=0.96\columnwidth]{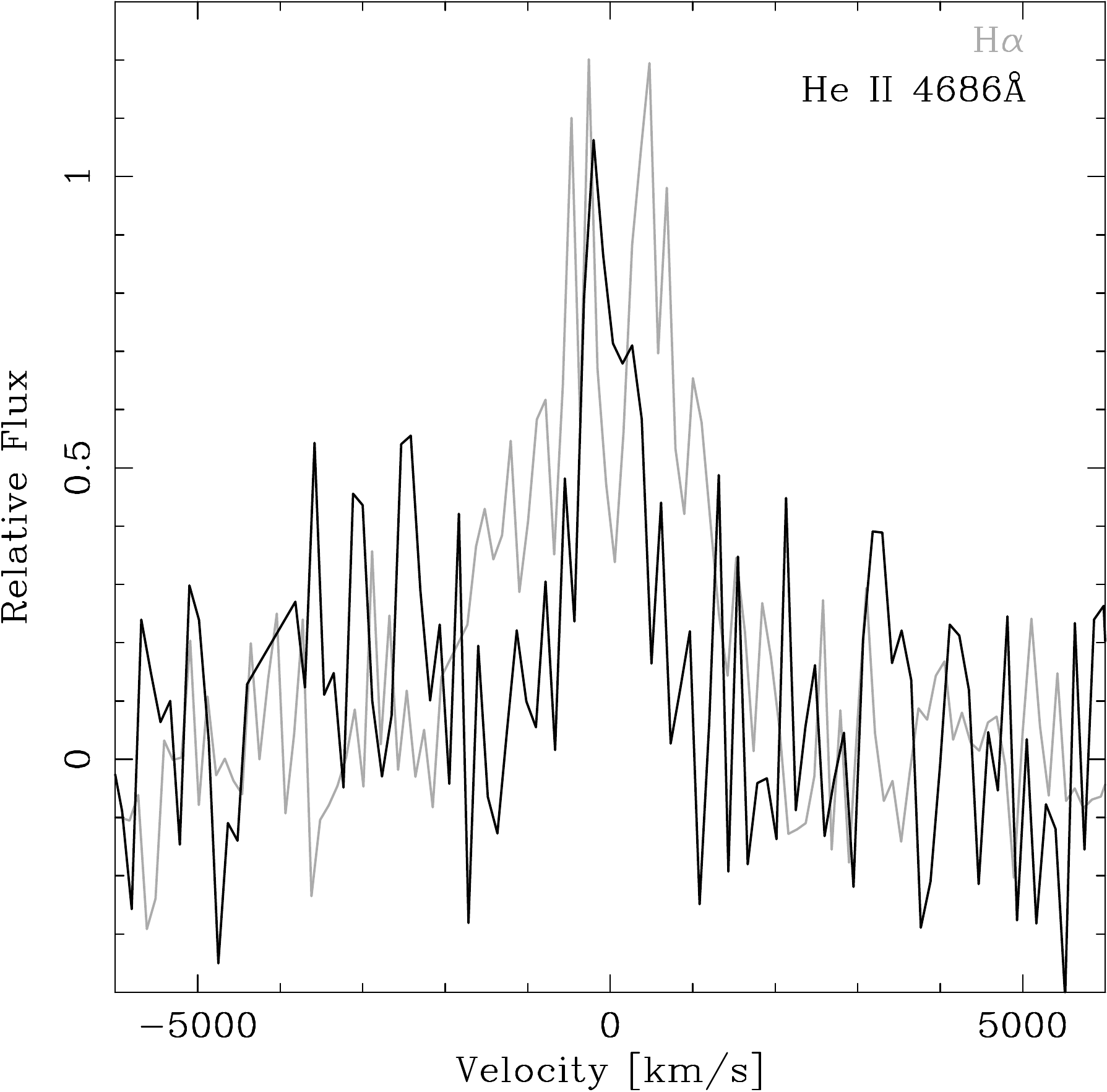}
\caption{Comparison of H$\alpha$ and He\,{\sc ii} 4686\,\AA\ emission lines in the $t=5.83$\,d ARC spectrum.\label{fig:apovel}}
\end{figure}

The 4.91\,day spectrum of the 2015 eruption shows H$\alpha$ and H$\beta$ emission.  By comparison, the 2016 4.52-day spectrum also shows a clear emission line from He\,{\sc ii} (4686\,\AA), consistent with the Bowen blend being dominated by He\,{\sc ii} at this stage of the eruption.  However, we note that this is unlikely to mark a significant difference between 2015 and 2016, as these late spectra typically have very low signal-to-noise ratios. The ARC spectra are shown in Figure~\ref{fig:apo}. The last of these spectra, taken 5.83\,d after eruption shows strong He\,{\sc ii} (4686\,\AA) emission. The S/N of the spectrum is relatively low, but the He\,{\sc ii} emission appears narrower than the H$\alpha$ line at the same epoch, as seen in Figure~\ref{fig:apovel}. At this stage of the eruption we calculate the FWHM of He~{\sc ii} (4686\,\AA) to be $930\pm150$\,km\,s$^{-1}$, compared to $2210\pm250$\,km\,s$^{-1}$ for H$\alpha$. The ARC spectra have a resolution of $R\sim1000$, so these two FWHM measurements are not greatly affected by instrumental broadening. Narrow He\,{\sc ii} emission has been observed in a number of other novae. It is seen in the Galactic RN U\,Sco from the time the SSS becomes visible \citep{2012A&A...544A.149M}. Those authors used the changes in the narrow lines with respect to the orbital motion (U\,Sco is an eclipsing system; \citealp{1990ApJ...355L..39S}) to argue that such emission arises from a reforming accretion disk. In the case of the 2016 eruption of \novak, we clearly observe the SSS at 5.8\,d, meaning this final ARC spectrum is taken during the SSS phase. This is consistent with the suggestion that, in \novak, the accretion disk survives the eruption largely intact (\hstphot). In this scenario, the optically thick ejecta prevent us seeing evidence of the disk in our early spectra. We note however, \citet{2014A&A...564A..76M} argued that in the case of KT\,Eri, there could be two sources of such narrow He\,{\sc ii} emission, initially being due to slower moving material in the ejecta, before becoming quickly dominated by emission from the binary itself (as in U\,Sco) as the SSS enters the plateau phase.

\hstphot\ presented a low S/N, post-SSS spectrum taken 18.8 days after the 2014 eruption of \novak.  This spectrum was consistent with that expected from an accretion disk, and H$\beta$ was seen in emission.  However, no evidence of the He\,{\sc ii} (4686\,\AA) line was seen in that spectrum.  It is possible that the strong He\,{\sc ii} line seen in the ARC spectrum arose from the disk but that the transition was excited by the on-going SSS at that time.

As with previous eruptions, the emission line profiles of individual lines showed significant evolution during the 2016 eruption. The FWHM of the main H$\alpha$ emission line (excluding the very high velocity material) narrows from $4540\pm300$\,km\,s$^{-1}$ on day 0.54 to $2210\pm250$\,km\,s$^{-1}$ on day 5.83. The velocity evolution of the 2016 eruption is compared to that of previous eruptions in Figure~\ref{fig:spec3}, and is largely consistent. The H$\alpha$ FWHM measurements of all 2016 eruption spectra are given in Table~\ref{tab:vel}

\begin{table}
\caption{FWHM Velocity Measurements of the H$\alpha$ Profile During the 2016 Eruption. \label{tab:vel}}
\begin{center}
\begin{tabular}{lll}
\hline
$\Delta t$ (days) &H$\alpha$ FWHM (km\,s$^{-1}$) &Instrument\\
\hline
0.54$\pm$0.01 &4540$\pm$300 &ALFOSC\\
0.61$\pm$0.06 &3880$\pm$220 &SPRAT\\
0.82$\pm$0.11 &3260$\pm$130 &DIS\\
1.29$\pm$0.02 &3010$\pm$90 &HFOSC\\
1.66$\pm$0.07 &3070$\pm$120 &SPRAT\\
1.80$\pm$0.08 &2910$\pm$80 &DIS\\
1.87$\pm$0.02 &2690$\pm$60 &LRS2-B\\
2.23$\pm$0.02 &2560$\pm$90 &HFOSC\\
2.58$\pm$0.05 &2820$\pm$170 &SPRAT\\
3.59$\pm$0.02 &2790$\pm$350 &SPRAT\\
4.53$\pm$0.02 &2850$\pm$540 &SPRAT\\
5.83$\pm$0.05 &2210$\pm$250 &DIS\\
\hline
\end{tabular}
\end{center}
\end{table}

\subsection{The \xmm EPIC spectra and their connection to the \swift XRT data}
\label{sec:xmm_spec}

\begin{figure}
\includegraphics[width=\columnwidth]{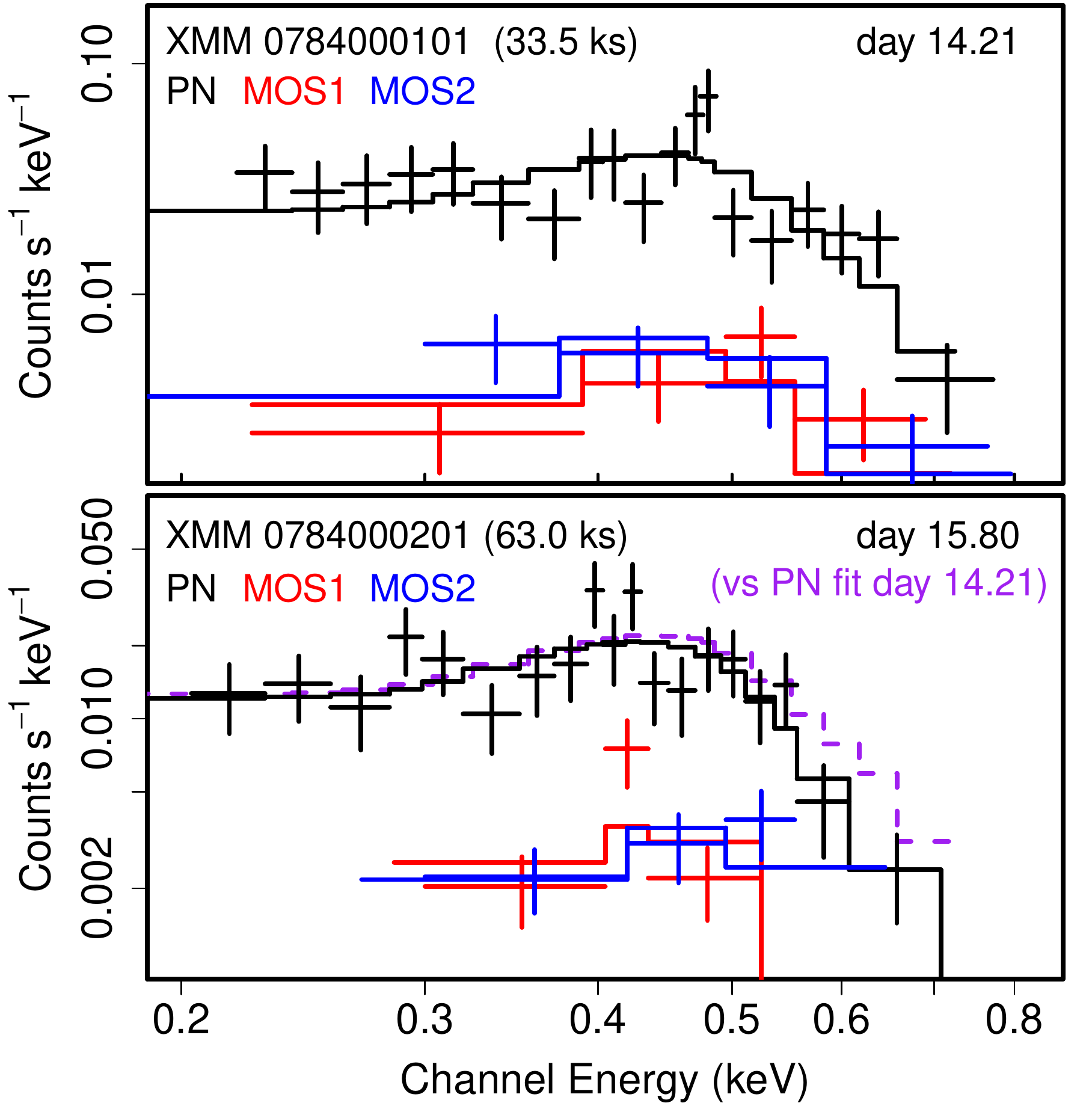}
\caption{\xmm EPIC spectra of \nova for the two pointings and the three individual (colour-coded) detectors (cf.\ Table~\ref{tab:xmm}). The blackbody fits are shown as solid lines. In the bottom panel the dashed purple line shows the scaled EPIC pn fit from the upper panel, indicating a tentative drop in temperature from  $kT = 58^{+8}_{-5}$~eV on day 14.21 to $kT = 45\pm5$~eV day 15.8. See Table~\ref{tab:xmm_spec} for details on the spectral fits.}
\label{fig:xmm_spec}
\end{figure}

The \xmm EPIC spectra for the two observations listed in Table~\ref{tab:xmm} were fitted with an absorbed blackbody model. The three detectors were modeled simultaneously, with only the normalizations free to vary independently. In Table~\ref{tab:xmm_spec} we summarize the best fit parameters and also include a simultaneous fit of all EPIC spectra. The binned spectra, with a minimum of 10 counts per bin, are plotted in Figure~\ref{fig:xmm_spec} together with the model curves. The binning is solely used for visualization here; the spectra were fitted with one-count bins and Poisson (fitting) statistics \citep{1979ApJ...228..939C}. The $\chi^2$ numbers were used as test statistics. 

\begin{table*}
\caption{Spectral Fits for \xmm Data.}
\label{tab:xmm_spec}
\begin{center}
\begin{tabular}{lrrrrrrr}\hline\hline \noalign{\smallskip}
 ObsID & $\Delta t^a$ & \nh & kT & red.\ $\chi^2$ & d.o.f. &  kT$_{0.7}^b$ & red.\ $\chi^{2\,\,b}$ \\
  & (d) & (\ohcm{21}) & (eV) & & & (eV) & \\ \hline \noalign{\smallskip}
 0784000101 & 14.21 & $2.2^{+0.6}_{-0.7}$ & $58^{+8}_{-5}$ & 1.29 & 149 & $77^{+4}_{-3}$ & 1.44 \\
 0784000201 & 15.80 & $2.7^{+0.6}_{-0.5}$ & $45\pm5$ & 1.06 & 140 & $68^{+4}_{-3}$ & 1.35 \\
 Both combined & 15.01 & $2.2\pm0.4$ & $53^{+5}_{-3}$ & 1.22 & 291 & $73^{+3}_{-2}$ & 1.42 \\
\hline
\end{tabular}
\end{center}
\noindent
\tablenotetext{a}{Time in days after the nova eruption (cf.\ Table~\ref{tab:xmm})}
\tablenotetext{b}{The blackbody temperature (and the reduced $\chi^2$ of the fit) when assuming a fixed \nh = $0.7$ \hcm{21} for comparison with the \swift XRT temperature evolution (see Fig.\,\ref{fig:xrt_xmm_lc}).}
\end{table*}

In Table~\ref{tab:xmm_spec} and Figure~\ref{fig:xmm_spec} we immediately see that the two spectra are (a) very similar and (b) contain relatively few spectral counts, leading to a low spectral resolution. The latter point is mainly due to the unexpectedly low flux at the time of the observations, but is also exacerbated by the strong background flaring (cf.\ Table~\ref{tab:xmm}).

In Table~\ref{tab:xmm_spec} we also list a second set of blackbody temperature values (kT$_{0.7}$) for the assumption of a fixed \nh = $0.7$ \hcm{21}. The purpose of this is to compare these temperatures to the \swift XRT models which share the same assumption (cf.\ Section~\ref{sec:xrt_lc}). In both sets of temperatures in Table~\ref{tab:xmm_spec} there is a slight trend toward higher temperatures in the first observation (day 14.21) compared to the second one (day 15.80). While the binned spectra in Figure~\ref{fig:xmm_spec} give a similar impression, which would be consistent with a gradually cooling WD, it needs to be emphasized that this gradient has no high significance because the two (sets of) temperatures are consistent within their $2\sigma-3\sigma$ uncertainties. In fact, the combined fit in Table~\ref{tab:xmm_spec} has reduced $\chi^2$ statistics and parameter uncertainties that are similar (the latter even slightly lower) than those of the individual fits.

In Figure~\ref{fig:xrt_xmm_lc} the \xmm data points are added to the \swift light curve and temperature evolution. For the conversion from pn to XRT count rate we used the HEASarc WebPIMMS tool \citep[based on PIMMS v4.8d,][]{1993Legac...3...21M} under the assumption of the best-fit blackbody parameters in the third and fourth column of Table~\ref{tab:xmm_spec}.

While the equivalent count rates as well as the temperatures are consistent with the XRT trend of a fading and cooling source there appear to be systematic differences between the XRT and pn rates. This could simply be due to systematic calibration uncertainties between the EPIC pn and the XRT \citep{2017AJ....153....2M}. Another reason might be the ongoing flux variability (see Section~\ref{sec:xmm_lc}). However, it is also possible that deficiencies in the spectral model are preventing a closer agreement between both instruments. We refrain from an attempt to align the pn and XRT count rates because currently there are too many free parameters (e.g., the potential absorption or emission features discussed in \othreek) and insufficient constraints on them. We hope that a future \xmm observation will be able to catch this enigmatic source in a brighter state to shine more (collected) light on its true spectral properties.

\section{Discussion}\label{sec:discussion}

\subsection{The relative light curve evolution and the exact eruption date}\label{sec:disc_date}
\smallskip

The precision of the eruption dates for previous outbursts was improved by aligning their light curves, specifically the early, quasi-linear decline (\othreek). For the 2016 eruption, a priori we cannot be certain that this decline phase would be expected to align with previous years because the bright optical peak (Figure~\ref{fastphot} left) constitutes an obvious deviation from the established pattern. However, in Figure~\ref{optical_lc} we find that after the peak feature, most filters appear to decline in the same way as during the previous years. Therefore, we conclude that our estimated eruption date of $\mathrm{MJD}= 57734.32 \pm 0.17$ (2016-12-12.32 UT) is precise to within the uncertainties -- and this brings about a natural alignment of the light curves.

\subsection{The peculiarities of the 2016 eruption and their description by theoretical models}
\label{sec:disc_pecul}
From the combined optical and X-ray light curves in Figures~\ref{optical_lc} and \ref{fig:xrt_xmm_lc} it can be seen that in 2016 (i) the optical peak may have been brighter and (ii) the SSS phase was intrinsically shorter than the previous three eruptions (but began at the same time after eruption). In addition, the gap between the 2015 and 2016 eruptions was longer than usual. Below we study these discrepancies in detail and describe them with updated theoretical model calculations. The following discussion ignores the impact of a possible half-year recurrence (cf.~\halfk), the potential dates of which are currently not well constrained (except for the first half of 2016; Henze et al.\ 2018, in prep.).

The critical advantage of studying a statistically significant number of eruptions from the same nova system is that we can reasonably assume parameters like (accretion and eruption) geometry, metallicity of the accreted material, as well as WD mass, spin, and composition to remain (sufficiently) constant. Therefore, \nova plays a unique role in understanding the variations in nova eruption parameters. 

\subsubsection{A brighter peak after a longer gap?}\label{sec:disc_peak}

This section aims to understand the surprising increase in the optical peak luminosity (the `cusp') by relating it to the delayed eruption date through the theoretical models of \citet{2006ApJS..167...59H, 2014ApJ...793..136K, 2017ApJ...838..153K}. While the specifics of our arguments are derived from this particular set of models, we note that all current nova light curve simulations agree on the general line of reasoning \citep[e.g.][]{2005ApJ...623..398Y, 2013ApJ...777..136W}.  We also note that \hstphot\ found an elevated mass accretion rate to that employed by \citet{2014ApJ...793..136K, 2017ApJ...838..153K}, but again the general trends discussed below do not depend on the absolute value of the assumed mass accretion rate.

\begin{figure}
\includegraphics[width=\columnwidth]{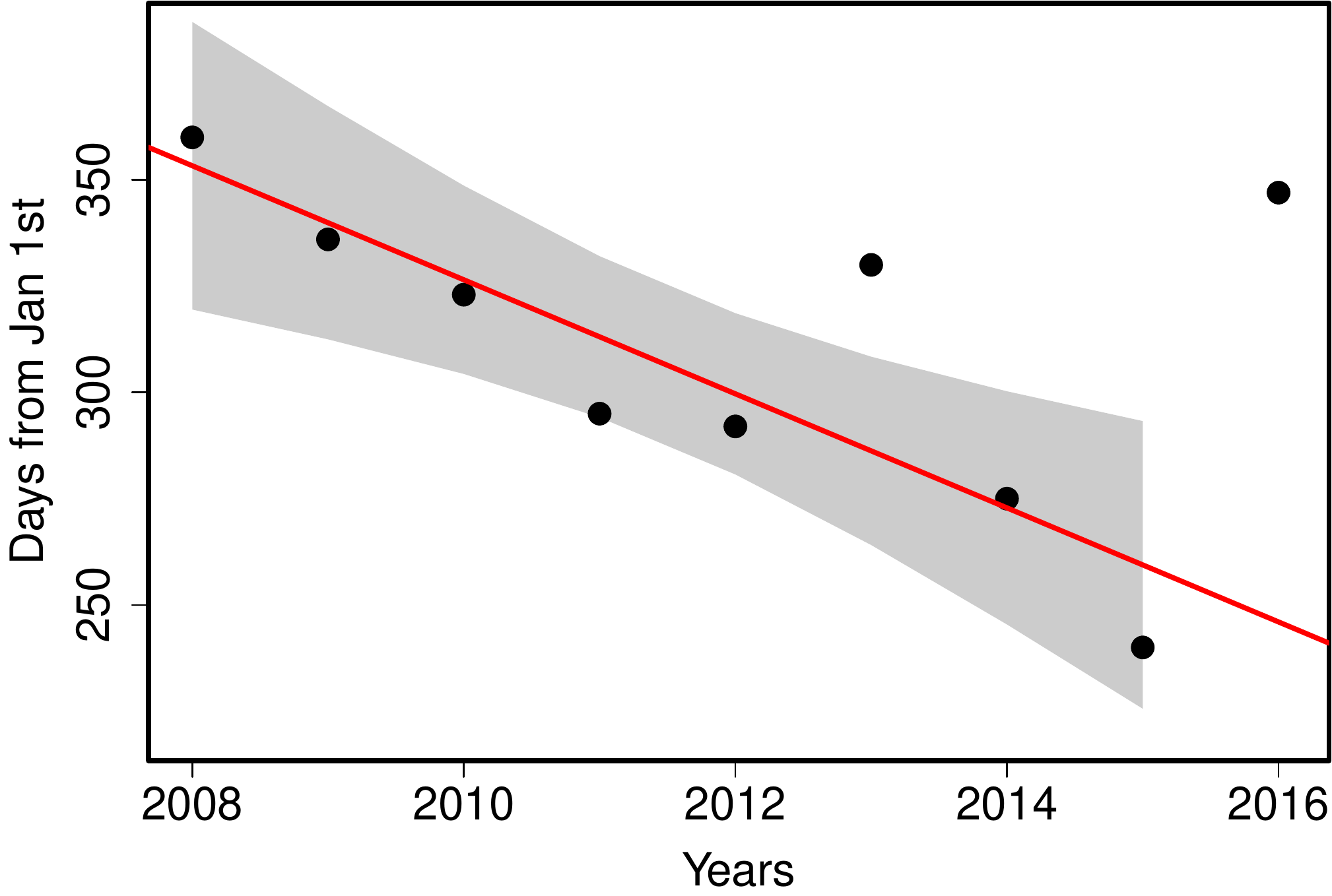}
\caption{Eruption dates (in days of the year) vs the year from 2008 onward. Individual uncertainties are smaller than the symbols. The best linear model for the \textit{2008-2015 eruptions} in shown in red with the 95\% uncertainties plotted in gray (cf.\ \othreek). The .}
\label{fig:rec_time}
\end{figure}

The gap between the 2015 and 2016 eruptions was 472\,d. This is 162\,d longer than the 310\,d between the 2013 and 2014 eruptions (see Table~\ref{eruption_history} and Figure~\ref{fig:rec_time}) and about 35\% longer than the median gap (347\,d) between the successive eruptions from 2008 to 2015. The well-observed 2015 eruption was very similar to the eruptions in 2013 and 2014 (\othreek) and did not show any indications that would have hinted at a delay in the 2016 eruption (also see \hstphot). This section compares the peculiar 2016 eruption specifically to the 2014 outburst, because we know that the latter was preceded and followed by a ``regular'' eruption (see Figures\,\ref{optical_lc} and \ref{fig:xrt_xmm_lc}, and \othreek). In general, we know that the peak brightness of a nova is higher for a more massive envelope if free-free emission dominates the SED \citep{2006ApJS..167...59H}.

We consider two specific cases:\ (1) the mean mass accretion-rate onto the WD ($\dot{M}_\mathrm{acc}$) was constant but hydrogen ignition occurs in a certain range around the theoretically expected time and, as a result, the elapsed inter-eruption time was longer in 2016 due to stochastic variance. Alternatively, (2) the mean mass accretion-rate leading up to the 2016 eruption was lower than typical and, as a result, the elapsed time was longer.

(1) If the mean accretion rates prior to the 2014 and 2016 eruptions were the same, then the mass accreted by the WD in 2016 was $\Delta t_{\rm rec}\times \dot M_{\rm acc} = 162\,{\rm days} \times 1.6 \times 10^{-7}\,M_\sun$\,yr$^{-1} = 0.71\times10^{-7}\,M_\sun$ larger than in 2014. Here we used the mass accretion rate of the $1.38\,M_\sun$ model proposed for \nova by \citet{2017ApJ...838..153K}. The authors obtained the relation between a wind mass-loss rate and the photospheric temperature (see their Figure 12). The wind mass-loss rate is larger for a lower-temperature envelope, which corresponds to a more extended and more massive envelope. 

In Figure 12 of \citet{2017ApJ...838..153K}, the rightmost point on the red line corresponds to the peak luminosity of the 2014 eruption. If at this point the envelope mass is higher by $0.71\times 10^{-7}\,M_\sun$, then the wind mass-loss rate should increase by $\Delta \log \dot M_{\rm wind} \sim 0.08$. 
 
For the free-free emission of novae the optical/IR luminosity is proportional to the square of the wind mass-loss rate \citep[see e.g.][]{2006ApJS..167...59H}. Thus, the peak magnitude of the optical/IR free-free emission is $2.5 \times (\Delta \log \dot M_{\rm wind}) \times 2 = 2.5 \times 0.08 \times 2 = 0.4$~mag brighter, which is roughly consistent with the increase in the peak magnitudes observed in 2016 in the $V$ and $u'$ bands (Figure~\ref{optical_lc}).

However, the time from the optical maximum to $t_{\rm on}$ of the SSS phase should become longer by 
\begin{multline*}
\Delta t = \frac{\Delta M_{\rm env}} { \Delta \dot M_{\rm wind} + \dot M_{\rm wind}}\\
= \frac{M_{\rm env}}{\dot M_{\rm wind}} \frac{\Delta M_{\rm env}/M_{\rm env}}{\Delta \dot M_{\rm wind}/ \dot M_{\rm wind} + 1} \sim \frac{6 \times 0.35}{0.2+1}=1.75 \text{ days},
\end{multline*}
\noindent where $M_{\rm env}$ is the hydrogen-rich envelope mass. This is not consistent with the $t_{\rm on}\sim6$ days in the 2016 (and 2013--2015) eruptions.

In general, all models agree that a higher-mass envelope would lead to a stronger, brighter eruption with a larger ejected mass \citep[e.g.][]{1998MNRAS.296..502S,2005ApJ...623..398Y, 2006ApJS..167...59H, 2013ApJ...777..136W}

(2) For the other case of a lower mean accretion rate, we have estimated the ignition mass of the hydrogen-rich envelope, based on the calculations of \citet{2016ApJ...830...40K, 2017ApJ...838..153K}, to be larger by 9\% for the 1.35 times longer recurrence period ($0.91\times 1.35=1.23$ yr). Then, the peak magnitude of the free-free emission is $2.5 \times (\Delta \log \dot M_{\rm wind}) \times 2 = 2.5\times 0.02\times 2 = 0.1$ mag brighter, but the time from the optical maximum to $t_{\rm on}$ of the SSS phase is longer by only
\begin{multline*}
\Delta t = \frac{\Delta M_{\rm env}} { \Delta \dot M_{\rm wind} + \dot M_{\rm wind}}\\
= \frac{M_{\rm env}}{\dot M_{\rm wind}} \frac{\Delta M_{\rm env}/M_{\rm env}}{\Delta \dot M_{\rm wind}/ \dot M_{\rm wind} + 1} \sim \frac{6 \times 0.09}{0.05+1}=0.5 \text{ days}.
\end{multline*}
\noindent The peak brightness of the 2016 outburst is about 0.5 days sooner than those in the 2013, 2014, and 2015 eruptions (see Figure\,\ref{fastphot} left).  These two features, the $\sim 0.1$ mag brighter and 0.5 days earlier peak, are roughly consistent with the 2016 eruption except for the $\sim 1$ mag brighter cusp (Figure\,\ref{fastphot} left).

Observationally, we have shown that the expansion velocities of the 2016 eruption were comparable to previous outbursts (Section~\ref{sec:vis_spec}). Together with the comparable SSS turn-on time scale (Section~\ref{sec:xrt_lc}) this strongly suggests that a similar amount of material was ejected. Therefore, scenario (2) would be preferred here.

It should be emphasized that neither scenario addresses the short-lived, cuspy nature of the peak in contrast to the relatively similar light curves before or after it occurred. The models of \citet{2017ApJ...838..153K} and their earlier studies would predict a smooth light curve with brighter peak and different rise and decline rates.

Ultimately, scenario (2) would also require an explanation of what caused the accretion rate to decrease.  The late decline photometry of the 2015 eruption indicated that the accretion disk survived that eruption (\hstphot), however, we have no data from 2013 or 2014 with which to compare the end of that eruption.  The similarities of the 2013--2015 eruptions would imply that there was nothing untoward about the 2015 eruption that affected the disk in a different manner to the previous eruptions. Therefore the `blame' probably lies with the donor.

The mass transfer rate in cataclysmic variable stars is known to be variable on time scales from minutes to years \citep[e.g.,][and references therein]{1995cvs..book.....W}. The shortest period variations (so called ``flickering"), with typical amplitudes of tenths of a magnitude, are believed to be caused by propagating fluctuations in the local mass accretion rate within the accretion disk \citep{2014MNRAS.438.1233S}. The longer time scale variations that may be relevant to \nova can cause much larger variations in luminosity. In some cases, as in the VY\,Sculptoris stars, the mass transfer from the secondary star can cease altogether for an extended period of time \citep[e.g.,][]{1981ApJ...251..611R,1985ApJ...290..707S}. The VY\,Scl phenomena is believed to be caused by disruptions in the mass transfer rate caused by star spots on the secondary star drifting underneath the L1 point \citep[e.g.,][]{1994ApJ...427..956L,1998ApJ...499..348K, 2004AJ....128.1279H}. It might be possible that a similar mechanism may be acting in \novak, resulting in mass transfer rate variations sufficient to cause the observed small-scale variability in the recurrence time and potentially even larger ``outliers'' as in 2016.

\subsubsection{A shorter SSS phase}
\label{sec:disc_sss}
In this section we aim to explain the significantly shorter duration of the 2016 SSS phase in comparison with previous eruptions and with the help of the theoretical X-ray light curve models of \citet{2017ApJ...838..153K}.

While a high initial accreted mass at the time of ignition leads to a brighter optical peak (as discussed in the previous section), it does not change the duration of the SSS phase, assuming that the WD envelope settles down to a thermal equilibrium when any wind phase stops. For the same WD mass, a larger accreted mass results in a higher wind mass-loss rate but does not affect the evolution after the maximum photospheric radius has been reached \citep[e.g.,][]{2006ApJS..167...59H}. The shorter SSS duration and thus the shorter duration of the total outburst compared to previous years (Figure~\ref{fig:xrt_xmm_lc}) therefore needs an additional explanation.

\citet{2017ApJ...838..153K} presented a $1.38~M_\sun$ WD model with a mean mass accretion-rate of $1.6\times 10^{-7}\,M_\sun$\,yr$^{-1}$ for \novak. They assumed that the mass-accretion resumes immediately after the wind stops, i.e., at the beginning of the SSS phase. The accretion supplies fresh H-rich matter to the WD and substantially lengthens the SSS lifetime, ``re-feeding'' the SSS, because the mass-accretion rate is the same order as the proposed steady hydrogen shell-burning rate of $\sim 5 \times 10^{-7}\,M_\sun$\,yr$^{-1}$.  If the accretion does not resume during the SSS phase, or only with a reduced rate, then the SSS duration becomes shorter. This effect is model-independent.

\begin{figure}
\includegraphics[width=\columnwidth]{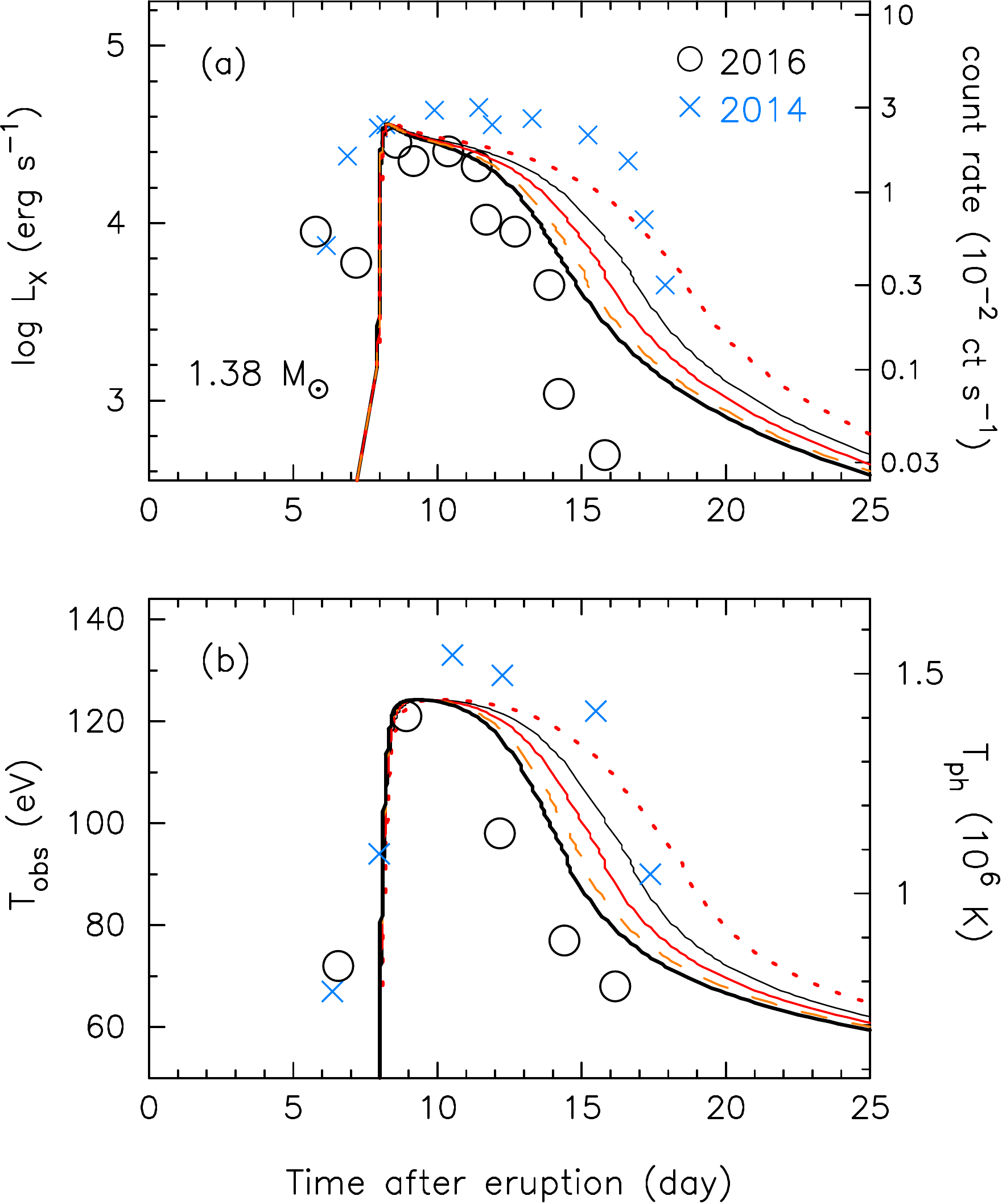}
\caption{Comparison of the theoretical light curve models with the observational data of the 2016 (open black circles, cf.\ Tables~\ref{tab:swift} and \ref{tab:xmm}) and 2014 (blue crosses; cf.\ \xtwok) eruptions. The 2014 temperatures were re-analyzed assuming the updated \nh = $0.7$ \hcm{21} (cf.\ Section~\ref{sec:xrt_lc}). The theoretical model light curves are based on a $1.38\,M_\sun$ WD with a mass accretion rate of $1.6\times 10^{-7}\,M_\sun$\,yr$^{-1}$ \citep{2017ApJ...838..153K}. The five theoretical curves correspond to the cases of no accretion (thick black lines), and factors of 0.3 (dashed orange lines), 0.65 (solid red lines), 1.0 (thin solid black lines), and 1.5 (dotted red lines) times the original mass accretion rate of $1.6\times 10^{-7}\,M_\sun$\,yr$^{-1}$. (a) Theoretical model X-ray light curves (0.3--1.0\,keV). (b) Theoretical model photospheric blackbody temperature. There is a clear trend towards a shorter SSS phase for weaker accretion. Improved models are needed to fit the observations with higher accuracy.}
\label{fig:sss_model}
\end{figure}

To give a specific example, we calculate the SSS light curves and photospheric temperature evolution for various, post-eruption, mass-accretion rates and plot them in Figure~\ref{fig:sss_model}. Those are not fits to the data but models that serve the purpose of illustrating the observable effect of a gradually dimished post-eruption re-feeding. The thick solid black lines denote the case of no post-eruption accretion (during the SSS phase). The thin solid black lines represent the case that the mass-accretion resumes post-eruption with $1.6\times 10^{-7}\,M_\sun$\,yr$^{-1}$, just after the optically thick winds stop. The orange dashed, solid red, dotted red lines correspond to the mass-accretion rates of 0.3, 0.65, and 1.5 times the original mass-accretion rate of $1.6\times 10^{-7}\,M_\sun$\,yr$^{-1}$, respectively.

It is clearly shown that a higher post-eruption mass-accretion rate produces a longer SSS phase. Figure~\ref{fig:sss_model}a shows the X-ray count rates in the 2014 (blue crosses) and 2016 (open black circles) eruptions. The ordinate of the X-ray count rate is vertically shifted to match the theoretical X-ray light curves (cf.\ Figure~\ref{fig:xrt_xmm_lc}). The model X-ray flux drops earlier for a lower mass-accretion rate, which could (as a trend) explain the shorter duration of the 2016 SSS phase. 

Figure~\ref{fig:sss_model}b shows the evolution of the blackbody temperature obtained from the \swift spectra with the neutral hydrogen column density of \nh = $0.7$ \hcm{21} (cf.\ Figure~\ref{fig:xrt_xmm_lc} and Section~\ref{sec:xrt_lc}). The lines show the photospheric temperature of our models. The model temperature decreases earlier for a lower mass-accretion rate. This trend is also consistent with the difference between the 2014 and 2016 eruptions. 

Thus, the more rapid evolution of the SSS phase in the 2016 eruption can be partly understood if mass-accretion does not resume soon after the wind stops (zero accretion, thick black line in Figure~\ref{fig:sss_model}). Note, that the observed change in SSS duration clearly has a larger magnitude than the models (Figure~\ref{fig:sss_model}). This could indicate deficiencies in the current models and/or that additional effects contributed to the shortening of the 2016 SSS phase. One factor that has an impact on the SSS duration is the chemical composition of the envelope \citep[e.g.,][]{2005A&A...439.1061S}. However, it would be difficult to explain why the abundances of the accreted material would suddenly change from one eruption to the next. In any case, our observations make a strong case for a discontinued re-feeding of the SSS simply by comparing the observed parameters of the 2016 eruption to previous outbursts. The models are consistent with the general trend but need to be improved to be able to simulate the magnitude of the effect.

\hstphot\ presented evidence that the accretion disk survives eruptions of \novak, the 2015 eruption specifically. In Section~\ref{sec:disc_peak} we found that the accretion rate prior to the 2016 eruption might have been lower. If this lower accretion rate was caused by a lower mass-transfer rate from the companion, which is a reasonable possibility, then this would lead to a less massive disk (which was potentially less luminous; see Henze et al.\ 2018, in prep.). Thus, even if the eruption itself was not stronger than in previous years, as evidenced by the consistent ejection velocities (Section~\ref{sec:vis_spec}) and SSS turn-on time scale (Section~\ref{sec:xrt_lc}), it could still lead to a greater disruption of such a less massive disk. A part of the inner disk mass may be lost, which could prevent or hinder the reestablishment of mass accretion while the SSS is still active.

This scenario can consistently explain the trends toward a brighter optical peak and a shorter SSS phase for the delayed 2016 eruption. Understanding the \textit{quantitative} magnitude of these changes, and fitting the theoretical light curves more accurately to the observed fluxes, requires additional models that can be tested in future eruptions of \novak. In addition, we strongly encourage the community to contribute alternative interpretations and models that could help us to understand the peculiar 2016 outburst properties.

\subsubsection{Similar features in archival data?}\label{similar}

\label{sec:disc_arch}
Intriguingly, there is tentative evidence that the characteristic features of the 2016 eruption, namely the bright optical peak and the short SSS phase, might have been present in previous eruptions. Here we discuss briefly the corresponding observational data.

Recall that in X-rays there were two serendipitous detections with ROSAT \citep{1982AdSpR...2..241T} in early 1992 and 1993 (see Table~\ref{eruption_history}). \citet{1995ApJ...445L.125W} studied the resulting light curves and spectra in detail. Their Figure~2 shows that in both years the ROSAT coverage captured the beginning of the SSS phase. By chance, the time-axis zero points in these plots are shifted by almost exactly one day with respect to the eruption date as inferred from the rise of the SSS flux; This means that, for example, their day 5 corresponds to day 4 after eruption.

While the 1992 X-ray light curve stops around day eight, the 1993 coverage extends towards day 13 \citep{1995ApJ...445L.125W}. Both light curves show the early SSS variability expected from \nova (cf.\ Figure~\ref{fig:xrt_split}), but in 1993 the last two data points, near days 12 and 13, have lower count rates than expected from a ``regular'', 2015-type eruption (cf.\ Figure~\ref{fig:xrt_xmm_lc}). At this stage of the eruption, we would expect the light curve variations to become significantly lower (see also \othreek).

Of course, these are only two data points. However, the corresponding count rate uncertainties are relatively small and at face value these points are more consistent with the 2016-style early X-ray decline than with the 2015 SSS phase which was still bright at this stage (Figure~\ref{fig:xrt_xmm_lc}). Thus, it is possible that the 1993 eruption had a similarly short SSS phase as the 2016 eruption. The $\sim341$\,d between the 1992 and 1993 eruptions (Table~\ref{eruption_history}), however, are well consistent with the 2008--2015 median of 347\,d and suggest no significant delay.

The short-lived, bright, optical cuspy peak seen from the $I$-band to the UV (see Figures\,\ref{optical_lc}, \ref{optical_zoom}, and \ref{fastphot} left) from the 2016 eruption may have also been seen in 2010.  The 2010 eruption of \novak\ was not discovered in real-time, but was instead recovered from archival observations (\halfk).  The 2010 eruption was only detected in two observations taken just 50\,minutes apart, but it appeared up to 0.6\,mag brighter than the 2013 and 2014 eruptions (and subsequently 2015).  As the 2010 observations were unfiltered, \halfk\ noted that the uncertainties on those observations were possibly dominated by calibration systematics -- the relative change in brightness is significant.  The 2010 photometry is compared with the 2016 photometry in Figure~\ref{2010comp} (right), the epoch of the 2010 data was arbitrarily marked as $t=0.7$\,d.  It is clear from Figure~\ref{2010comp} (right), that the bright peak seen in 2016 is not inconsistent with the data from 2010.  But it is also clear from Figure~\ref{2010comp} (right) that the unfiltered data again illustrate that, other than the cusp itself, the 2016 light curve is similar to those of the 2013--15 eruptions.  Indeed, these unfiltered data have much less of a gap around the $t=1$\,d peak (as seen in 2013--15) than the filtered data do (see Figures~\ref{optical_lc} and \ref{optical_zoom}). 

However, despite this tentative evidence of a previous `cusp', the 2010 eruption fits the original recurrence period model very well.  In fact, it was the eruption that confirmed that original model.  So the 2010 eruption appears to have behaved `normally' -- but we do note the extreme sparsity of data from 2010.  So we must question whether the two deviations from the norm in 2016, the bright cuspy peak, and the X-ray behavior are causally related.

Additionally, we must ask whether the short-lived bright cuspy peak is normal behavior.  Figure~\ref{fastphot} (left) demonstrates this conundrum well.  As noted in Section~\ref{sec:disc_date}, the epoch of the 2016 eruption has been identified simply by the availability of pre-/post-eruption data, $t=0$ has not been tuned (as in 2013--2015) to minimize light curve deviations or based on any other factors.  The final rise light curve data from 2013--2015 is sparse, indeed much more data have been collected during this phase in 2016 than in 2013--2015 combined, including the two-color fast-photometry run from the INT -- in fact, improving the final rise data coverage was a specified pre-eruption goal for 2016.  Figure~\ref{fastphot} (left) indicates that should such a short-lived bright peak have occurred in any of 2013, 2014, or 2015, and given our light curve coverage of those eruptions, we may not have detected it.  Under the assumption that the eruption times of the 2013--2016 eruptions have been correctly accounted for, we would not have detected a `2016 cuspy maximum' in each of 2013, 2014, or 2015.  It is also worth noting that the final rise of the 2016 eruption was poorly covered in the $B$-band (as in all filters in previous years), and no sign of this cuspy behavior is seen in that band!  The UV data may shed more light, but we note the unfortunate inconsistency of filters.  

In conclusion, we currently don't have enough final rise data to securely determine whether the 2016 cuspy peak is unusual. However, the planned combination of rapid follow-up and high cadence observations of future eruptions are specifically designed to explore the early time evolution of the eruptions.

\subsection{What caused the cusp?}\label{cusp}

Irrespective of any causal connection between the late 2016 eruption and the newly observed bright cusp, the smooth light curve models can not explain the nature of this new feature.  As the cusp `breaks' the previously smooth presentation of the observed light curve and the inherently smooth nature of the model light curves, it must be due to an additional, unconsidered, parameter of the system.  Here we briefly discuss a number of possible causes in no particular order.

The cusp could in principle be explained as the shock-breakout associated with the initial thermonuclear runaway, but with evidence of a slower light curve evolution preceding the cusp (see Figure~\ref{fastphot} left) the timescales would appear incompatible. 

An additional consideration would be the interaction between the ejecta and the donor.  Under the assumption of a Roche lobe-filling donor, \hstphot\ proposed a range of WD--donor orbital separations of $25-44\,R_\odot$,  those authors also indicated that much larger separations were viable if accretion occured from the wind of the donor.  Assuming Roche lobe overflow and typical ejecta velocities at the epoch of the cusp of $\sim4000$\,km\,s$^{-1}$ (see the bottom right plot of Figure~\ref{fig:spec3}), one would expect an ejecta--donor interaction to occur 0.02--0.06\,days post-eruption (here we have also accounted for the radius of the donor, $R\simeq14\,R_\odot$; \hstphot).   With the cusp seemingly occurring 0.65\,days post-eruption, the orbital separation would need to be $\sim330\,R_\odot$ ($\sim1.6$\,au). From this we would infer an orbital period in the range $350-490$\,days (i.e., $\gtrsim P_\mathrm{rec}$), depending on the donor mass, and mass transfer would occur by necessity through wind accretion.  We note that the eruption time uncertainty ($\pm0.17$\,d) has little effect on the previous discussion. \othreek, \hstspec, and \hstphot\ all argued that the system inclination must be low, despite this it is still possible that the observation of such an ejecta--donor interaction may depend upon the orbital phase (with respect to the observer) at the time of eruption.

As a final discussion point, we note that \othreek\ and \hstspec\ both presented evidence of highly asymmetric ejecta; proposing an equatorial component almost in the plane of the sky, and a freely expanding higher-velocity -- possibly collimated -- polar outflow directed close to the line-of-sight.  We also note that the velocity difference between these components may be a factor of three or higher.  If we treat these components as effectively independent ejecta, we would therefore expect their associated light curves to evolve at different rates, with the polar component showing the more rapid evolution.  Therefore, we must ask whether the `normal' (2013--2015) light curve is that of the `bulk' equatorial ejecta, and the `cusp' is the first photometric evidence of the faster evolving polar ejecta?  We note that such proposals have also been put forward to explain multi-peak light curves from other phenomena, for example, kilonovae \citep[see][and the references therein]{2017ApJ...851L..21V}.

\subsection{Predicting the date of the next eruption(s)}
\label{sec:disc_next}

A consequence of the delayed 2016 eruption is that the dates of the next few eruptions are much more difficult to predict than previously thought. Figure~\ref{fig:rec_time} demonstrates how much this surprising delay disrupted the apparently stable trend toward eruptions occurring successively earlier in the year (and Section~\ref{sec:disc_pecul} discusses the possible reasons).

Currently, detailed examinations of the statistical properties of the recurrence period distribution are hampered by the relatively small number of nine eruptions, and thereby eight different gaps, since 2008 (cf.\ Table~\ref{eruption_history}). \nova is the only known nova for which we will overcome this limitation in the near future. For now, we cannot reject the hypothesis that the gaps follow a Gaussian distribution, with Lilliefors (Kolmogorov-Smirnov) test p-value $\sim 0.11$, even with the long delay between 2015 and 2016. The distribution mean (median) is 363\,d (347\,d), with a standard deviation of 52\,d. Thereby, the 472 days prior to the 2016 eruption could indicate a genuine outlier, a skewed distribution, or simply an extreme variation from the mean. It is too early to tell.

In addition, all these gaps of roughly 1~yr length would be affected by the presence of an underlying 6-month period which could dampen the more extreme swings. Of course, the original prediction of a half-year period by \halfk\ was partly based on the apparently stable trend toward earlier eruptions since 2008. Comparing this recent trend to the dates of historical X-ray detections in 1992, 1993, and 2001 (\xonek), \halfk\ found that the most parsimonious explanation for the observed discrepancies between the two regimes would be a 6-month shift. However, the putative 6-month eruption still remains to be found (Henze et al.\ 2018, in prep.). At present, a single eruption deviating from this pattern does not present sufficient evidence to discard the 6-month scenario. The next (few) eruption date(s) will be crucial in evaluating the recurrence period statistics.

While this manuscript was with the referee, the next eruption was discovered on 2017 Dec 31 \citep{2017ATel11116....1B}. The $\sim384$\,d gap between the 2016 and 2017 eruptions is consistent with the pre-2016 eruption pattern. A comprehensive multi-wavelength analysis of the new eruption will be presented in a subsequent work.

\section{Summary \& Conclusions}\label{sec:conclusions}

\begin{enumerate}

\item The 2016 eruption occurred on December 12.32 UT, which was 472 days after the 2015 eruption. Thereby, it appeared to interrupt the general trend of eruptions since 2008 occurring slightly earlier in the year (with $t_\mathrm{rec} = 347\pm10$\,d).

\item The 2016 eruption light curve exhibited a short lived `cuspy' peak between $0.7\leq t \leq 0.9$\,days post-eruption, around 0.5 magnitudes brighter than the smooth peak at $t\simeq1$\,d observed in previous eruptions.  This aside, the optical and UV light curve developed in a very similar manner to the 2013/2014/2015 eruptions.

\item The cuspy peak occurs during a previously unsampled portion of the light curve.  Therefore we cannot rule out this being a `normal' feature that has previously been missed.  There is tentative evidence of a similar occurrence during the 2010 eruption.

\item The first 2016 outburst spectrum, taken 0.54\,d after the eruption, was one of the earliest spectra taken of any \nova eruption. From this we identified P\,Cygni profiles in the optical spectrum of \nova for the first time, indicating an expansion velocity of $\sim6200$\,km\,s$^{-1}$. In addition, a late spectrum taken 5.83\,d after eruption revealed narrow He~{\sc ii} emission, possibly arising from the surviving accretion disk. There is however no evidence that the spectroscopic evolution of the 2016 eruption deviated significantly from the behavior in previous years.

\item The \swift XRT light curve deviated significantly from the previous behavior. The flux started to decline around day 11 which is several days earlier than expected. In a consistent way, the evolution of the effective temperature was similar to the 2013--2015 eruptions until day 11 but afterwards decreased significantly earlier. A 100\,ks \xmm ToO observation, split into two pointings, managed to characterize the decaying SSS flux and temperature to be consistent with the XRT data and discovered surprising, strong variability at a stage that had previously suggested only marginal variation.

\item The tendency of the changes in recurrence period, optical peak brightness, and SSS duration can be consistently described in early theoretical model calculations. When we assume a lower accretion rate we find that this (i) increases the time between eruptions, (ii) leads to a less-massive disk the disruption of which delays the onset of mass-accretion and shortens the SSS phase, and (iii) increases the ignition mass and thereby the peak magnitude. This scenario will need to be explored in more detail in the future. We also strongly encourage alternative models and interpretations.

\end{enumerate}

\acknowledgements
\textit{We are deeply indebted to the late Swift PI Neil Gehrels for his long-term support of our project and for giving our community the game-changing Swift observatory. This paper is dedicated to his memory.}

We thank the anonymous referee for their constructive comments that helped to improve the paper.

We are, as always, grateful to the \swift Team for making the ToO observations possible, in particular the duty scientists as well as the science planners. This research made use of data supplied by the UK Swift Science Data Centre at the University of Leicester.

Based on observations obtained with \xmmk, an ESA science mission with instruments and contributions directly funded by ESA Member States and NASA.

Based on observations made with the NASA/ESA Hubble Space Telescope, obtained from the Data Archive at the Space Telescope Science Institute (STScI), which is operated by the Association of Universities for Research in Astronomy, Inc., under NASA contract NAS 5-26555.  These observations are associated with program \#14651.  Support for program \#14651 was provided by NASA through a grant from STScI.

The Liverpool Telescope is operated on the island of La Palma by Liverpool John Moores University (LJMU) in the Spanish Observatorio del Roque de los Muchachos of the Instituto de Astrof\'{i}sica de Canarias with financial support from STFC.  

This work makes use of observations from the LCO network. 

Based (in part) on data collected with the Danish 1.54-m telescope at the ESO La Silla Observatory.

The data presented here were obtained in part with ALFOSC, which is provided by the Instituto de Astrofisica de Andalucia (IAA) under a joint agreement with the University of Copenhagen and NOTSA. 

The Hobby-Eberly Telescope (HET) is a joint project of the University of Texas at Austin, the Pennsylvania State University, Stanford University, Ludwig-Maximilians-Universit\"at M\"unchen, and Georg-August-Universit\"at G\"ottingen. The HET is named in honor of its principal benefactors, William P.\ Hobby and Robert E.\ Eberly.

The LBT is an international collaboration among institutions in the United States, Italy and Germany. LBT Corporation partners are: The University of Arizona on behalf of the Arizona Board of Regents; Istituto Nazionale di Astrofisica, Italy; LBT Beteiligungsgesellschaft, Germany, representing the Max-Planck Society, The Leibniz Institute for Astrophysics Potsdam, and Heidelberg University; The Ohio State University, and The Research Corporation, on behalf of The University of Notre Dame, University of Minnesota and University of Virginia.

The Pirka telescope is operated by Graduate School of Science, Hokkaido University, and it also participates in the Optical \& Near-Infrared Astronomy Inter-University Cooperation Program, supported by the MEXT of Japan.

We acknowledge with thanks the variable star observations from the AAVSO International Database contributed by observers worldwide and used in this research.

The Institute for Gravitation and the Cosmos is supported by the Eberly College of Science and the Office of the Senior Vice President for Research at the Pennsylvania State University.

We wish to thank G.\ Mansir for sharing her observing time at the 1.54\,m Danish telescope on December 15th. We acknowledge G.\ Zeimann who reduced the HET spectra. We thank T.\ Johnson for assisting in observations at MLO. We wish to acknowledge Luc\'\i a Su\'arez-Andr\'es  (ING) for obtaining the INT observations on a Director's Discretionary Time night, generously awarded by Marc Balcells (ING) and Cecilia Fari\~na (INT) to our collaboration. 

M.\ Henze acknowledges the support of the Spanish Ministry of Economy and Competitiveness (MINECO) under the grant FDPI-2013-16933, the support of the Generalitat de Catalunya/CERCA programme, and the hospitality of the Liverpool John Moores University during collaboration visits.

S.~C.\ Williams acknowledges a visiting research fellowship at Liverpool John Moores University. 

M.\ Kato and I.\ Hachisu acknowledge support in part by Grants-in-Aid for Scientific Research (15K05026, 16K05289) of the Japan Society for the Promotion of Science. 

G.\ Anupama and M.\ Pavana thank the HCT observers who spared part of their time for the observations.

K.\ Chinetti acknowledges support by the GROWTH project funded by the National Science Foundation under Grant No.\ 1545949.

P.\ Godon wishes to thank William (Bill) P.\ Blair for his kind hospitality in the Rowland Department of Physics \& Astronomy at the Johns Hopkins University.

M.\ Hernanz acknowledges MINECO support under the grant ESP2015\_66134\_R as well as the support of the Generalitat de Catalunya/CERCA program.

K.\ Hornoch, H.\ Ku\v{c}\'akov\'a and J.\ Vra\v{s}til were supported by the project RVO:67985815.

R.\ Hounsell acknowledges support from the HST grant 14651.

E.\ Paunzen acknowledges support by the Ministry of Education of the Czech Republic (grant LG15010).

V.\ Ribeiro acknowledges financial support from Funda\c{c}\~{a}o para a Ci\^encia e a Tecnologia (FCT) in the form of an exploratory project of reference IF/00498/2015, from Center for Research \& Development in Mathematics and Applications (CIDMA) strategic project UID/MAT/04106/2013 and supported by Enabling Green E-science for the Square Kilometer Array Research Infrastructure (ENGAGESKA), POCI-01-0145-FEDER-022217, funded by Programa Operacional Competitividade e Internacionaliza\c{c}\~ao (COMPETE 2020) and FCT, Portugal.

P. Rodr\'\i guez-Gil acknowledges support by a Ram\'on y Cajal fellowship (RYC2010--05762). The use of Tom Marsh's {\sc pamela} package is gratefully acknowledged.

K.~L.\ Page and J.~P.\ Osborne acknowledge the support of the UK Space Agency.

T. Oswalt acknowledges support from U.S. NSF grant AST-1358787.

S.\ Starrfield acknowledges partial support to ASU from NASA and HST grants.

This research has made use of ``Aladin sky atlas" developed at CDS, Strasbourg Observatory, France.

This research has made use of the SIMBAD database, operated at CDS, Strasbourg, France.

PyRAF is a product of the Space Telescope Science Institute, which is operated by AURA for NASA.

We wish to thank the Observatorio Astrofisico de Javalambre Data Processing and Archiving Unit (UPAD) for reducing and calibrating the JAST/T80 data.

\facilities{AAVSO, ARC, Danish 1.54m Telescope, FTN, HCT, HET, HST (WFC3), ING:Newton, LBT, LCO, Liverpool:2m, Mayall, MLO:1m, NOT, OAO:0.5m, OO:0.65, PO:1.2m, ERAU:0.6m, ERAU:1m, Swift, XMM}

\software{AIP4WIN, Aladin \citep[v9;][]{2000A&AS..143...33B,2014ASPC..485..277B}, APAS \citep{2007ApJ...661L..45D}, APHOT \citep{1994ExA.....5..375P}, {\tt calwf3} \citep[v3.4;][]{2012wfci.book.....D}, DOLPHOT \citep[v2.0;][]{2000PASP..112.1383D}, FotoDif (v3.95), HEASOFT (v6.16), IRAF \citep[v2.16 and v2.16.1,][]{1993ASPC...52..173T}, MaxIm DL (v5.23), Mira Pro x64 (v8.011 and v8.012), Panacea, PGPLOT (v5.2), PyRAF, R \citep{R_manual}, SAOImage DS9, Starlink \citep[v2016A,][]{1982QJRAS..23..485D}, Source Extractor \citep[v2.8.6;][]{1996A&AS..117..393B}, SWarp \citep[v2.19.1][]{2002ASPC..281..228B}, XIMAGE (v4.5.1), XMM-SAS (v15.0.0), XSELECT (v2.4c), XSPEC \citep[v12.8.2;][]{1996ASPC..101...17A}}

\bibliographystyle{aasjournal}
\bibliography{refs} 

\begin{thebibliography}{}
\expandafter\ifx\csname natexlab\endcsname\relax\def\natexlab#1{#1}\fi

\bibitem[{{Arnaud}(1996)}]{1996ASPC..101...17A}
{Arnaud}, K.~A. 1996, in Astronomical Society of the Pacific Conference Series,
  Vol. 101, Astronomical Data Analysis Software and Systems V, ed.
  {G.~H.~Jacoby \& J.~Barnes}, 17--+

\bibitem[{{Balucinska-Church} \& {McCammon}(1992)}]{1992ApJ...400..699B}
{Balucinska-Church}, M., \& {McCammon}, D. 1992, \apj, 400, 699

\bibitem[{{Barsukova} {et~al.}(2011){Barsukova}, {Fabrika}, {Hornoch},
  {Fatkhullin}, {Sholukhova}, \& {Pietsch}}]{2011ATel.3725....1B}
{Barsukova}, E., {Fabrika}, S., {Hornoch}, K., {et~al.} 2011, ATel, 3725, 1

\bibitem[{{Bertin} \& {Arnouts}(1996)}]{1996A&AS..117..393B}
{Bertin}, E., \& {Arnouts}, S. 1996, \aaps, 117, 393

\bibitem[{{Bertin} {et~al.}(2002){Bertin}, {Mellier}, {Radovich}, {Missonnier},
  {Didelon}, \& {Morin}}]{2002ASPC..281..228B}
{Bertin}, E., {Mellier}, Y., {Radovich}, M., {et~al.} 2002, in Astronomical
  Society of the Pacific Conference Series, Vol. 281, Astronomical Data
  Analysis Software and Systems XI, ed. {D.~A.~Bohlender, D.~Durand, \&
  T.~H.~Handley}, 228

\bibitem[{{Boch} \& {Fernique}(2014)}]{2014ASPC..485..277B}
{Boch}, T., \& {Fernique}, P. 2014, in Astronomical Society of the Pacific
  Conference Series, Vol. 485, Astronomical Data Analysis Software and Systems
  XXIII, ed. N.~{Manset} \& P.~{Forshay}, 277

\bibitem[{{Bode} \& {Evans}(2008)}]{2008clno.book.....B}
{Bode}, M.~F., \& {Evans}, A., eds. 2008, Cambridge Astrophysics Series,
  Vol.~43, {Classical Novae, 2nd Edition} (Cambridge: Cambridge University
  Press)

\bibitem[{{Bode} {et~al.}(2016){Bode}, {Darnley}, {Beardmore}, {Osborne},
  {Page}, {Walter}, {Krautter}, {Melandri}, {Ness}, {O'Brien}, {Orio},
  {Schwarz}, {Shara}, \& {Starrfield}}]{2016ApJ...818..145B}
{Bode}, M.~F., {Darnley}, M.~J., {Beardmore}, A.~P., {et~al.} 2016, \apj, 818,
  145

\bibitem[{{Bonnarel} {et~al.}(2000){Bonnarel}, {Fernique}, {Bienaym{\'e}},
  {Egret}, {Genova}, {Louys}, {Ochsenbein}, {Wenger}, \&
  {Bartlett}}]{2000A&AS..143...33B}
{Bonnarel}, F., {Fernique}, P., {Bienaym{\'e}}, O., {et~al.} 2000, \aaps, 143,
  33

\bibitem[{{Boyd} {et~al.}(2017){Boyd}, {Hornoch}, {Henze}, {Darnley},
  {Shafter}, {Kafka}, {Kato}, \& {et al.}}]{2017ATel11116....1B}
{Boyd}, D., {Hornoch}, K., {Henze}, M., {et~al.} 2017, The Astronomer's
  Telegram, 11116

\bibitem[{{Breeveld} {et~al.}(2011){Breeveld}, {Landsman}, {Holland}, {Roming},
  {Kuin}, \& {Page}}]{2011AIPC.1358..373B}
{Breeveld}, A.~A., {Landsman}, W., {Holland}, S.~T., {et~al.} 2011, in American
  Institute of Physics Conference Series, Vol. 1358, American Institute of
  Physics Conference Series, ed. J.~E. {McEnery}, J.~L. {Racusin}, \&
  N.~{Gehrels}, 373--376

\bibitem[{{Brown} {et~al.}(2013){Brown}, {Baliber}, {Bianco}, {Bowman},
  {Burleson}, {Conway}, {Crellin}, {Depagne}, {De Vera}, {Dilday}, {Dragomir},
  {Dubberley}, {Eastman}, {Elphick}, {Falarski}, {Foale}, {Ford}, {Fulton},
  {Garza}, {Gomez}, {Graham}, {Greene}, {Haldeman}, {Hawkins}, {Haworth},
  {Haynes}, {Hidas}, {Hjelstrom}, {Howell}, {Hygelund}, {Lister}, {Lobdill},
  {Martinez}, {Mullins}, {Norbury}, {Parrent}, {Paulson}, {Petry}, {Pickles},
  {Posner}, {Rosing}, {Ross}, {Sand}, {Saunders}, {Shobbrook}, {Shporer},
  {Street}, {Thomas}, {Tsapras}, {Tufts}, {Valenti}, {Vander Horst}, {Walker},
  {White}, \& {Willis}}]{2013PASP..125.1031B}
{Brown}, T.~M., {Baliber}, N., {Bianco}, F.~B., {et~al.} 2013, \pasp, 125, 1031

\bibitem[{{Burke} {et~al.}(2016){Burke}, {Kaur}, {Oswalt}, {Erdman},
  {Hartmann}, {Henze}, {Darnley}, {Hornoch}, \&
  {Kucakova}}]{2016ATel.9861....1B}
{Burke}, D., {Kaur}, A., {Oswalt}, T., {et~al.} 2016, ATel, 9861

\bibitem[{{Burrows} {et~al.}(2005){Burrows}, {Hill}, {Nousek}, {Kennea},
  {Wells}, {Osborne}, {Abbey}, {Beardmore}, {Mukerjee}, {Short}, {Chincarini},
  {Campana}, {Citterio}, {Moretti}, {Pagani}, {Tagliaferri}, {Giommi},
  {Capalbi}, {Tamburelli}, {Angelini}, {Cusumano}, {Br{\"a}uninger}, {Burkert},
  \& {Hartner}}]{2005SSRv..120..165B}
{Burrows}, D.~N., {Hill}, J.~E., {Nousek}, J.~A., {et~al.} 2005, \ssr, 120, 165

\bibitem[{{Cash}(1979)}]{1979ApJ...228..939C}
{Cash}, W. 1979, \apj, 228, 939

\bibitem[{{Chandrasekhar}(1931)}]{1931ApJ....74...81C}
{Chandrasekhar}, S. 1931, \apj, 74, 81

\bibitem[{{Chonis} {et~al.}(2014){Chonis}, {Hill}, {Lee}, {Tuttle}, \&
  {Vattiat}}]{2014SPIE.9147E..0AC}
{Chonis}, T.~S., {Hill}, G.~J., {Lee}, H., {Tuttle}, S.~E., \& {Vattiat}, B.~L.
  2014, in \procspie, Vol. 9147, Ground-based and Airborne Instrumentation for
  Astronomy V, 91470A

\bibitem[{{Chonis} {et~al.}(2016){Chonis}, {Hill}, {Lee}, {Tuttle}, {Vattiat},
  {Drory}, {Indahl}, {Peterson}, \& {Ramsey}}]{2016SPIE.9908E..4CC}
{Chonis}, T.~S., {Hill}, G.~J., {Lee}, H., {et~al.} 2016, in \procspie, Vol.
  9908, Ground-based and Airborne Instrumentation for Astronomy VI, 99084C

\bibitem[{{Darnley}(2016)}]{2016ATel.9910....1D}
{Darnley}, M.~J. 2016, ATel, 9910

\bibitem[{{Darnley}(2017)}]{2017ASPC..509..515D}
{Darnley}, M.~J. 2017, in Astronomical Society of the Pacific Conference
  Series, Vol. 509, 20th European White Dwarf Workshop, ed. P.-E. {Tremblay},
  B.~{Gaensicke}, \& T.~{Marsh}, 515

\bibitem[{{Darnley} {et~al.}(2018){Darnley}, {Healy}, {Henze}, \&
  {Williams}}]{2018ATel11149....1D}
{Darnley}, M.~J., {Healy}, M.~W., {Henze}, M., \& {Williams}, S.~C. 2018, The
  Astronomer's Telegram, 11149

\bibitem[{{Darnley} {et~al.}(2015{\natexlab{a}}){Darnley}, {Henze}, {Shafter},
  \& {Kato}}]{2015ATel.7964....1D}
{Darnley}, M.~J., {Henze}, M., {Shafter}, A.~W., \& {Kato}, M.
  2015{\natexlab{a}}, ATel, 7964, 1

\bibitem[{{Darnley} {et~al.}(2015{\natexlab{b}}){Darnley}, {Henze}, {Shafter},
  \& {Kato}}]{2015ATel.7965....1D}
---. 2015{\natexlab{b}}, ATel, 7965, 1

\bibitem[{{Darnley} \& {Hounsell}(2016)}]{2016ATel.9874....1D}
{Darnley}, M.~J., \& {Hounsell}, R. 2016, ATel, 9874

\bibitem[{{Darnley} {et~al.}(2016{\natexlab{a}}){Darnley}, {Kuin}, {Page},
  {Osborne}, {Schwarz}, {Shore}, {Starrfield}, \&
  {Williams}}]{2016ATel.8587....1D}
{Darnley}, M.~J., {Kuin}, N.~P.~M., {Page}, K.~L., {et~al.} 2016{\natexlab{a}},
  ATel, 8587

\bibitem[{{Darnley} {et~al.}(2016{\natexlab{b}}){Darnley}, {Rodriguez-Gil}, \&
  {Prieto-Arranz}}]{2016ATel.9852....1D}
{Darnley}, M.~J., {Rodriguez-Gil}, P., \& {Prieto-Arranz}, J.
  2016{\natexlab{b}}, ATel, 9852

\bibitem[{{Darnley} {et~al.}(2014){Darnley}, {Williams}, {Bode}, {Henze},
  {Ness}, {Shafter}, {Hornoch}, \& {Votruba}}]{2014A&A...563L...9D}
{Darnley}, M.~J., {Williams}, S.~C., {Bode}, M.~F., {et~al.} 2014, \aap, 563,
  L9

\bibitem[{{Darnley} {et~al.}(2016{\natexlab{c}}){Darnley}, {Williams}, {Henze},
  {Shafter}, {Kafka}, \& {Kato}}]{2016ATel.9906....1D}
{Darnley}, M.~J., {Williams}, S.~C., {Henze}, M., {et~al.} 2016{\natexlab{c}},
  ATel, 9906

\bibitem[{{Darnley} {et~al.}(2007){Darnley}, {Kerins}, {Newsam}, {Duke},
  {Gould}, {Han}, {Ibrahimov}, {Im}, {Jeon}, {Karimov}, {Lee}, \&
  {Park}}]{2007ApJ...661L..45D}
{Darnley}, M.~J., {Kerins}, E., {Newsam}, A., {et~al.} 2007, \apjl, 661, L45

\bibitem[{{Darnley} {et~al.}(2015{\natexlab{c}}){Darnley}, {Henze}, {Steele},
  {Bode}, {Ribeiro}, {Rodr{\'{\i}}guez-Gil}, {Shafter}, {Williams}, {Baer},
  {Hachisu}, {Hernanz}, {Hornoch}, {Hounsell}, {Kato}, {Kiyota}, {Ku{\v
  c}{\'a}kov{\'a}}, {Maehara}, {Ness}, {Piascik}, {Sala}, {Skillen}, {Smith},
  \& {Wolf}}]{2015A&A...580A..45D}
{Darnley}, M.~J., {Henze}, M., {Steele}, I.~A., {et~al.} 2015{\natexlab{c}},
  \aap, 580, A45

\bibitem[{{Darnley} {et~al.}(2016{\natexlab{d}}){Darnley}, {Henze}, {Bode},
  {Hachisu}, {Hernanz}, {Hornoch}, {Hounsell}, {Kato}, {Ness}, {Osborne},
  {Page}, {Ribeiro}, {Rodr{\'{\i}}guez-Gil}, {Shafter}, {Shara}, {Steele},
  {Williams}, {Arai}, {Arcavi}, {Barsukova}, {Boumis}, {Chen}, {Fabrika},
  {Figueira}, {Gao}, {Gehrels}, {Godon}, {Goranskij}, {Harman}, {Hartmann},
  {Hosseinzadeh}, {Horst}, {Itagaki}, {Jos{\'e}}, {Kabashima}, {Kaur}, {Kawai},
  {Kennea}, {Kiyota}, {Ku{\v c}{\'a}kov{\'a}}, {Lau}, {Maehara}, {Naito},
  {Nakajima}, {Nishiyama}, {O'Brien}, {Quimby}, {Sala}, {Sano}, {Sion},
  {Valeev}, {Watanabe}, {Watanabe}, {Williams}, \& {Xu}}]{2016ApJ...833..149D}
{Darnley}, M.~J., {Henze}, M., {Bode}, M.~F., {et~al.} 2016{\natexlab{d}},
  \apj, 833, 149

\bibitem[{{Darnley} {et~al.}(2017{\natexlab{a}}){Darnley}, {Hounsell},
  {O'Brien}, {Rodr{\'{\i}}guez-Gil}, {Shafter}, {Shara}, {Henze}, {Bode},
  {Galera-Rosillo}, {Harman}, {Ness}, {Ribeiro}, {Vaytet}, \&
  {Williams}}]{2017arXiv171204872D}
{Darnley}, M.~J., {Hounsell}, R., {O'Brien}, T.~J., {et~al.}
  2017{\natexlab{a}}, ArXiv e-prints, arXiv:1712.04872

\bibitem[{{Darnley} {et~al.}(2017{\natexlab{b}}){Darnley}, {Hounsell}, {Godon},
  {Perley}, {Henze}, {Kuin}, {Williams}, {Williams}, {Bode}, {Harman},
  {Hornoch}, {Link}, {Ness}, {Ribeiro}, {Sion}, {Shafter}, \&
  {Shara}}]{2017ApJ...849...96D}
{Darnley}, M.~J., {Hounsell}, R., {Godon}, P., {et~al.} 2017{\natexlab{b}},
  \apj, 849, 96

\bibitem[{{Darnley} {et~al.}(2017{\natexlab{c}}){Darnley}, {Hounsell}, {Godon},
  {Perley}, {Henze}, {Kuin}, {Williams}, {Williams}, {Bode}, {Harman},
  {Hornoch}, {Link}, {Ness}, {Ribeiro}, {Sion}, {Shafter}, \&
  {Shara}}]{2017ApJ...847...35D}
---. 2017{\natexlab{c}}, \apj, 847, 35

\bibitem[{{Disney} \& {Wallace}(1982)}]{1982QJRAS..23..485D}
{Disney}, M.~J., \& {Wallace}, P.~T. 1982, \qjras, 23, 485

\bibitem[{{Dolphin}(2000)}]{2000PASP..112.1383D}
{Dolphin}, A.~E. 2000, \pasp, 112, 1383

\bibitem[{{Dressel}(2012)}]{2012wfci.book.....D}
{Dressel}, L. 2012, {Wide Field Camera 3 Instrument Handbook for Cycle 21 v.
  5.0} (Baltimore, MD: STScI)

\bibitem[{{Ederoclite} {et~al.}(2016){Ederoclite}, {Henze}, {Aguado},
  {Allende}, {Williams}, {Darnley}, {Sala}, {Shafter}, \&
  {Hornoch}}]{2016ATel.9281....1E}
{Ederoclite}, A., {Henze}, M., {Aguado}, D., {et~al.} 2016, ATel, 9281

\bibitem[{{Erdman} {et~al.}(2016){Erdman}, {Kaur}, {Oswalt}, {Burke},
  {Hartmann}, {Henze}, {Darnley}, {Shafter}, \& {Horst}}]{2016ATel.9857....1E}
{Erdman}, P., {Kaur}, A., {Oswalt}, T., {et~al.} 2016, ATel, 9857

\bibitem[{{Evans} {et~al.}(2009){Evans}, {Beardmore}, {Page}, {Osborne},
  {O'Brien}, {Willingale}, {Starling}, {Burrows}, {Godet}, {Vetere}, {Racusin},
  {Goad}, {Wiersema}, {Angelini}, {Capalbi}, {Chincarini}, {Gehrels}, {Kennea},
  {Margutti}, {Morris}, {Mountford}, {Pagani}, {Perri}, {Romano}, \&
  {Tanvir}}]{2009MNRAS.397.1177E}
{Evans}, P.~A., {Beardmore}, A.~P., {Page}, K.~L., {et~al.} 2009, \mnras, 397,
  1177

\bibitem[{{Fabrika} {et~al.}(2017){Fabrika}, {Sholukhova}, {Vinokurov},
  {Valeev}, {Sarkisyan}, {Hornoch}, {Henze}, \&
  {Shafter}}]{2017ATel.9942....1F}
{Fabrika}, S., {Sholukhova}, O., {Vinokurov}, A., {et~al.} 2017, ATel, 9942

\bibitem[{{Fabrika} {et~al.}(2016){Fabrika}, {Sholukhova}, {Valeev},
  {Burenkov}, {Makarov}, {Henze}, {Williams}, {Darnley}, {Shafter}, {Hornoch},
  {Ederoclite}, \& {Sala}}]{2016ATel.9383....1F}
{Fabrika}, S., {Sholukhova}, O., {Valeev}, A., {et~al.} 2016, ATel, 9383

\bibitem[{{Fedorov} {et~al.}(2009){Fedorov}, {Myznikov}, \&
  {Akhmetov}}]{2009MNRAS.393..133F}
{Fedorov}, P.~N., {Myznikov}, A.~A., \& {Akhmetov}, V.~S. 2009, \mnras, 393,
  133

\bibitem[{{Freedman} \& {Madore}(1990)}]{1990ApJ...365..186F}
{Freedman}, W.~L., \& {Madore}, B.~F. 1990, \apj, 365, 186

\bibitem[{{Gehrels} {et~al.}(2004){Gehrels}, {Chincarini}, {Giommi}, {Mason},
  {Nousek}, {Wells}, {White}, {Barthelmy}, {Burrows}, {Cominsky}, {Hurley},
  {Marshall}, {M{\'e}sz{\'a}ros}, {Roming}, {Angelini}, {Barbier}, {Belloni},
  {Campana}, {Caraveo}, {Chester}, {Citterio}, {Cline}, {Cropper}, {Cummings},
  {Dean}, {Feigelson}, {Fenimore}, {Frail}, {Fruchter}, {Garmire}, {Gendreau},
  {Ghisellini}, {Greiner}, {Hill}, {Hunsberger}, {Krimm}, {Kulkarni}, {Kumar},
  {Lebrun}, {Lloyd-Ronning}, {Markwardt}, {Mattson}, {Mushotzky}, {Norris},
  {Osborne}, {Paczynski}, {Palmer}, {Park}, {Parsons}, {Paul}, {Rees},
  {Reynolds}, {Rhoads}, {Sasseen}, {Schaefer}, {Short}, {Smale}, {Smith},
  {Stella}, {Tagliaferri}, {Takahashi}, {Tashiro}, {Townsley}, {Tueller},
  {Turner}, {Vietri}, {Voges}, {Ward}, {Willingale}, {Zerbi}, \&
  {Zhang}}]{2004ApJ...611.1005G}
{Gehrels}, N., {Chincarini}, G., {Giommi}, P., {et~al.} 2004, \apj, 611, 1005

\bibitem[{{Hachisu} \& {Kato}(2006)}]{2006ApJS..167...59H}
{Hachisu}, I., \& {Kato}, M. 2006, \apjs, 167, 59

\bibitem[{{Henden} {et~al.}(2016){Henden}, {Templeton}, {Terrell}, {Smith},
  {Levine}, \& {Welch}}]{2016yCat.2336....0H}
{Henden}, A.~A., {Templeton}, M., {Terrell}, D., {et~al.} 2016, VizieR Online
  Data Catalog, 2336

\bibitem[{{Henze} {et~al.}(2015{\natexlab{a}}){Henze}, {Darnley}, {Kabashima},
  {Nishiyama}, {Itagaki}, \& {Gao}}]{2015A&A...582L...8H}
{Henze}, M., {Darnley}, M.~J., {Kabashima}, F., {et~al.} 2015{\natexlab{a}},
  \aap, 582, L8

\bibitem[{{Henze} {et~al.}(2016{\natexlab{a}}){Henze}, {Darnley}, {Shafter},
  {Kafka}, \& {Kato}}]{2016ATel.9853....1D}
{Henze}, M., {Darnley}, M.~J., {Shafter}, A.~W., {Kafka}, S., \& {Kato}, M.
  2016{\natexlab{a}}, ATel, 9853

\bibitem[{{Henze} {et~al.}(2016{\natexlab{b}}){Henze}, {Darnley}, {Shafter},
  {Kafka}, \& {Kato}}]{2016ATel.9853....1H}
---. 2016{\natexlab{b}}, ATel, 9853

\bibitem[{{Henze} {et~al.}(2016{\natexlab{c}}){Henze}, {Darnley}, {Shafter},
  {Kafka}, \& {Kato}}]{2016ATel.9872....1H}
---. 2016{\natexlab{c}}, ATel, 9872

\bibitem[{{Henze} {et~al.}(2016{\natexlab{d}}){Henze}, {Darnley}, {Shafter},
  {Kafka}, \& {Kato}}]{2016ATel.9907....1H}
---. 2016{\natexlab{d}}, ATel, 9907

\bibitem[{{Henze} {et~al.}(2018{\natexlab{a}}){Henze}, {Darnley}, {Shafter},
  {Kafka}, {Kato}, {Williams}, \& {et al.}}]{2018ATel11121....1H}
{Henze}, M., {Darnley}, M.~J., {Shafter}, A.~W., {et~al.} 2018{\natexlab{a}},
  The Astronomer's Telegram, 11121

\bibitem[{{Henze} {et~al.}(2018{\natexlab{b}}){Henze}, {Darnley}, {Shafter},
  {Kafka}, {Kato}, {Williams}, \& {et al.}}]{2018ATel11130....1H}
---. 2018{\natexlab{b}}, The Astronomer's Telegram, 11130

\bibitem[{{Henze} {et~al.}(2014{\natexlab{a}}){Henze}, {Ness}, {Darnley},
  {Bode}, {Williams}, {Shafter}, {Kato}, \& {Hachisu}}]{2014A&A...563L...8H}
{Henze}, M., {Ness}, J.-U., {Darnley}, M.~J., {et~al.} 2014{\natexlab{a}},
  \aap, 563, L8

\bibitem[{{Henze} {et~al.}(2016{\natexlab{e}}){Henze}, {Williams}, {Darnley},
  {Ederoclite}, {Sala}, {Shafter}, \& {Hornoch}}]{2016ATel.9280....1H}
{Henze}, M., {Williams}, S.~C., {Darnley}, M.~J., {et~al.} 2016{\natexlab{e}},
  ATel, 9280

\bibitem[{{Henze} {et~al.}(2015{\natexlab{b}}){Henze}, {Williams}, {Darnley},
  {Shafter}, \& {Hornoch}}]{2015ATel.8290....1H}
{Henze}, M., {Williams}, S.~C., {Darnley}, M.~J., {Shafter}, A.~W., \&
  {Hornoch}, K. 2015{\natexlab{b}}, ATel, 8290

\bibitem[{{Henze} {et~al.}(2015{\natexlab{c}}){Henze}, {Williams}, {Darnley},
  {Shafter}, \& {Hornoch}}]{2015ATel.8235....1H}
---. 2015{\natexlab{c}}, ATel, 8235, 1

\bibitem[{{Henze} {et~al.}(2009){Henze}, {Pietsch}, {Podigachoski}, {Burwitz},
  {Haberl}, {Updike}, {Hartmann}, {Milne}, {Williams}, {Papamastorakis},
  {Reig}, \& {Strigachev}}]{2009ATel.2286....1H}
{Henze}, M., {Pietsch}, W., {Podigachoski}, P., {et~al.} 2009, ATel, 2286, 1

\bibitem[{{Henze} {et~al.}(2010){Henze}, {Pietsch}, {Haberl}, {Hernanz},
  {Sala}, {Della Valle}, {Hatzidimitriou}, {Rau}, {Hartmann}, {Greiner},
  {Burwitz}, \& {Fliri}}]{2010A&A...523A..89H}
{Henze}, M., {Pietsch}, W., {Haberl}, F., {et~al.} 2010, \aap, 523, A89

\bibitem[{{Henze} {et~al.}(2011){Henze}, {Pietsch}, {Haberl}, {Hernanz},
  {Sala}, {Hatzidimitriou}, {Della Valle}, {Rau}, {Hartmann}, \&
  {Burwitz}}]{2011A&A...533A..52H}
---. 2011, \aap, 533, A52

\bibitem[{{Henze} {et~al.}(2014{\natexlab{b}}){Henze}, {Pietsch}, {Haberl},
  {Della Valle}, {Sala}, {Hatzidimitriou}, {Hofmann}, {Hernanz}, {Hartmann}, \&
  {Greiner}}]{2014A&A...563A...2H}
---. 2014{\natexlab{b}}, \aap, 563, A2

\bibitem[{{Henze} {et~al.}(2015{\natexlab{d}}){Henze}, {Ness}, {Darnley},
  {Bode}, {Williams}, {Shafter}, {Sala}, {Kato}, {Hachisu}, \&
  {Hernanz}}]{2015A&A...580A..46H}
{Henze}, M., {Ness}, J.-U., {Darnley}, M.~J., {et~al.} 2015{\natexlab{d}},
  \aap, 580, A46

\bibitem[{{Henze} {et~al.}(2015{\natexlab{e}}){Henze}, {Darnley}, {Shafter},
  {Kato}, {Hachisu}, {Bode}, {Ness}, {Osborne}, {Kennea}, \&
  {Gehrels}}]{2015ATel.7984....1H}
{Henze}, M., {Darnley}, M.~J., {Shafter}, A.~W., {et~al.} 2015{\natexlab{e}},
  ATel, 7984, 1

\bibitem[{{Henze} {et~al.}(2016{\natexlab{f}}){Henze}, {Williams}, {Darnley},
  {Ederoclite}, {Sala}, {Shafter}, {Chinetti}, {Jose}, \&
  {Hernanz}}]{2016ATel.9276....1H}
{Henze}, M., {Williams}, S.~C., {Darnley}, M.~J., {et~al.} 2016{\natexlab{f}},
  ATel, 9276

\bibitem[{{Hernanz} \& {Jos{\'e}}(2008)}]{2008NewAR..52..386H}
{Hernanz}, M., \& {Jos{\'e}}, J. 2008, \nar, 52, 386

\bibitem[{{Hillman} {et~al.}(2016){Hillman}, {Prialnik}, {Kovetz}, \&
  {Shara}}]{2016ApJ...819..168H}
{Hillman}, Y., {Prialnik}, D., {Kovetz}, A., \& {Shara}, M.~M. 2016, \apj, 819,
  168

\bibitem[{{Honeycutt} \& {Kafka}(2004)}]{2004AJ....128.1279H}
{Honeycutt}, R.~K., \& {Kafka}, S. 2004, \aj, 128, 1279

\bibitem[{{Hornoch} {et~al.}(2016){Hornoch}, {Paunzen}, {Vrastil}, {Kucakova},
  {Darnley}, \& {Henze}}]{2016ATel.9883....1H}
{Hornoch}, K., {Paunzen}, E., {Vrastil}, J., {et~al.} 2016, ATel, 9883

\bibitem[{{Hornoch} \& {Shafter}(2015)}]{2015ATel.7116....1H}
{Hornoch}, K., \& {Shafter}, A.~W. 2015, ATel, 7116

\bibitem[{{Hornoch} \& {Vrastil}(2012)}]{2012ATel.4364....1H}
{Hornoch}, K., \& {Vrastil}, J. 2012, ATel, 4364, 1

\bibitem[{{Hubble}(1929)}]{1929ApJ....69..103H}
{Hubble}, E.~P. 1929, \apj, 69, 103

\bibitem[{{Itagaki}(2016)}]{2016Ita}
{Itagaki}, K. 2016, {CBAT},  {IAU},
  \url{http://www.cbat.eps.harvard.edu/unconf/followups/J00452888+4154097.html}

\bibitem[{{Itagaki} {et~al.}(2016){Itagaki}, {Gao}, {Darnley}, {Henze},
  {Shafter}, {Williams}, {Kafka}, \& {Kato}}]{2016ATel.9848....1I}
{Itagaki}, K., {Gao}, X., {Darnley}, M.~J., {et~al.} 2016, ATel, 9848

\bibitem[{{Jansen} {et~al.}(2001){Jansen}, {Lumb}, {Altieri}, {Clavel}, {Ehle},
  {Erd}, {Gabriel}, {Guainazzi}, {Gondoin}, {Much}, {Munoz}, {Santos},
  {Schartel}, {Texier}, \& {Vacanti}}]{2001A&A...365L...1J}
{Jansen}, F., {Lumb}, D., {Altieri}, B., {et~al.} 2001, \aap, 365, L1

\bibitem[{{Jos\'e}(2016)}]{Jos16}
{Jos\'e}, J. 2016, {Stellar Explosions: Hydrodynamics and Nucleosynthesis}
  (CRC/Taylor and Francis, Boca Raton, FL, USA), doi:10.1201/b19165

\bibitem[{{Kato} {et~al.}(2015){Kato}, {Saio}, \&
  {Hachisu}}]{2015ApJ...808...52K}
{Kato}, M., {Saio}, H., \& {Hachisu}, I. 2015, \apj, 808, 52

\bibitem[{{Kato} {et~al.}(2017){Kato}, {Saio}, \&
  {Hachisu}}]{2017ApJ...838..153K}
---. 2017, \apj, 838, 153

\bibitem[{{Kato} {et~al.}(2014){Kato}, {Saio}, {Hachisu}, \&
  {Nomoto}}]{2014ApJ...793..136K}
{Kato}, M., {Saio}, H., {Hachisu}, I., \& {Nomoto}, K. 2014, \apj, 793, 136

\bibitem[{{Kato} {et~al.}(2016){Kato}, {Saio}, {Henze}, {Ness}, {Osborne},
  {Page}, {Darnley}, {Bode}, {Shafter}, {Hernanz}, {Gehrels}, {Kennea}, \&
  {Hachisu}}]{2016ApJ...830...40K}
{Kato}, M., {Saio}, H., {Henze}, M., {et~al.} 2016, \apj, 830, 40

\bibitem[{{Kaur} {et~al.}(2016){Kaur}, {Erdman}, {Oswalt}, {Burke}, {Hartmann},
  {Henze}, \& {Darnley}}]{2016ATel.9881....1K}
{Kaur}, A., {Erdman}, P., {Oswalt}, T., {et~al.} 2016, ATel, 9881

\bibitem[{{King} \& {Cannizzo}(1998)}]{1998ApJ...499..348K}
{King}, A.~R., \& {Cannizzo}, J.~K. 1998, \apj, 499, 348

\bibitem[{{Korotkiy} \& {Elenin}(2011)}]{2011Kor}
{Korotkiy}, S., \& {Elenin}, L. 2011, {CBAT},  {IAU},
  \url{http://www.cbat.eps.harvard.edu/unconf/followups/J00452885+4154094.html}

\bibitem[{{Kotani} {et~al.}(2005){Kotani}, {Kawai}, {Yanagisawa}, {Watanabe},
  {Arimoto}, {Fukushima}, {Hattori}, {Inata}, {Izumiura}, {Kataoka}, {Koyano},
  {Kubota}, {Kuroda}, {Mori}, {Nagayama}, {Ohta}, {Okada}, {Okita}, {Sato},
  {Serino}, {Shimizu}, {Shimokawabe}, {Suzuki}, {Toda}, {Ushiyama}, {Yatsu},
  {Yoshida}, \& {Yoshida}}]{2005NCimC..28..755K}
{Kotani}, T., {Kawai}, N., {Yanagisawa}, K., {et~al.} 2005, Nuovo Cimento C
  Geophysics Space Physics C, 28, 755

\bibitem[{{Kraft} {et~al.}(1991){Kraft}, {Burrows}, \&
  {Nousek}}]{1991ApJ...374..344K}
{Kraft}, R.~P., {Burrows}, D.~N., \& {Nousek}, J.~A. 1991, \apj, 374, 344

\bibitem[{{Krautter}(2008)}]{2008ASPC..401..139K}
{Krautter}, J. 2008, in Astronomical Society of the Pacific Conference Series,
  Vol. 401, RS Ophiuchi (2006) and the Recurrent Nova Phenomenon, ed.
  A.~{Evans}, M.~F. {Bode}, T.~J. {O'Brien}, \& M.~J. {Darnley}, 139

\bibitem[{{Kuin} {et~al.}(2018){Kuin}, {Page}, {Mroz}, {Darnley}, {Osborn},
  {Walter}, {di Mille}, {Morell}, {Munari}, {Shore}, {Starrfield}, \&
  {Williams}}]{KuinPaper}
{Kuin}, N.~P.~M., {Page}, K.~L., {Mroz}, P., {et~al.} 2018, in preparation for
  submission to MNRAS

\bibitem[{{Livio} \& {Pringle}(1994)}]{1994ApJ...427..956L}
{Livio}, M., \& {Pringle}, J.~E. 1994, \apj, 427, 956

\bibitem[{{Madsen} {et~al.}(2017){Madsen}, {Beardmore}, {Forster}, {Guainazzi},
  {Marshall}, {Miller}, {Page}, \& {Stuhlinger}}]{2017AJ....153....2M}
{Madsen}, K.~K., {Beardmore}, A.~P., {Forster}, K., {et~al.} 2017, \aj, 153, 2

\bibitem[{{Marsh}(1989)}]{1989PASP..101.1032M}
{Marsh}, T.~R. 1989, \pasp, 101, 1032

\bibitem[{{Mason} {et~al.}(2012){Mason}, {Ederoclite}, {Williams}, {Della
  Valle}, \& {Setiawan}}]{2012A&A...544A.149M}
{Mason}, E., {Ederoclite}, A., {Williams}, R.~E., {Della Valle}, M., \&
  {Setiawan}, J. 2012, \aap, 544, A149

\bibitem[{{Mason} {et~al.}(2001){Mason}, {Breeveld}, {Much}, {Carter},
  {Cordova}, {Cropper}, {Fordham}, {Huckle}, {Ho}, {Kawakami}, {Kennea},
  {Kennedy}, {Mittaz}, {Pandel}, {Priedhorsky}, {Sasseen}, {Shirey}, {Smith},
  \& {Vreux}}]{2001A&A...365L..36M}
{Mason}, K.~O., {Breeveld}, A., {Much}, R., {et~al.} 2001, \aap, 365, L36

\bibitem[{{Massey} {et~al.}(2006){Massey}, {Olsen}, {Hodge}, {Strong},
  {Jacoby}, {Schlingman}, \& {Smith}}]{2006AJ....131.2478M}
{Massey}, P., {Olsen}, K.~A.~G., {Hodge}, P.~W., {et~al.} 2006, \aj, 131, 2478

\bibitem[{{Mroz} \& {Udalski}(2016)}]{2016ATel.8578....1M}
{Mroz}, P., \& {Udalski}, A. 2016, ATel, 8578

\bibitem[{{Mukai}(1993)}]{1993Legac...3...21M}
{Mukai}, K. 1993, Legacy, vol.~3, p.21-31, 3, 21

\bibitem[{{Munari} {et~al.}(2014){Munari}, {Mason}, \&
  {Valisa}}]{2014A&A...564A..76M}
{Munari}, U., {Mason}, E., \& {Valisa}, P. 2014, \aap, 564, A76

\bibitem[{{Naito} {et~al.}(2016){Naito}, {Watanabe}, {Sano}, {Kuramoto},
  {Itagaki}, {Kiyota}, {Arai}, {Matsumoto}, {Kojiguchi}, {Sugiura}, {Maehara},
  {Nishiyama}, {Kabashima}, {Henze}, {Darnley}, {Shafter}, \&
  {Kato}}]{2016ATel.9891....1N}
{Naito}, H., {Watanabe}, F., {Sano}, Y., {et~al.} 2016, ATel, 9891

\bibitem[{{Newsham} {et~al.}(2014){Newsham}, {Starrfield}, \&
  {Timmes}}]{2014ASPC..490..287N}
{Newsham}, G., {Starrfield}, S., \& {Timmes}, F.~X. 2014, in Astronomical
  Society of the Pacific Conference Series, Vol. 490, Stellar Novae: Past and
  Future Decades, ed. P.~A. {Woudt} \& V.~A.~R.~M. {Ribeiro}, 287

\bibitem[{{Nishiyama} \& {Kabashima}(2008)}]{2008Nis}
{Nishiyama}, K., \& {Kabashima}, F. 2008, {CBAT},  {IAU},
  \url{http://www.cbat.eps.harvard.edu/iau/CBAT\_M31.html\#2008-12a}

\bibitem[{{Nishiyama} \& {Kabashima}(2012)}]{2012Nis}
---. 2012, {CBAT},  {IAU},
  \url{http://www.cbat.eps.harvard.edu/unconf/followups/J00452884+4154095.html}

\bibitem[{{Oke}(1990)}]{1990AJ.....99.1621O}
{Oke}, J.~B. 1990, \aj, 99, 1621

\bibitem[{{Osborne}(2015)}]{2015JHEAp...7..117O}
{Osborne}, J.~P. 2015, Journal of High Energy Astrophysics, 7, 117

\bibitem[{{Pavana} \& {Anupama}(2016)}]{2016ATel.9865....1P}
{Pavana}, M., \& {Anupama}, G.~C. 2016, ATel, 9865

\bibitem[{{Piascik} {et~al.}(2014){Piascik}, {Steele}, {Bates}, {Mottram},
  {Smith}, {Barnsley}, \& {Bolton}}]{2014SPIE.9147E..8HP}
{Piascik}, A.~S., {Steele}, I.~A., {Bates}, S.~D., {et~al.} 2014, in \procspie,
  Vol. 9147, Ground-based and Airborne Instrumentation for Astronomy V, 91478H

\bibitem[{{Pietsch}(2010)}]{2010AN....331..187P}
{Pietsch}, W. 2010, Astronomische Nachrichten, 331, 187

\bibitem[{{Pietsch} {et~al.}(2007){Pietsch}, {Haberl}, {Sala}, {Stiele},
  {Hornoch}, {Riffeser}, {Fliri}, {Bender}, {B{\"u}hler}, {Burwitz}, {Greiner},
  \& {Seitz}}]{2007A&A...465..375P}
{Pietsch}, W., {Haberl}, F., {Sala}, G., {et~al.} 2007, \aap, 465, 375

\bibitem[{{Poole} {et~al.}(2008){Poole}, {Breeveld}, {Page}, {Landsman},
  {Holland}, {Roming}, {Kuin}, {Brown}, {Gronwall}, {Hunsberger}, {Koch},
  {Mason}, {Schady}, {vanden Berk}, {Blustin}, {Boyd}, {Broos}, {Carter},
  {Chester}, {Cucchiara}, {Hancock}, {Huckle}, {Immler}, {Ivanushkina},
  {Kennedy}, {Marshall}, {Morgan}, {Pandey}, {de Pasquale}, {Smith}, \&
  {Still}}]{2008MNRAS.383..627P}
{Poole}, T.~S., {Breeveld}, A.~A., {Page}, M.~J., {et~al.} 2008, \mnras, 383,
  627

\bibitem[{{Pravec} {et~al.}(1994){Pravec}, {Hudec}, {Sold{\'a}n}, {Sommer}, \&
  {Schenkl}}]{1994ExA.....5..375P}
{Pravec}, P., {Hudec}, R., {Sold{\'a}n}, J., {Sommer}, M., \& {Schenkl}, K.~H.
  1994, Experimental Astronomy, 5, 375

\bibitem[{{R Development Core Team}(2011)}]{R_manual}
{R Development Core Team}. 2011, R: A Language and Environment for Statistical
  Computing, R Foundation for Statistical Computing, Vienna, Austria, {ISBN}
  3-900051-07-0

\bibitem[{{Ritchey}(1917)}]{1917PASP...29..210R}
{Ritchey}, G.~W. 1917, \pasp, 29, 210

\bibitem[{{Robinson} {et~al.}(1981){Robinson}, {Barker}, {Cochran}, {Cochran},
  \& {Nather}}]{1981ApJ...251..611R}
{Robinson}, E.~L., {Barker}, E.~S., {Cochran}, A.~L., {Cochran}, W.~D., \&
  {Nather}, R.~E. 1981, \apj, 251, 611

\bibitem[{{Roming} {et~al.}(2005){Roming}, {Kennedy}, {Mason}, {Nousek}, {Ahr},
  {Bingham}, {Broos}, {Carter}, {Hancock}, {Huckle}, {Hunsberger}, {Kawakami},
  {Killough}, {Koch}, {McLelland}, {Smith}, {Smith}, {Soto}, {Boyd},
  {Breeveld}, {Holland}, {Ivanushkina}, {Pryzby}, {Still}, \&
  {Stock}}]{2005SSRv..120...95R}
{Roming}, P.~W.~A., {Kennedy}, T.~E., {Mason}, K.~O., {et~al.} 2005, \ssr, 120,
  95

\bibitem[{{Rosino}(1973)}]{1973A&AS....9..347R}
{Rosino}, L. 1973, \aaps, 9, 347

\bibitem[{{Sako} {et~al.}(2012){Sako}, {Aoki}, {Doi}, {Ienaka}, {Kobayashi},
  {Matsunaga}, {Mito}, {Miyata}, {Morokuma}, {Nakada}, {Soyano}, {Tarusawa},
  {Miyazaki}, {Nakata}, {Okada}, {Sarugaku}, \&
  {Richmond}}]{2012SPIE.8446E..6LS}
{Sako}, S., {Aoki}, T., {Doi}, M., {et~al.} 2012, in \procspie, Vol. 8446,
  Ground-based and Airborne Instrumentation for Astronomy IV, 84466L

\bibitem[{{Sala} \& {Hernanz}(2005)}]{2005A&A...439.1061S}
{Sala}, G., \& {Hernanz}, M. 2005, \aap, 439, 1061

\bibitem[{{Scaringi}(2014)}]{2014MNRAS.438.1233S}
{Scaringi}, S. 2014, \mnras, 438, 1233

\bibitem[{{Schaefer}(1990)}]{1990ApJ...355L..39S}
{Schaefer}, B.~E. 1990, \apjl, 355, L39

\bibitem[{{Schaefer}(2010)}]{2010ApJS..187..275S}
---. 2010, \apjs, 187, 275

\bibitem[{{Schwarz} {et~al.}(2011){Schwarz}, {Ness}, {Osborne}, {Page},
  {Evans}, {Beardmore}, {Walter}, {Helton}, {Woodward}, {Bode}, {Starrfield},
  \& {Drake}}]{2011ApJS..197...31S}
{Schwarz}, G.~J., {Ness}, J.-U., {Osborne}, J.~P., {et~al.} 2011, \apjs, 197,
  31

\bibitem[{{Shafter} {et~al.}(2012){Shafter}, {Hornoch}, {Ciardullo}, {Darnley},
  \& {Bode}}]{2012ATel.4503....1S}
{Shafter}, A.~W., {Hornoch}, K., {Ciardullo}, J.~V.~R., {Darnley}, M.~J., \&
  {Bode}, M.~F. 2012, ATel, 4503, 1

\bibitem[{{Shafter} {et~al.}(2016){Shafter}, {Horst}, {Igarashi}, \&
  {Johnson}}]{2016ATel.9864....1S}
{Shafter}, A.~W., {Horst}, J., {Igarashi}, A., \& {Johnson}, T. 2016, ATel,
  9864

\bibitem[{{Shafter} {et~al.}(1985){Shafter}, {Szkody}, {Liebert}, {Penning},
  {Bond}, \& {Grauer}}]{1985ApJ...290..707S}
{Shafter}, A.~W., {Szkody}, P., {Liebert}, J., {et~al.} 1985, \apj, 290, 707

\bibitem[{{Shafter} {et~al.}(2015){Shafter}, {Henze}, {Rector}, {Schweizer},
  {Hornoch}, {Orio}, {Pietsch}, {Darnley}, {Williams}, {Bode}, \&
  {Bryan}}]{2015ApJS..216...34S}
{Shafter}, A.~W., {Henze}, M., {Rector}, T.~A., {et~al.} 2015, \apjs, 216, 34

\bibitem[{{Shore} {et~al.}(1991){Shore}, {Sonneborn}, {Starrfield}, {Hamuy},
  {Williams}, {Cassatella}, \& {Drechsel}}]{1991ApJ...370..193S}
{Shore}, S.~N., {Sonneborn}, G., {Starrfield}, S.~G., {et~al.} 1991, \apj, 370,
  193

\bibitem[{{Sin} {et~al.}(2017){Sin}, {Henze}, {Sala}, {Ederoclite}, {Hernanz},
  {Jose}, {Hornoch}, {Conseil}, \& {Kucakova}}]{2017ATel10001....1S}
{Sin}, P., {Henze}, M., {Sala}, G., {et~al.} 2017, ATel, No.~10001, 1

\bibitem[{{Stark} {et~al.}(1992){Stark}, {Gammie}, {Wilson}, {Bally}, {Linke},
  {Heiles}, \& {Hurwitz}}]{1992ApJS...79...77S}
{Stark}, A.~A., {Gammie}, C.~F., {Wilson}, R.~W., {et~al.} 1992, \apjs, 79, 77

\bibitem[{{Starrfield} {et~al.}(2016){Starrfield}, {Iliadis}, \&
  {Hix}}]{2016PASP..128e1001S}
{Starrfield}, S., {Iliadis}, C., \& {Hix}, W.~R. 2016, \pasp, 128, 051001

\bibitem[{{Starrfield} {et~al.}(1985){Starrfield}, {Sparks}, \&
  {Truran}}]{1985ApJ...291..136S}
{Starrfield}, S., {Sparks}, W.~M., \& {Truran}, J.~W. 1985, \apj, 291, 136

\bibitem[{{Starrfield} {et~al.}(1998){Starrfield}, {Truran}, {Wiescher}, \&
  {Sparks}}]{1998MNRAS.296..502S}
{Starrfield}, S., {Truran}, J.~W., {Wiescher}, M.~C., \& {Sparks}, W.~M. 1998,
  \mnras, 296, 502

\bibitem[{{Steele} {et~al.}(2004){Steele}, {Smith}, {Rees}, {Baker}, {Bates},
  {Bode}, {Bowman}, {Carter}, {Etherton}, {Ford}, {Fraser}, {Gomboc}, {Lett},
  {Mansfield}, {Marchant}, {Medrano-Cerda}, {Mottram}, {Raback}, {Scott},
  {Tomlinson}, \& {Zamanov}}]{2004SPIE.5489..679S}
{Steele}, I.~A., {Smith}, R.~J., {Rees}, P.~C., {et~al.} 2004, in Society of
  Photo-Optical Instrumentation Engineers (SPIE) Conference Series, Vol. 5489,
  Ground-based Telescopes, ed. J.~M. {Oschmann}, Jr., 679--692

\bibitem[{{Str{\"u}der} {et~al.}(2001){Str{\"u}der}, {Briel}, {Dennerl},
  {Hartmann}, {Kendziorra}, {Meidinger}, {Pfeffermann}, {Reppin}, {Aschenbach},
  {Bornemann}, {Br{\"a}uninger}, {Burkert}, {Elender}, {Freyberg}, {Haberl},
  {Hartner}, {Heuschmann}, {Hippmann}, {Kastelic}, {Kemmer}, {Kettenring},
  {Kink}, {Krause}, {M{\"u}ller}, {Oppitz}, {Pietsch}, {Popp}, {Predehl},
  {Read}, {Stephan}, {St{\"o}tter}, {Tr{\"u}mper}, {Holl}, {Kemmer}, {Soltau},
  {St{\"o}tter}, {Weber}, {Weichert}, {von Zanthier}, {Carathanassis}, {Lutz},
  {Richter}, {Solc}, {B{\"o}ttcher}, {Kuster}, {Staubert}, {Abbey}, {Holland},
  {Turner}, {Balasini}, {Bignami}, {La Palombara}, {Villa}, {Buttler},
  {Gianini}, {Lain{\'e}}, {Lumb}, \& {Dhez}}]{2001A&A...365L..18S}
{Str{\"u}der}, L., {Briel}, U., {Dennerl}, K., {et~al.} 2001, \aap, 365, L18

\bibitem[{{Tan} {et~al.}(2016){Tan}, {Gao}, \& {Henze}}]{2016ATel.9885....1T}
{Tan}, H., {Gao}, X., \& {Henze}, M. 2016, ATel, 9885

\bibitem[{{Tang} {et~al.}(2013){Tang}, {Cao}, \&
  {Kasliwal}}]{2013ATel.5607....1T}
{Tang}, S., {Cao}, Y., \& {Kasliwal}, M.~M. 2013, ATel, 5607, 1

\bibitem[{{Tang} {et~al.}(2014){Tang}, {Bildsten}, {Wolf}, {Li}, {Kong}, {Cao},
  {Cenko}, {De Cia}, {Kasliwal}, {Kulkarni}, {Laher}, {Masci}, {Nugent},
  {Perley}, {Prince}, \& {Surace}}]{2014ApJ...786...61T}
{Tang}, S., {Bildsten}, L., {Wolf}, W.~M., {et~al.} 2014, \apj, 786, 61

\bibitem[{{Tody}(1993)}]{1993ASPC...52..173T}
{Tody}, D. 1993, in Astronomical Society of the Pacific Conference Series,
  Vol.~52, Astronomical Data Analysis Software and Systems II, ed. R.~J.
  {Hanisch}, R.~J.~V. {Brissenden}, \& J.~{Barnes}, 173

\bibitem[{{Tr{\"u}mper}(1982)}]{1982AdSpR...2..241T}
{Tr{\"u}mper}, J. 1982, Advances in Space Research, 2, 241

\bibitem[{{Turner} {et~al.}(2001){Turner}, {Abbey}, {Arnaud}, {Balasini},
  {Barbera}, {Belsole}, {Bennie}, {Bernard}, {Bignami}, {Boer}, {Briel},
  {Butler}, {Cara}, {Chabaud}, {Cole}, {Collura}, {Conte}, {Cros}, {Denby},
  {Dhez}, {Di Coco}, {Dowson}, {Ferrando}, {Ghizzardi}, {Gianotti}, {Goodall},
  {Gretton}, {Griffiths}, {Hainaut}, {Hochedez}, {Holland}, {Jourdain},
  {Kendziorra}, {Lagostina}, {Laine}, {La Palombara}, {Lortholary}, {Lumb},
  {Marty}, {Molendi}, {Pigot}, {Poindron}, {Pounds}, {Reeves}, {Reppin},
  {Rothenflug}, {Salvetat}, {Sauvageot}, {Schmitt}, {Sembay}, {Short},
  {Spragg}, {Stephen}, {Str{\"u}der}, {Tiengo}, {Trifoglio}, {Tr{\"u}mper},
  {Vercellone}, {Vigroux}, {Villa}, {Ward}, {Whitehead}, \&
  {Zonca}}]{2001A&A...365L..27T}
{Turner}, M.~J.~L., {Abbey}, A., {Arnaud}, M., {et~al.} 2001, \aap, 365, L27

\bibitem[{{Villar} {et~al.}(2017){Villar}, {Guillochon}, {Berger}, {Metzger},
  {Cowperthwaite}, {Nicholl}, {Alexander}, {Blanchard}, {Chornock},
  {Eftekhari}, {Fong}, {Margutti}, \& {Williams}}]{2017ApJ...851L..21V}
{Villar}, V.~A., {Guillochon}, J., {Berger}, E., {et~al.} 2017, \apjl, 851, L21

\bibitem[{{Warner}(1995)}]{1995cvs..book.....W}
{Warner}, B. 1995, {Cataclysmic variable stars} (Cambridge Astrophysics Series,
  Cambridge, New York: Cambridge University Press, 1995)

\bibitem[{{Wenger} {et~al.}(2000){Wenger}, {Ochsenbein}, {Egret}, {Dubois},
  {Bonnarel}, {Borde}, {Genova}, {Jasniewicz}, {Lalo{\"e}}, {Lesteven}, \&
  {Monier}}]{2000A&AS..143....9W}
{Wenger}, M., {Ochsenbein}, F., {Egret}, D., {et~al.} 2000, \aaps, 143, 9

\bibitem[{{White} {et~al.}(1995){White}, {Giommi}, {Heise}, {Angelini}, \&
  {Fantasia}}]{1995ApJ...445L.125W}
{White}, N.~E., {Giommi}, P., {Heise}, J., {Angelini}, L., \& {Fantasia}, S.
  1995, \apjl, 445, L125

\bibitem[{{Williams} {et~al.}(2004){Williams}, {Garcia}, {Kong}, {Primini},
  {King}, {Di Stefano}, \& {Murray}}]{2004ApJ...609..735W}
{Williams}, B.~F., {Garcia}, M.~R., {Kong}, A.~K.~H., {et~al.} 2004, \apj, 609,
  735

\bibitem[{{Williams}(1992)}]{1992AJ....104..725W}
{Williams}, R.~E. 1992, \aj, 104, 725

\bibitem[{{Williams} {et~al.}(2015{\natexlab{a}}){Williams}, {Darnley},
  {Henze}, {Shafter}, \& {Hornoch}}]{2015ATel.8242....1W}
{Williams}, S.~C., {Darnley}, M.~J., {Henze}, M., {Shafter}, A.~W., \&
  {Hornoch}, K. 2015{\natexlab{a}}, ATel, 8242, 1

\bibitem[{{Williams} {et~al.}(2015{\natexlab{b}}){Williams}, {Shafter},
  {Hornoch}, {Henze}, \& {Darnley}}]{2015ATel.8234....1W}
{Williams}, S.~C., {Shafter}, A.~W., {Hornoch}, K., {Henze}, M., \& {Darnley},
  M.~J. 2015{\natexlab{b}}, ATel, 8234

\bibitem[{{Wilms} {et~al.}(2000){Wilms}, {Allen}, \&
  {McCray}}]{2000ApJ...542..914W}
{Wilms}, J., {Allen}, A., \& {McCray}, R. 2000, \apj, 542, 914

\bibitem[{{Wolf} {et~al.}(2013){Wolf}, {Bildsten}, {Brooks}, \&
  {Paxton}}]{2013ApJ...777..136W}
{Wolf}, W.~M., {Bildsten}, L., {Brooks}, J., \& {Paxton}, B. 2013, \apj, 777,
  136

\bibitem[{{Yaron} {et~al.}(2005){Yaron}, {Prialnik}, {Shara}, \&
  {Kovetz}}]{2005ApJ...623..398Y}
{Yaron}, O., {Prialnik}, D., {Shara}, M.~M., \& {Kovetz}, A. 2005, \apj, 623,
  398

\end{thebibliography}

\appendix

\section{Additional optical telescopes observing the 2016 eruption of \novak}\label{app:optical_photometry}

Among the numerous ground-based observatories that were monitoring the position of \nova for half a year, here we only include those that happened to have weather conditions suitable enough to obtain photometry immediately prior and during the eruption. Regardless of luck with the weather, we are immensely grateful for the hard work and persistence of the entire 2016 monitoring collaboration -- the members of which can be found in the author list of this paper.

Below we only list those facilities or telescopes that newly joined our observations of \novak. Details on those instruments that obtained photometry here and already in the 2015 eruption campaign can be found in the Appendix of \othreek. This includes the Ond\v{r}ejov Observatory \citep{2016ATel.9861....1B, 2016ATel.9883....1H}, the Mount Laguna Observatory \citep[MLO;][]{2016ATel.9857....1E, 2016ATel.9864....1S}, and the Nayoro Observatory 1.6\,m Pirka telescope \citep{2016ATel.9891....1N}.

\subsection{Itagaki 50\,cm telescope}

The 2016 eruption was discovered by \citet{2016ATel.9848....1I} using five images (480\,s total exposure time) obtained with the 0.5\,m f/6 telescope, with a BITRAN BN-52E(KAF-1001E) camera, at the Itagaki Astronomical Observatory, Japan. Additional light curve photometry was first reported in \citet{2016ATel.9891....1N}.

\subsection{Xingming Observatory Half-Meter-Telescope (HMT)}

The confirmation detection and follow-up photometry of \nova were obtained at the Half-Meter-Telescope of the Xingming Observatory, China \citet{2016ATel.9848....1I, 2016ATel.9885....1T}. The instrument is a 0.508\,m aperture, with a focal length of 2.052\,m using a QHY11 CCD camera. 
All images were calibrated using the standard procedure, including flat-field correction, and dark and bias frames using the Maxim DL software. The relative photometry was obtained in PyRAF with an aperture optimized to the seeing of each individual image. The final magnitudes were calibrated using comparison stars from the XPM catalogue \citep{2009MNRAS.393..133F}.

\subsection{Himalayan Chandra Telescope (HCT)}

The central 2k$\times$2k region of the Himalayan Faint Object Spectrograph and Camera (HFOSC) mounted on the 2m Himalayan Chandra Telescope (HCT) is used for imaging and gives a field of view of $10^\prime\times10^\prime$ at a scale of $0^{\prime\prime}\!\!.296$\,pixel$^{-1}$.

Photometric observations were made on 2016 December 14.74 UT in the $VRI$ bands, and in the $BVRI$ bands on December 15.67. The images were bias subtracted, and flat field corrected using twilight flats. Instrumental magnitudes were obtained using aperture photometry. An aperture of radius three times FWHM was used. Differential photometry was performed with respect to the stars in the field (\othreek) to estimate the magnitude of the nova. 

\subsection{Embry Riddle Aeronautical University (ERAU)}

Photometry of \nova was obtained at the Embry Riddle Aeronautical University, Florida, with (a) a 24\,inch CDK Cassegrain telescope equipped with a SBIG STX 16803 detector, and (b) a 1\,m RC telescope equipped with an identical detector.  Both telescopes took series of 600s images through Omega SDSS $g'$, $r'$, and $i'$ filters. The total exposure time for each reported magnitude varied between 1--4 hours. The magnitudes were extracted using standard aperture photometric techniques in IRAF (v2.16.1) and calibrated using the \othreek\ standard stars in the field. The photometry was first reported in \citet{2016ATel.9857....1E,2016ATel.9861....1B, 2016ATel.9881....1K}.

\subsection{Danish 1.54\,m La Silla}

Late-time optical photometric data was collected with the Danish 1.54\,m telescope at the ESO La Silla Observatory, operated remotely from Ond\v{r}ejov, using the Danish Faint Object Spectrograph and Camera (DFOSC) instrument \citep{2016ATel.9883....1H}. For each epoch, a series of ten 90s exposures was taken. Standard reduction procedures for raw CCD images were applied (bias subtraction and flat field correction) using the APHOT software \citep{1994ExA.....5..375P}. Reduced images within the same series were co-added to improve the signal-to-noise ratio and the gradient of the galaxy background was flattened using a spatial median filter via the SIPS program. Photometric measurements of the nova were then performed using aperture photometry in APHOT. Five nearby secondary standard stars from \citep{2006AJ....131.2478M} were used to photometrically calibrate the magnitudes included in Table~\ref{optical_photometry_table}. 

\subsection{Kiso Observatory}

We obtained V-band CCD images with the 1.05\,m Schmidt telescope equipped with the Kiso Wide Field Camera \citep{2012SPIE.8446E..6LS} of the Kiso Observatory, University of Tokyo, Japan. Typical, we took 3 images with 60\,s exposure per night. The dark-subtraction and flat-fielding were performed with IRAF (v2.16.1), before image stacking by using SWarp \citep[v2.19.1][]{2002ASPC..281..228B}. Photometry of the stacked images was performed via the aperture photometry package in Source Extractor \citep[v2.8.6;][]{1996A&AS..117..393B}. We used nearby stars in SDSS, APASS \citep{2016yCat.2336....0H} and \othreek\ for the photometric calibration. The data were first reported in \citet{2016ATel.9891....1N}.

\subsection{Okayama Astrophysical Observatory}

Additional $g'$, $R_\mathrm{C}$, and $I_\mathrm{C}$ band upper limits reported by \citet{2016ATel.9891....1N} were obtained using the 0.5\,m, f/6.5 MITSuME telescope \citep{2005NCimC..28..755K}, equipped with an Apogee Alta U6 camera, of the Okayama Astrophysical Observatory, Japan. We took 10 images with
60\,s exposure per night for each of the three bands. Image calibration and photometry followed the same procedure as for the Kiso observatory above.

\subsection{Osaka Kyoiku University}

\citet{2016ATel.9891....1N} first reported pre-eruption upper limits and light curve photometry obtained by the 0.51\,m, f/12 telescope with an Andor DW936N-BV camera of the Osaka Kyoiku University, Japan. These observations were obtained using an $R_\mathrm{C}$ filter with 300\,s exposure per image. A stack of 14 images were combined using the IRAF task \texttt{imcombine}. We carried out aperture photometry \texttt{apphot} and PSF photometry \texttt{daophot} within the IRAF environment. The source \#11 in \othreek\ was used as a comparison star.

\subsection{Miyaki-Argenteus observatory}

Light curve monitoring was performed using a 0.5m f/6.8 telescope, equipped with a SBIG STL1001E camera, at the Miyaki-Argenteus Observatory, Japan \citep{2016ATel.9891....1N}.

\subsection{Nayoro Observatory - 0.4\,m Meili telescope}

We performed observations at Nayoro Observatory, Nayoro, Japan, using the 0.4\,m Meili telescope (Meade Schmidt-Cassegrain Telescope) with a SBIG STL-1001E CCD camera (unfiltered or with R-band filter). The obtained images were reduced in a standard manner and stacked using the StellaImage (v6.5) software. Photometry was conducted using Makali'i, a free software provided by National Astronomical Observatory of Japan and AstroArts Inc. for education and research. Magnitudes are measured using an ensemble of comparison stars listed in \othreek. The limiting magnitudes correspond to an S/N of 3 \citep{2016ATel.9891....1N}. The 1.6\,m Pirka telescope of the same observatory was also used in this campaign and is described in \othreek.

\subsection{New Mexico Skies + AstroCamp Observatory}

Additional monitoring data was reported first by \citep{2016ATel.9891....1N} based on remote observations with the following instruments: (a) a 0.5\,m f/4.5 CDK astrograph, with a FLI-PL11002M CCD, at the New Mexico Skies site (Mayhill, NM, USA), (b) a 0.43\,m f/6.8 CDK astrograph plus SBIG STL-11000M CCD at the AstroCamp Observatory hosting site (Nerpio, Spain); (c) a 0.32\,m f/8.0 CDK astrograph, equipped with a SBIG STXL-6303E CCD, at the AstroCamp Observatory.

\subsection{Hankasalmi Observatory (AAVSO OAR)}

Pre-eruption upper limits were obtained at Hankasalmi Observatory, Finland using a 0.4\,m RC (RCOS) telescope equipped with a SBIG STL-1001E CCD. Typically 25 to 100 unfiltered images with 60\,s exposure were obtained per night and stacked using MaxImDL (v4.61). The stacked image was checked for a nova detection by visually using SAOImage DS9 and photometrically using a custom software, with an aperture radius of 6\arcsec and a background annulus of 12\arcsec--18\arcsec. Upper limits were estimated according to the formula $m + 2.5 \log{s}/3$, where m and s are the magnitude and signal-to-noise, respectively, of comparison star \#12 in \othreek.

\subsection{CBA Concord Observatory (AAVSO COO)}

We observed \nova with the CBA Concord PF29 telescope -- a prime focus 0.74\,m f/4.36 reflector on an English Cradle mount -- located in suburban Concord, CA, USA. Two cameras have been used during this project:\ an SBIG STL1001E with a clear filter ($1\farcs52$ pixel) and an SBIG STF 8300M (unfiltered, $0\farcs34$ pixel$^{-1}$, $2\times2$ binning).

Unfiltered groups of 40--50 images of 15 or 20\,s duration were taken and median-combined using the AIP4Win\footnote{\url{http://www.stargazing.net/david/aip4win/}} software tool. Typically, 2--4 sets of these groups were averaged within the AAVSO VPHOT\footnote{\url{https://www.aavso.org/vphot}} online photometry tool. The minimum number of sub-frames was almost always $>100$, usually $\sim200$. The unfiltered or clear-filter measurements were referenced to the $V$-band comparison stars (see \othreek).

\subsection{iTelescope.net T24/T11 (AAVSO COO)}

We obtained remote observations with iTelescope.net utilizing (i) the T24 telescope, a Planewave 0.61m CDK Telescope f/6.5 and a FLI PL-9000 CCD camera at the hosting site in Sierra Remote Observatory (SRO), Auberry, CA, USA; and (ii) the T11 telescope, a Planewave 0.5m CDK with a FLI PL-11002M CCD camera at the New Mexico Skies hosting site at Mayfield, NM, USA. Typically three 5 min frames (T24) or three 3 min frames (T11) were obtained in the Luminance filter (a clear filter with UV and IR cut-off). Images were median-combined in AAVSO VPHOT. The detection limits (S/N = 4) are typically 20.7 mag (T24) or 20.2 mag (T11), calibrated using the $R$, $I$ and $V$-band standards in (\othreek). The photometry was estimated in the same way as for the Concord Observatory above.

\subsection{Newcastle Observatory (AAVSO CMJA)}

We obtained data from the Newcastle Observatory in Newcastle, Ontario, Canada using a 0.40\,m Meade Schmidt-Cassegrain (ACF) Telescope working at f/7 and a QSI 516ws CCD camera. The images were obtained in the filters Johnson $V$, Cousins $I_\mathrm{C}$, or unfiltered. Most images were a stack of 6 frames, median-combined to minimize cosmic ray effects. At least one imaging run was obtained per night when weather permitted. Occasionally, a second imaging run in the same night (before dawn) was attempted.

All individual images were automatically put through an image processing pipeline for bias, dark, and flat-field calibration, as well as plate-solved to include WCS coordinates before being stacked for analysis. The stack image is viewed in Aladin \citep[v9;][]{2000A&AS..143...33B,2014ASPC..485..277B} with the SIMBAD database \citep{2000A&AS..143....9W} loaded to accurately locate the target. The detection of the target was compared using the comparison star \#8 in \othreek\ which has a $V$-band magnitude of 19.087. If the target was not detected (S/N$<3$), the limit was reported as fainter than 19.1\,mag.

\subsection{Large Binocular Telescope (LBT)}

We obtained optical images and photometry of \nova on 2017-01-08.12 UT with the 8.4\,m Large Binocular Telescope and Multi-Object Double Spectrograph (MODS2).  Images were obtained in the standard SDSS $u'g'r'i'z'$ filters with a total integration time of 300\,s in each of the $g'r'i'z'$ filters and 600\,s in the $u'$ filter at an image scale of 0\farcs125 per pixel with a field of view of about $6\arcmin \times 6\arcmin$.  Image quality was typically 0\farcs8 to 1\farcs0 under non--photometric conditions.  Bias and twilight--sky flat--field images were obtained in each of the $u'g'r'i'z'$ filters to facilitate the data reduction.  All reductions were performed using IRAF (v2.16).

\subsection{West Challow Observatory (AAVSO BDG)}

We obtained observations at West Challow Observatory, Oxfordshire, UK, on most clear nights using a 0.35\,m Meade Schmidt-Cassegrain Telescope working at f/6.3 with a clear filter and a Starlight Xpress SXVR-H9 CCD camera. Typically 20-30 images with 60\,s exposure were recorded, dark-subtracted, flat-fielded and stacked using Astrometrica. Having determined that the nova was not visible in the stacked image at the expected position, the magnitude of the faintest clearly recognizable stellar object in the vicinity of the nova as determined by Astrometrica was reported as the faint magnitude limit for that night. When detected, the magnitude of the nova was measured using the AIP4WIN software\footnote{\url{http://www.willbell.com/aip4win/aip.htm}} and an ensemble of the $V$-band comparison stars listed in \othreek.

\subsection{Bernezzo Observatory (AAVSO MAND)}

Light curve photometry was obtained at Bernezzo Observatory, Italy, using a 0.25\,m f/4 reflector with an Atik 314L CCD and a scale of $1\farcs33$ per pixel. We stacked 19 individual $V$-band images with 120\,s exposure each for a S/N=29.6 detection listed in Table~\ref{optical_photometry_table}. The astrometric solution was calibrated through the Astrometrica software. The photometry was extracted using the software FotoDif (v3.95)\footnote{\url{http://www.astrosurf.com/orodeno/fotodif/index.htm}} and calibrated via the AAVSO Variable Star Plotter\footnote{\url{https://www.aavso.org/apps/vsp/}}, which uses comparison starts from \othreek.

\subsection{AAVSO PXR}

We observed the nova using a 0.4\,m SCT telescope, equipped with an SBIG 6303 CCD, located on Haleakala, Hawaii, as part of the LCO group\footnote{
\url{https://lco.global/}}. The exposure times were 60\,s with no stacking, flats and darks were applied by LCO.  The filter was Johnson $V$ and the photometry used was AIP4WIN using the aperture function. The calibration stars were taken from the APASS catalogue \citep{2016yCat.2336....0H}.

\subsection{AAVSO HBB}

We obtained light curve photometry using a Meade 0.4-m SCT, with an Astrondon V-band photometric filter and a FLI Proline CCD camera (with 1kx1k back illuminated SITe chip), located at New Smyrna Beach, Florida. Typically, 20--30 sets of 60-s exposures were stacked. The image capture and photometry used the MaximDL v6.14 software. The photometry was calibrated using the comparison starts from \othreek.

\subsection{Polaris Observatory}

Images were obtained at the Polaris Observatory, Budapest, Hungary, using a 0.25 m f/4 Newtonian reflector with a V filter and an ALCCD5.2 (QHY6) CCD camera. All raw images were processed with gcx v1.3 (dark subtraction and flat field correction and stacking). The integration times were 12 x 180 s. The stacked image was plate-solved with the \texttt{solve-field} tool of astrometry.net. The aperture photometry was performed using IRAF (v2.16.1) and calibrated using the V-band reference stars of \othreek\ via the AAVSO VSP\footnote{\url{https://www.aavso.org/apps/vsp/}}.

\subsection{Javalambre Observatory (OAJ)}

One set of two 400s H$\alpha$ images (central wavelength 6600\,\AA; FWHM 145\,\AA) was obtained during the eruption with the JAST/T80 telescope at the Observatorio Astrofisico de Javalambre, in Teruel, owned, managed and operated by the Centro de Estudios de Fisica del Cosmos de Aragon. The aperture photometry was derived using the Source Extractor software \citep[v2.8.6;][]{1996A&AS..117..393B} and calibrated with $R$-band data of the Local Group Galaxies Survey \citep{2006AJ....131.2478M}.

\subsection{Observatoire de Haute Provence (AAVSO HDR)}

Light curve photometry was obtained via remote observations at the ROTAT and SATINO-2 telescopes, both located at the Observatoire de Haute Provence, France. The telescopes are remotely operated by the ``Foundation Interactive Astronomy and Astrophysics'', Germany. ROTAT is a 0.60\,m f/3.2 Newtonian reflector used with a clear filter and a SBIG 11000 STL CCD camera. SATINO-2 is a 0.30\,m f/6 Schmidt-Cassegrain reflector used with a clear filter and a SBIG ST8-E CCD camera. ROTAT photometry is estimated from a calibrated 600\,s guided exposure, while SATINO photometry is based on 19 calibrated and summed 300\,s exposures. The photometric analysis was carried out with the MIRA PRO x64 software (v8.012). The photometry was calibrated using the $R$-band magnitudes of the \othreek\ comparison stars \#11 and \#12.

\newpage

\section{Observations of the 2016 eruption of \novak}

Tables~\ref{optical_photometry_table} and \ref{tab:swift_split} provide full details of the observations of the 2016 eruption of \novak.

\begin{deluxetable}{llllllll}
\tablecaption{Complete Dataset of the Visible and Near Infrared Photometric Observations of the 2016 Eruption of \novak.\label{optical_photometry_table}}
\tablehead{
\colhead{Date} & \colhead{$\Delta t$\tablenotemark{\dag}} & \multicolumn{2}{c}{MJD 57,000+} & \colhead{Telescope \&} & \colhead{Exposure} & \colhead{Filter} & \colhead{Photometry} \\
\colhead{(UT)} & \colhead{(days)} & \colhead{Start} & \colhead{End} & \colhead{Instrument} & \colhead{(secs)}}
\startdata
2016-12-07.110 & \toe{729.110} & 729.085 & 729.134 & AAVSO COO & $198\times15$ & -- & $>20.2$\\
2016-12-09.807 & \toe{731.821} & 731.807 &731.834 & AAVSO OAR & $25\times60$ & -- & $>20.3$\\
2016-12-10.164 & \toe{732.164} & 732.154 & 732.174 & AAVSO COO & $7\times180$ & -- & $>18.3$\\
2016-12-10.707 & \toe{732.721} & 732.707 & 732.734 & AAVSO OAR & $25\times60$ & -- & $>20.3$\\
2016-12-10.945 & \toe{732.945} & 732.942 & 732.948 & AAVSO BDG & $8\times60$ & -- & $>19.5$\\
2016-12-11.139 & \toe{733.139} & 733.130 & 733.175 & AAVSO COO & $6\times180$ & -- & $>18.4$\\
2016-12-11.338 & \toe{733.338} & 733.330 & 733.346 & Meili 0.4m & $22\times30$ & -- & $>19.0$\tablenotemark{a}\\
2016-12-11.479 & \toe{733.479} & 733.478 & 733.480 & Miyaki-Argenteus & $2\times60$ & -- & $>19.0$\tablenotemark{a}\\
2016-12-12.070 & \toe{734.070} & 734.060 & 734.079 & AAVSO COO & $5\times180$ & -- & $>19.2$\\
2016-12-12.091 & \toe{734.091} & & & New Mexico Skies & $5\times120$ & -- & $>18.3$\tablenotemark{a}\\
2016-12-12.487 & \toe{734.487} & & & Itagaki 50cm & 480 & -- & 18.2\tablenotemark{b}\\
2016-12-12.536 & \toe{734.536} & & & Xingming HMT & 60 & -- & 17.9\tablenotemark{b}\\
2016-12-12.537 & \toe{734.537} & & & Xingming HMT & 60 & -- & 18.1\tablenotemark{b}\\
2016-12-12.538 & \toe{734.538} & & & Xingming HMT & 60 & -- & 18.1\tablenotemark{b}\\
2016-12-12.539 & \toe{734.539} & & & Xingming HMT & 60 & -- & 18.0\tablenotemark{b}\\
2016-12-12.540 & \toe{734.540} & & & Xingming HMT & 60 & -- & 18.3\tablenotemark{b}\\
2016-12-12.782 & \toe{734.782} & & & AstroCamp Observatory & $11\times120$ & -- & $17.85\pm0.10$\tablenotemark{a}\\
2016-12-12.946 & \toe{734.946} & & & ROTAT & 600 & -- & $17.80\pm0.22$\\
2016-12-12.954 & \toe{734.954} & & & ROTAT & 600 & -- & $17.50\pm0.22$\\
2016-12-12.961 & \toe{734.961} & & & ROTAT & 600 & -- & $17.65\pm0.22$\\
2016-12-12.968 & \toe{734.968} & & & ROTAT & 600 & -- & $17.59\pm0.20$\\
2016-12-12.978 & \toe{734.978} & & & AstroCamp Observatory & $15\times120$ & -- & $17.91\pm0.14$\tablenotemark{a}\\
2016-12-13.133 & \toe{735.133} & 735.038 & 735.228 & AAVSO COO & $23\times180$ & -- & $18.254\pm0.136$\\
2016-12-13.109 & \toe{735.109} & & & New Mexico Skies & $15\times120$ & -- & $18.06\pm0.07$\tablenotemark{a}\\
2016-12-13.756 & \toe{735.756} & & & SATINO-2 & $19\times300$ & -- & $18.27\pm0.24$\\
2016-12-13.798 & \toe{735.798} & & & AstroCamp Observatory & $23\times120$ & -- & $18.79\pm0.23$\tablenotemark{a}\\
2016-12-14.139 & \toe{736.139} & & & New Mexico Skies & $29\times120$ & -- & $19.11\pm0.10$\tablenotemark{a}\\
2016-12-14.378 & \toe{736.378} & & & Itagaki Observatory & \nodata & -- & $19.3\pm0.2$\tablenotemark{a}\\
2016-12-14.756 & \toe{736.756} & 736.744 & 736.765 & AAVSO BDG & $30\times60$ & -- & $19.71\pm0.23$\\
2016-12-14.792 & \toe{736.792} & & & ROTAT & 600 & -- & $18.75\pm0.47$\\
2016-12-15.525 & \toe{737.525} & & & Xingming HMT & $4\times90$ & -- & $19.6\pm0.2$\tablenotemark{c}\\
2016-12-16.449 & \toe{738.449} & 738.448 & 738.450 & Miyaki-Argenteus & $2\times90$ & -- & $>20.2$\tablenotemark{a}\\
2016-12-16.506 & \toe{738.506} & & & Xingming HMT & $13\times90$ & -- & $20.1\pm0.2$\tablenotemark{c}\\
2016-12-17.133 & \toe{739.133} & 739.083 & 739.183 & AAVSO COO & $297\times15$ & -- & $>20.5$ \\
2016-12-17.467 & \toe{739.467} & & &Itagaki Observatory & \nodata & -- & $>20.5$\tablenotemark{a}\\
2016-12-17.539 & \toe{739.539} & & & Xingming HMT & $10\times90$ & -- & $>20.3$\tablenotemark{c}\\
2016-12-19.110 & \toe{741.110} & 741.098 & 741.121 & AAVSO COO & $334\times15$ & -- & $>20.4$\\
2016-12-19.161 & \toe{741.161} & 741.125 & 741.204 & AAVSO COO & $17\times300$ & -- & $>20.7$\\
2016-12-19.462 & \toe{741.462} & 741.415 & 741.508 & Meili 0.4m & $91\times30$ & -- & $>20.5$\tablenotemark{a}\\
2016-12-28.118 & \toe{750.118} & 750.116 & 750.120 & AAVSO COO & 300 & -- & $>19.9$\\
2016-12-29.110 & \toe{751.110} & 751.048 & 751.172 & AAVSO COO & $334\times15$ & -- & $>20.8$\\
\hline
2016-12-13.097 & \toe{735.097} & & & MLO & 180 & $B$ & $18.50\pm0.10$\tablenotemark{d}\\
2016-12-13.289 & \toe{735.289} & & & MLO & 600 & $B$ & $18.65\pm0.10$\tablenotemark{d}\\
2016-12-14.129 & \toe{736.129} & & & MLO & 1200 & $B$ & $19.37\pm0.10$\tablenotemark{d}\\
2016-12-15.679 & \toe{737.679} & & & HCT HFOSC & $2\times900$ & $B$ & $20.73\pm0.09$ \\
\hline
2016-12-11.368 & \toe{733.368} & 733.365 & 733.371 & Kiso Observatory & $3\times60$ & $V$ & $>19.1$\tablenotemark{a}\\
2016-12-12.826 & \toe{734.826} & & & Polaris Observatory & $12\times180$ & $V$ & $18.094\pm0.113$ \\
2016-12-12.847 & \toe{734.847} & & & AAVSO MAND & $19\times120$ & $V$ & $17.511\pm0.015$ \\
2016-12-12.90 & \toe{734.90} & & & Ond\v{r}ejov 0.65m & 1080 & $V$ & $17.95\pm0.09$\tablenotemark{e}\\
2016-12-12.969 & \toe{734.969} & & & AAVSO PXR & 60 & $V$ & $17.918\pm0.477$ \\
2016-12-12.970 & \toe{734.970} & & & AAVSO PXR & 60 & $V$ & $17.457\pm0.323$ \\
2016-12-13.012 & \toe{735.012} & & & AAVSO HBB & $10\times60$ & $V$ & $17.759\pm0.090$ \\
2016-12-13.014 & \toe{735.014} & & & AAVSO HBB & $10\times60$ & $V$ & $17.670\pm0.093$ \\
2016-12-13.016 & \toe{735.014} & & & AAVSO HBB & $10\times60$ & $V$ & $17.926\pm0.095$ \\
2016-12-13.038 & \toe{735.038} & & & AAVSO HBB & $10\times60$ & $V$ & $18.125\pm0.117$ \\
2016-12-13.047 & \toe{735.047} & & & AAVSO HBB & $10\times60$ & $V$ & $18.054\pm0.108$ \\
2016-12-13.055 & \toe{735.055} & & & AAVSO HBB & $10\times60$ & $V$ & $18.255\pm0.123$ \\
2016-12-13.066 & \toe{735.066} & & & AAVSO HBB & $10\times60$ & $V$ & $18.234\pm0.111$ \\
2016-12-13.079 & \toe{735.079} & & & AAVSO HBB & $10\times60$ & $V$ & $18.218\pm0.141$ \\
2016-12-13.092 & \toe{735.092} & & & AAVSO HBB & $10\times60$ & $V$ & $18.267\pm0.116$ \\
2016-12-13.105 & \toe{735.105} & & & AAVSO HBB & $10\times60$ & $V$ & $18.244\pm0.130$ \\
2016-12-13.112 & \toe{735.112} & & & MLO & 300 & $V$ & $18.35\pm0.08$\tablenotemark{d}\\
2016-12-13.118 & \toe{735.118} & & & AAVSO HBB & $10\times60$ & $V$ & $18.225\pm0.127$ \\
2016-12-13.130 & \toe{735.130} & & & AAVSO HBB & $10\times60$ & $V$ & $18.304\pm0.140$ \\
2016-12-13.142 & \toe{735.142} & & & AAVSO HBB & $10\times60$ & $V$ & $18.126\pm0.132$ \\
2016-12-13.227 & \toe{735.227} & & & AAVSO CMJA & $3\times300$ & $V$ & $18.394\pm0.197$ \\
2016-12-13.297 & \toe{735.297} & & & MLO & 600 & $V$ & $18.60\pm0.08$\tablenotemark{d}\\
2016-12-14.115 & \toe{736.115} & & & MLO & 1200 & $V$ & $19.20\pm0.10$\tablenotemark{d}\\
2016-12-14.435 & \toe{736.435} & 736.433 & 736.437 & Kiso Observatory & $2\times60$ & $V$ & $>18.2$\tablenotemark{a}\\
2016-12-14.755 & \toe{736.755} & & & HCT HFOSC & $2\times300$ & $V$ & $20.03\pm0.02$ \\
2016-12-15.367 & \toe{737.367} & 737.362 & 737.372 & Kiso Observatory & $3\times60$ & $V$ & $>20.0$\tablenotemark{a}\\
2016-12-15.660 & \toe{737.660} & & & HCT HFOSC & $3\times420$ & $V$ & $20.98\pm0.03$ \\
\hline
2016-12-11.382 & \toe{733.382} & 733.379 & 733.386 & Okayama Astrophysical Observatory & $9\times60$ & $R$ & $>18.0$\tablenotemark{a}\\
2016-12-11.389 & \toe{733.389} & 733.38 & 733.40 & Osaka Kyoiku University & $5\times150$ & $R$ & $>18.8$\tablenotemark{a}\\
2016-12-12.083 & \toe{734.083} & & & Palomar $48^{\prime\prime}$ & 60 & $R$ & $>19.9$\\
2016-12-12.90 & \toe{734.90} & & & Ond\v{r}ejov 0.65m & 1260 & $R$ & $17.76\pm0.05$\tablenotemark{e}\\
2016-12-13.125 & \toe{735.125} & & & MLO & 300 & $R$ & $17.97\pm0.05$\tablenotemark{d}\\
2016-12-13.204 & \toe{735.204} & & & MLO & 600 & $R$ & $18.00\pm0.05$\tablenotemark{d}\\
2016-12-14.088 & \toe{736.088} & & & MLO & 600 & $R$ & $18.57\pm0.05$\tablenotemark{d}\\
2016-12-14.188 & \toe{736.188} & 736.150 & 736.228 & AAVSO COO & $30\times180$ & $R$ & $19.045\pm0.329$\\
2016-12-14.383 & \toe{736.383} & 736.380 & 736.387 & Okayama Astrophysical Observatory & $9\times60$ & $R$ & $>16.3$\tablenotemark{a}\\
2016-12-14.47 & \toe{736.47} & 736.38 & 736.54 & Osaka Kyoiku University & $14\times300$ & $R$ & $19.1\pm0.1$\tablenotemark{a}\\
2016-12-14.722 & \toe{736.722} & & & HCT HFOSC & $3\times150$ & $R$ & $19.09\pm0.03$ \\
2016-12-15.031 & \toe{737.031} & & & Danish 1.54m & 900 & $R$ & $19.66\pm0.07$\tablenotemark{f}\\
2016-12-15.087 & \toe{737.087} & & & Palomar $48^{\prime\prime}$ & 60 & $R$ & $19.94\pm0.25$\\
2016-12-15.107 & \toe{737.107} & & & Palomar $48^{\prime\prime}$ & 60 & $R$ & $>19.8$\\
2016-12-15.370 & \toe{737.370} & 737.337 & 737.403 & Meili 0.4m & $76\times60$ & $R$ & $>20.2$\tablenotemark{a}\\
2016-12-15.504 & \toe{737.504} & 737.433 & 737.523 & Okayama Astrophysical Observatory & $12\times60$ & $R$ & $>18.0$\tablenotemark{a}\\
2016-12-15.638 & \toe{737.638} & & & HCT HFOSC & $3\times300$ & $R$ & $20.06\pm0.03$ \\
2016-12-16.45 & \toe{738.50} & 738.41 & 738.49& Osaka Kyoiku University & $20\times300$ & $R$ & $>18.7$\tablenotemark{a}\\
2016-12-16.693 & \toe{738.693} & & & Ond\v{r}ejov 0.65m & 1800 & $R$ & $20.4\pm0.2$\tablenotemark{f}\\
2016-12-17.040 & \toe{739.040} & & & Danish 1.54m & 900 & $R$ & $20.74\pm0.08$\tablenotemark{f}\\
2016-12-17.50 & \toe{739.52} & 739.46 & 739.54& Osaka Kyoiku University & $19\times300$ & $R$ & $>19.3$\tablenotemark{a}\\
2016-12-18.035 & \toe{740.035} & & & Danish 1.54m & 900 & $R$ & $20.82\pm0.09$\tablenotemark{f}\\
2016-12-18.474 & \toe{740.474} & 740.444 & 740.503 & Pirka 1.6m & $5\times300$ & $R$ & $>20.6$\tablenotemark{a}\\
2016-12-19.041 & \toe{741.041} & & & Danish 1.54m & 900 & $R$ & $21.19\pm0.15$\tablenotemark{f}\\
2016-12-19.455 & \toe{741.455} & 741.434 & 741.476 & Pirka 1.6m & $7\times300$ & $R$ & $>20.5$\tablenotemark{a}\\
2016-12-20.035 & \toe{742.035} & & & Danish 1.54m & 900 & $R$ & $21.5\pm0.2$\tablenotemark{f}\\
\hline
2016-12-12.90 & \toe{734.90} & & & Ond\v{r}ejov 0.65m & 1080 & $I$ & $17.68\pm0.08$\tablenotemark{e}\\
2016-12-13.140 & \toe{735.140} & & & MLO & 300 & $I$ & $17.76\pm0.05$\tablenotemark{d}\\
2016-12-13.283 & \toe{735.283} & & & MLO & 600 & $I$ & $17.85\pm0.05$\tablenotemark{d}\\
2016-12-14.105 & \toe{736.105} & & & MLO & 600 & $I$ & $18.56\pm0.08$\tablenotemark{d}\\
2016-12-14.738 & \toe{736.738} & & & HCT HFOSC & $3\times150$ & $I$ & $18.80\pm0.03$ \\
2016-12-15.437 & \toe{737.437} & 737.426 & 737.447 & Pirka 1.6m & $3\times600$ & $I$ & $>21.0$\tablenotemark{a}\\
2016-12-15.647 & \toe{737.647} & & & HCT HFOSC & $3\times180$ & $I$ & $19.42\pm0.02$ \\
2016-12-19.484 & \toe{741.484} & 741.477 & 741.484 & Pirka 1.6m & $2\times300$ & $I$ & $>21.8$\tablenotemark{a}\\
2016-12-28.074 & \toe{750.074} & 750.072 & 750.076 & AAVSO COO & 300 & $I$ & $>17.4$\\
\hline
2016-12-10.850 & \toe{732.850} & 732.845 & 732.856 & LT IO:O & $3\times300$ & $u'$ & $>22.795$\\
2016-12-10.925 & \toe{732.925} & 732.920 & 732.931 & LT IO:O & $3\times300$ & $u'$ & $>22.435$\\
2016-12-11.008 & \toe{733.008} & 733.002 & 733.013 & LT IO:O & $3\times300$ & $u'$ & $>22.343$\\
2016-12-11.903 & \toe{733.903} & 733.897 & 733.908 & LT IO:O & $3\times300$ & $u'$ & $>22.192$\tablenotemark{b}\\
2016-12-12.862 & \toe{734.862} & 734.858 & 734.865 & LT IO:O & $3\times180$ & $u'$ & $17.923\pm0.041$\\
2016-12-12.910 & \toe{734.910} & 734.906 & 734.913 & LT IO:O & $3\times180$ & $u'$ & $17.901\pm0.039$\\
2016-12-12.947 & \toe{734.947} & 734.944 & 734.951 & LT IO:O & $3\times180$ & $u'$ & $17.854\pm0.041$\\
2016-12-13.055 & \toe{735.055} & 735.052 & 735.059 & LT IO:O & $3\times180$ & $u'$ & $18.119\pm0.046$\\
2016-12-13.071 & \toe{735.071} & 735.067 & 735.074 & LT IO:O & $3\times180$ & $u'$ & $18.215\pm0.042$\\
2016-12-13.786 & \toe{735.786} & 735.783 & 735.789 & LT IO:O & $3\times180$ & $u'$ & $18.518\pm0.060$\\
2016-12-13.800 & \toe{735.800} & 735.797 & 735.803 & LT IO:O & $3\times180$ & $u'$ & $18.566\pm0.017$\\
2016-12-13.884 & \toe{735.884} & 735.880 & 735.887 & LT IO:O & $3\times180$ & $u'$ & $18.589\pm0.025$\\
2016-12-13.975 & \toe{735.975} & 735.972 & 735.979 & LT IO:O & $3\times180$ & $u'$ & $18.693\pm0.023$\\
2016-12-14.000 & \toe{736.000} & 735.997 & 736.004 & LT IO:O & $3\times180$ & $u'$ & $18.623\pm0.026$\\
2016-12-14.839 & \toe{736.839} & 736.835 & 736.842 & LT IO:O & $3\times180$ & $u'$ & $19.425\pm0.018$\\
2016-12-14.919 & \toe{736.919} & 736.916 & 736.922 & LT IO:O & $3\times180$ & $u'$ & $19.617\pm0.036$\\
2016-12-15.787 & \toe{737.787} & 737.784 & 737.791 & LT IO:O & $3\times180$ & $u'$ & $19.574\pm0.095$\\
2016-12-15.854 & \toe{737.854} & 737.853 & 737.855 & LT IO:O & 180 & $u'$ & $19.833\pm0.036$\\
2016-12-15.885 & \toe{737.885} & 737.882 & 737.889 & LT IO:O & $3\times180$ & $u'$ & $19.958\pm0.030$\\
2016-12-15.946 & \toe{737.946} & 737.943 & 737.950 & LT IO:O & $3\times180$ & $u'$ & $20.004\pm0.052$\\
2016-12-15.959 & \toe{737.959} & 737.955 & 737.962 & LT IO:O & $3\times180$ & $u'$ & $19.844\pm0.041$\\
2016-12-15.966 & \toe{737.966} & 737.963 & 737.970 & LT IO:O & $3\times180$ & $u'$ & $19.856\pm0.051$\\
2016-12-15.973 & \toe{737.973} & 737.971 & 737.975 & LT IO:O & $2\times180$ & $u'$ & $19.670\pm0.045$\\
2016-12-15.990 & \toe{737.990} & 737.986 & 737.993 & LT IO:O & $3\times180$ & $u'$ & $19.767\pm0.052$\\
2016-12-15.997 & \toe{737.997} & 737.994 & 738.000 & LT IO:O & $3\times180$ & $u'$ & $19.758\pm0.055$\\
2016-12-16.005 & \toe{738.005} & 738.001 & 738.008 & LT IO:O & $3\times180$ & $u'$ & $19.808\pm0.057$\\
2016-12-16.012 & \toe{738.012} & 738.009 & 738.016 & LT IO:O & $3\times180$ & $u'$ & $19.795\pm0.054$\\
2016-12-16.024 & \toe{738.024} & 738.020 & 738.027 & LT IO:O & $3\times180$ & $u'$ & $19.771\pm0.078$\\
2016-12-16.041 & \toe{738.041} & 738.038 & 738.044 & LT IO:O & $3\times180$ & $u'$ & $19.942\pm0.078$\\
2016-12-16.058 & \toe{738.058} & 738.055 & 738.062 & LT IO:O & $3\times180$ & $u'$ & $19.877\pm0.128$\\
2016-12-27.887 & \toe{749.887} & 749.876 & 749.897 & LT IO:O & $3\times600$ & $u'$ & $23.397\pm0.204$\tablenotemark{g,h}\\
2016-12-29.856 & \toe{751.856} & 751.845 & 751.867 & LT IO:O & $3\times600$ & $u'$ & $>18.6$\\
2016-12-30.843 & \toe{752.843} & 752.832 & 752.854 & LT IO:O & $3\times600$ & $u'$ & $>21.8$\\
2017-01-08.121 & \toe{761.121} & & & LBT MODS2R & $6\times100$ & $u'$ & $>23.3$ \\
\hline
2016-12-13.010 & \toe{735.010} & & & INT WFC & 60 & $g'$ & $18.342\pm0.033$\\
2016-12-13.012 & \toe{735.012} & & & INT WFC & 60 & $g'$ & $18.381\pm0.034$\\
2016-12-13.015 & \toe{735.015} & & & INT WFC & 60 & $g'$ & $18.399\pm0.038$\\
2016-12-13.017 & \toe{735.017} & & & INT WFC & 60 & $g'$ & $18.354\pm0.036$\\
2016-12-13.020 & \toe{735.020} & & & INT WFC & 60 & $g'$ & $18.367\pm0.037$\\
2016-12-13.022 & \toe{735.022} & & & INT WFC & 60 & $g'$ & $18.358\pm0.035$\\
2016-12-13.024 & \toe{735.024} & & & INT WFC & 60 & $g'$ & $18.404\pm0.039$\\
2016-12-13.027 & \toe{735.027} & & & INT WFC & 60 & $g'$ & $18.408\pm0.037$\\
2016-12-13.029 & \toe{735.029} & & & INT WFC & 60 & $g'$ & $18.324\pm0.036$\\
2016-12-13.032 & \toe{735.032} & & & INT WFC & 60 & $g'$ & $18.394\pm0.039$\\
2016-12-13.034 & \toe{735.034} & & & INT WFC & 60 & $g'$ & $18.369\pm0.038$\\
2016-12-13.036 & \toe{735.036} & & & INT WFC & 60 & $g'$ & $18.372\pm0.042$\\
2016-12-13.039 & \toe{735.039} & & & INT WFC & 60 & $g'$ & $18.343\pm0.041$\\
2016-12-13.041 & \toe{735.041} & & & INT WFC & 60 & $g'$ & $18.414\pm0.039$\\
2016-12-13.044 & \toe{735.044} & & & INT WFC & 60 & $g'$ & $18.396\pm0.040$\\
2016-12-13.046 & \toe{735.046} & & & INT WFC & 60 & $g'$ & $18.486\pm0.047$\\
2016-12-13.051 & \toe{735.051} & & & INT WFC & 60 & $g'$ & $18.444\pm0.040$\\
2016-12-13.053 & \toe{735.053} & & & INT WFC & 60 & $g'$ & $18.520\pm0.045$\\
2016-12-13.055 & \toe{735.055} & & & INT WFC & 60 & $g'$ & $18.454\pm0.046$\\
2016-12-13.058 & \toe{735.058} & & & INT WFC & 60 & $g'$ & $18.402\pm0.041$\\
2016-12-13.060 & \toe{735.060} & & & INT WFC & 60 & $g'$ & $18.396\pm0.041$\\
2016-12-13.062 & \toe{735.062} & & & INT WFC & 60 & $g'$ & $18.415\pm0.041$\\
2016-12-13.065 & \toe{735.065} & & & INT WFC & 60 & $g'$ & $18.430\pm0.042$\\
2016-12-13.067 & \toe{735.067} & & & INT WFC & 60 & $g'$ & $18.289\pm0.039$\\
2016-12-13.070 & \toe{735.070} & & & INT WFC & 60 & $g'$ & $18.379\pm0.042$\\
2016-12-13.072 & \toe{735.072} & & & INT WFC & 60 & $g'$ & $18.435\pm0.045$\\
2016-12-13.075 & \toe{735.075} & & & INT WFC & 60 & $g'$ & $18.404\pm0.045$\\
2016-12-13.078 & \toe{735.078} & & & INT WFC & 60 & $g'$ & $18.503\pm0.051$\\
2016-12-13.081 & \toe{735.081} & & & INT WFC & 60 & $g'$ & $18.372\pm0.046$\\
2016-12-13.084 & \toe{735.084} & & & INT WFC & 60 & $g'$ & $18.360\pm0.047$\\
2016-12-13.086 & \toe{735.086} & & & INT WFC & 60 & $g'$ & $18.471\pm0.056$\\
2016-12-13.089 & \toe{735.089} & & & INT WFC & 60 & $g'$ & $18.462\pm0.053$\\
2016-12-13.092 & \toe{735.092} & & & INT WFC & 60 & $g'$ & $18.450\pm0.058$\\
2016-12-13.096 & \toe{735.096} & & & INT WFC & 60 & $g'$ & $18.497\pm0.063$\\
2016-12-13.170 & \toe{735.170} & & & ERAU & $2\times600$ & $g'$ & $18.42\pm0.03$\\
2016-12-14.053 & \toe{736.053} & & & ERAU & $6\times600$ & $g'$ & $19.38\pm0.07$\\
2017-01-08.112 & \toe{761.112} & & & LBT MODS2R& $3\times100$ & $g'$ & $>23.1$ \\
\hline
2016-12-12.083 & \toe{734.083} & & & Palomar $48^{\prime\prime}$ & 60 & $r'$ & $>19.9$\\
2016-12-13.007 & \toe{735.007} & & & INT WFC & 60 & $r'$ & $18.152\pm0.043$\\
2016-12-13.008 & \toe{735.008} & & & INT WFC & 60 & $r'$ & $18.136\pm0.034$\\
2016-12-13.011 & \toe{735.011} & & & INT WFC & 60 & $r'$ & $18.104\pm0.032$\\
2016-12-13.014 & \toe{735.014} & & & INT WFC & 60 & $r'$ & $18.163\pm0.034$\\
2016-12-13.016 & \toe{735.016} & & & INT WFC & 60 & $r'$ & $18.092\pm0.033$\\
2016-12-13.018 & \toe{735.018} & & & INT WFC & 60 & $r'$ & $18.168\pm0.033$\\
2016-12-13.021 & \toe{735.021} & & & INT WFC & 60 & $r'$ & $18.089\pm0.033$\\
2016-12-13.023 & \toe{735.023} & & & INT WFC & 60 & $r'$ & $18.158\pm0.033$\\
2016-12-13.026 & \toe{735.026} & & & INT WFC & 60 & $r'$ & $18.147\pm0.033$\\
2016-12-13.028 & \toe{735.028} & & & INT WFC & 60 & $r'$ & $18.187\pm0.034$\\
2016-12-13.030 & \toe{735.030} & & & INT WFC & 60 & $r'$ & $18.135\pm0.033$\\
2016-12-13.033 & \toe{735.033} & & & INT WFC & 60 & $r'$ & $18.123\pm0.034$\\
2016-12-13.035 & \toe{735.035} & & & INT WFC & 60 & $r'$ & $18.115\pm0.033$\\
2016-12-13.038 & \toe{735.038} & & & INT WFC & 60 & $r'$ & $18.177\pm0.035$\\
2016-12-13.040 & \toe{735.040} & & & INT WFC & 60 & $r'$ & $18.146\pm0.041$\\
2016-12-13.042 & \toe{735.042} & & & INT WFC & 60 & $r'$ & $18.202\pm0.037$\\
2016-12-13.045 & \toe{735.045} & & & INT WFC & 60 & $r'$ & $18.158\pm0.034$\\
2016-12-13.047 & \toe{735.047} & & & INT WFC & 60 & $r'$ & $18.185\pm0.034$\\
2016-12-13.052 & \toe{735.052} & & & INT WFC & 60 & $r'$ & $18.112\pm0.028$\\
2016-12-13.054 & \toe{735.054} & & & INT WFC & 60 & $r'$ & $18.065\pm0.032$\\
2016-12-13.056 & \toe{735.056} & & & INT WFC & 60 & $r'$ & $18.118\pm0.030$\\
2016-12-13.059 & \toe{735.059} & & & INT WFC & 60 & $r'$ & $18.116\pm0.032$\\
2016-12-13.061 & \toe{735.061} & & & INT WFC & 60 & $r'$ & $18.181\pm0.034$\\
2016-12-13.064 & \toe{735.064} & & & INT WFC & 60 & $r'$ & $18.158\pm0.033$\\
2016-12-13.066 & \toe{735.066} & & & INT WFC & 60 & $r'$ & $18.153\pm0.031$\\
2016-12-13.069 & \toe{735.069} & & & INT WFC & 60 & $r'$ & $18.086\pm0.031$\\
2016-12-13.071 & \toe{735.071} & & & INT WFC & 60 & $r'$ & $18.137\pm0.033$\\
2016-12-13.073 & \toe{735.073} & & & INT WFC & 60 & $r'$ & $18.178\pm0.035$\\
2016-12-13.077 & \toe{735.077} & & & INT WFC & 60 & $r'$ & $18.190\pm0.034$\\
2016-12-13.079 & \toe{735.079} & & & INT WFC & 60 & $r'$ & $18.209\pm0.036$\\
2016-12-13.083 & \toe{735.083} & & & INT WFC & 60 & $r'$ & $18.240\pm0.040$\\
2016-12-13.085 & \toe{735.085} & & & INT WFC & 60 & $r'$ & $18.183\pm0.035$\\
2016-12-13.088 & \toe{735.088} & & & INT WFC & 60 & $r'$ & $18.229\pm0.038$\\
2016-12-13.090 & \toe{735.090} & & & INT WFC & 60 & $r'$ & $18.123\pm0.035$\\
2016-12-13.094 & \toe{735.094} & & & INT WFC & 60 & $r'$ & $18.258\pm0.041$\\
2016-12-13.103 & \toe{735.103} & & & ERAU & 600 & $r'$ & $18.02\pm0.03$\tablenotemark{i}\\
2016-12-13.110 & \toe{735.110} & & & ERAU & 600 & $r'$ & $18.00\pm0.04$\tablenotemark{i}\\
2016-12-13.117 & \toe{735.117} & & & ERAU & 600 & $r'$ & $18.00\pm0.04$\tablenotemark{i}\\
2016-12-13.125 & \toe{735.125} & & & ERAU & 600 & $r'$ & $18.08\pm0.05$\tablenotemark{i}\\
2016-12-13.132 & \toe{735.132} & & & ERAU & 600 & $r'$ & $18.11\pm0.05$\tablenotemark{i}\\
2016-12-13.139 & \toe{735.139} & & & ERAU & 600 & $r'$ & $17.98\pm0.04$\tablenotemark{i}\\
2016-12-13.146 & \toe{735.146} & & & ERAU & 600 & $r'$ & $18.13\pm0.05$\tablenotemark{i}\\
2016-12-13.154 & \toe{735.154} & & & ERAU & 600 & $r'$ & $18.03\pm0.05$\tablenotemark{i}\\
2016-12-13.161 & \toe{735.161} & & & ERAU & 600 & $r'$ & $18.12\pm0.05$\tablenotemark{i}\\
2016-12-13.201 & \toe{735.201} & & & ERAU & 600 & $r'$ & $18.39\pm0.07$\tablenotemark{i}\\
2016-12-13.208 & \toe{735.208} & & & ERAU & 600 & $r'$ & $18.37\pm0.07$\tablenotemark{i}\\
2016-12-13.216 & \toe{735.216} & & & ERAU & 600 & $r'$ & $18.19\pm0.06$\tablenotemark{i}\\
2016-12-13.223 & \toe{735.223} & & & ERAU & 600 & $r'$ & $18.31\pm0.07$\tablenotemark{i}\\
2016-12-13.230 & \toe{735.230} & & & ERAU & 600 & $r'$ & $18.37\pm0.07$\tablenotemark{i}\\
2016-12-13.97 & \toe{735.97} & & & ERAU & 600 & $r'$ & $18.8\pm0.12$\tablenotemark{e}\\ 
2016-12-13.99 & \toe{735.99} & & & ERAU & 600 & $r'$ & $18.9\pm0.11$\tablenotemark{e}\\
2016-12-14.01 & \toe{736.01} & & & ERAU & 600 & $r'$ & $18.9\pm0.11$\tablenotemark{e}\\
2016-12-14.030 & \toe{736.030} & & & ERAU & 600 & $r'$ & $18.85\pm0.04$\tablenotemark{j}\\
2016-12-14.04 & \toe{736.04} & & & ERAU & 600 & $r'$ & $18.8\pm0.10$\tablenotemark{e}\\
2016-12-14.06 & \toe{736.06} & & & ERAU & 600 & $r'$ & $18.9\pm0.12$\tablenotemark{e}\\
2016-12-14.07 & \toe{736.07} & & & ERAU & 600 & $r'$ & $18.9\pm0.10$\tablenotemark{e}\\
2016-12-14.09 & \toe{736.09} & & & ERAU & 600 & $r'$ & $18.7\pm0.10$\tablenotemark{e}\\
2016-12-14.10 & \toe{736.10} & & & ERAU & 600 & $r'$ & $19.0\pm0.13$\tablenotemark{e}\\
2016-12-14.12 & \toe{736.12} & & & ERAU & 600 & $r'$ & $19.0\pm0.13$\tablenotemark{e}\\
2016-12-14.13 & \toe{736.13} & & & ERAU & 600 & $r'$ & $18.9\pm0.10$\tablenotemark{e}\\
2016-12-14.15 & \toe{736.15} & & & ERAU & 600 & $r'$ & $19.5\pm0.20$\tablenotemark{e}\\
2016-12-15.01 & \toe{737.01} & & & ERAU & $6\times600$ & $r'$ & $19.96\pm0.07$\tablenotemark{j}\\
2016-12-15.06 & \toe{737.06} & & & ERAU & $11\times600$ & $r'$ & $19.83\pm0.08$\tablenotemark{j}\\
2016-12-16.01 & \toe{738.01} & & & ERAU & $6\times600$ & $r'$ & $20.30\pm0.10$\tablenotemark{j}\\
2016-12-16.99 & \toe{738.99} & & & ERAU & $10\times600$ & $r'$ & $20.85\pm0.13$\tablenotemark{j}\\
2016-12-17.08 & \toe{739.08} & & & ERAU & $13\times600$ & $r'$ & $21.08\pm0.15$\tablenotemark{j}\\
2016-12-18.00 & \toe{740.00} & & & ERAU & $19\times600$ & $r'$ & $21.35\pm0.14$\tablenotemark{j}\\
2017-01-08.112 & \toe{761.112} & & & LBT MODS2R & $3\times100$ & $r'$ & $>22.8$ \\
\hline
2016-12-13.185 & \toe{735.185} & & & ERAU & $2\times600$ & $i'$ & $18.08\pm0.05$\\
2016-12-13.99 & \toe{735.99} & & & ERAU & 600 & $i'$ & $19.0\pm0.14$\tablenotemark{e}\\
2016-12-14.01 & \toe{736.01} & & & ERAU & 600 & $i'$ & $18.8\pm0.12$\tablenotemark{e}\\
2016-12-14.04 & \toe{736.04} & & & ERAU & 600 & $i'$ & $18.9\pm0.13$\tablenotemark{e}\\
2016-12-14.06 & \toe{736.06} & & & ERAU & 600 & $i'$ & $19.01\pm0.14$\tablenotemark{e}\\
2016-12-14.07 & \toe{736.07} & & & ERAU & 600 & $i'$ & $18.8\pm0.11$\tablenotemark{e}\\
2016-12-14.09 & \toe{736.09} & & & ERAU & 600 & $i'$ & $18.8\pm0.12$\tablenotemark{e}\\
2016-12-14.10 & \toe{736.10} & & & ERAU & 600 & $i'$ & $19.1\pm0.14$\tablenotemark{e}\\
2016-12-14.12 & \toe{736.12} & & & ERAU & 600 & $i'$ & $18.7\pm0.10$\tablenotemark{e}\\
2016-12-14.13 & \toe{736.13} & & & ERAU & 600 & $i'$ & $19.0\pm0.14$\tablenotemark{e}\\
2016-12-14.15 & \toe{736.15} & & & ERAU & 600 & $i'$ & $19.1\pm0.17$\tablenotemark{e}\\
2017-01-08.118 & \toe{761.118} & & & LBT MODS2R & $3\times100$ & $i'$ & $>22.7$ \\
\hline
2017-01-08.123 & \toe{761.123} & & & LBT MODS2R& $3\times100$ & $z'$ & $>22.5$ \\
\hline
2016-12-07.976 & \toe{729.976} & 729.971 & 729.981 & \hst WFC3/UVIS & 898 & F657W & $23.074\pm0.092$\\
2016-12-08.041 & \toe{730.041} & 730.036 & 730.046 & \hst WFC3/UVIS & 898 & F657W & $23.350\pm0.107$\\
2016-12-08.053 & \toe{730.053} & 730.048 & 730.058 & \hst WFC3/UVIS & 898 & F657W & $22.986\pm0.088$\\
2016-12-09.300 & \toe{731.300} & 731.295 & 731.305 & \hst WFC3/UVIS & 898 & F657W & $23.527\pm0.130$\\
2016-12-09.312 & \toe{731.312} & 731.307 & 731.317 & \hst WFC3/UVIS & 898 & F657W & $23.634\pm0.143$\\
2016-12-09.324 & \toe{731.324} & 731.319 & 731.329 & \hst WFC3/UVIS & 898 & F657W & $23.343\pm0.117$\\
2016-12-10.293 & \toe{732.293} & 732.288 & 732.299 & \hst WFC3/UVIS & 898 & F657W & $23.356\pm0.113$\\
2016-12-10.305 & \toe{732.305} & 732.300 & 732.311 & \hst WFC3/UVIS & 898 & F657W & $23.542\pm0.122$\\
2016-12-10.317 & \toe{732.317} & 732.312 & 732.322 & \hst WFC3/UVIS & 898 & F657W & $23.361\pm0.102$\\
2016-12-11.022 & \toe{733.022} & 733.016 & 733.027 & \hst WFC3/UVIS & 898 & F657W & $23.390\pm0.109$\\
2016-12-11.087 & \toe{733.087} & 733.081 & 733.092 & \hst WFC3/UVIS & 898 & F657W & $23.237\pm0.098$\\
2016-12-11.099 & \toe{733.099} & 733.093 & 733.104 & \hst WFC3/UVIS & 898 & F657W & $23.352\pm0.110$\\
2016-12-12.794 & \toe{734.794} & & & OAJ & $2\times400$ & H$\alpha$ & $17.34\pm0.17$\\
2016-12-17.048 & \toe{739.048} & 739.042 & 739.053 & \hst WFC3/UVIS & 898 & F657W & $19.386\pm0.012$\tablenotemark{k}\\
2016-12-17.060 & \toe{739.060} & 739.055 & 739.065 & \hst WFC3/UVIS & 898 & F657W & $19.385\pm0.012$\tablenotemark{k}\\
2016-12-17.113 & \toe{739.113} & 739.108 & 739.118 & \hst WFC3/UVIS & 898 & F657W & $19.267\pm0.011$\tablenotemark{k}\\
\hline
2016-12-08.107 & \toe{730.107} & 730.102 & 730.113 & \hst WFC3/UVIS & 935 & F645W & $23.529\pm0.130$\\
2016-12-08.120 & \toe{730.120} & 730.114 & 730.125 & \hst WFC3/UVIS & 935 & F645W & $23.489\pm0.130$\\
2016-12-08.174 & \toe{730.174} & 730.168 & 730.179 & \hst WFC3/UVIS & 935 & F645W & $23.778\pm0.167$\\
2016-12-09.365 & \toe{731.365} & 731.360 & 731.371 & \hst WFC3/UVIS & 935 & F645W & $23.780\pm0.162$\\
2016-12-09.378 & \toe{731.378} & 731.372 & 731.383 & \hst WFC3/UVIS & 935 & F645W & $24.182\pm0.208$\\
2016-12-09.390 & \toe{731.390} & 731.385 & 731.396 & \hst WFC3/UVIS & 935 & F645W & $23.379\pm0.134$\\
2016-12-10.359 & \toe{732.359} & 732.353 & 732.364 & \hst WFC3/UVIS & 935 & F645W & $23.792\pm0.187$\\
2016-12-10.371 & \toe{732.371} & 732.365 & 732.376 & \hst WFC3/UVIS & 935 & F645W & $23.482\pm0.152$\\
2016-12-10.383 & \toe{732.383} & 732.378 & 732.389 & \hst WFC3/UVIS & 935 & F645W & $23.521\pm0.127$\\
2016-12-11.153 & \toe{733.153} & 733.148 & 733.158 & \hst WFC3/UVIS & 935 & F645W & $23.591\pm0.137$\\
2016-12-11.165 & \toe{733.165} & 733.160 & 733.171 & \hst WFC3/UVIS & 935 & F645W & $23.189\pm0.104$\\
2016-12-11.219 & \toe{733.219} & 733.214 & 733.225 & \hst WFC3/UVIS & 935 & F645W & $23.457\pm0.127$\\
2016-12-17.126 & \toe{739.126} & 739.120 & 739.131 & \hst WFC3/UVIS & 935 & F645W & $20.529\pm0.023$\tablenotemark{k}\\
2016-12-17.180 & \toe{739.180} & 739.174 & 739.185 & \hst WFC3/UVIS & 935 & F645W & $20.290\pm0.020$\tablenotemark{k}\\
2016-12-17.192 & \toe{739.192} & 739.187 & 739.197 & \hst WFC3/UVIS & 935 & F645W & $20.625\pm0.024$\tablenotemark{k}\\
\enddata
\tablecomments{(Includes all observations from $t\sim7$\,days before the eruption until $t\sim30$\,days post-eruption. This table is available in its entirety in machine-readable form.)}
\tablenotetext{\dag}{The time since eruption assumes an eruption date of 2016 December 12.32\,UT.}
\tablerefs{(a)~\citet{2016ATel.9891....1N},
(b)~\citet{2016ATel.9848....1I},
(c)~\citet{2016ATel.9885....1T},
(d)~\citet{2016ATel.9864....1S},
(e)~\citet{2016ATel.9861....1B},
(f)~\citet{2016ATel.9883....1H},
(g)~\citet{2016ATel.9906....1D},
(h)~\citet{2016ATel.9910....1D},
(i)~\citet{2016ATel.9857....1E},
(j)~\citet{2016ATel.9881....1K},
(k)~\citet{2016ATel.9874....1D}.}
\end{deluxetable}

\newpage

\begin{deluxetable}{rrrrrrrrrrrr}
\tablecaption{Individual \swift Snapshots of the Observations in Table~\ref{tab:swift}. Plotted in Figure~\ref{fig:xrt_split}.\label{tab:swift_split}}
\tablehead{
\colhead{ObsID\_part} & \colhead{Exp$^a$} & \colhead{Date$^b$} & \colhead{MJD$^b$} & \colhead{$\Delta t^c$} & \colhead{uvw2$^d$} & \colhead{Rate} \\
& \colhead{[ks]} & \colhead{[UT]} & \colhead{[d]} & \colhead{[d]} & \colhead{[mag]} & \colhead{[\power{-2} ct s$^{-1}$]}}
\startdata
00032613183\_1 & 2.26 & 2016-12-12.65 & 57734.65 & 0.33 & $16.56\pm0.08$ & $<0.5$ \\
00032613183\_2 & 1.73 & 2016-12-12.72 & 57734.72 & 0.40 & $16.61\pm0.08$ & $<0.6$ \\
00032613184\_1 & 0.95 & 2016-12-13.19 & 57735.19 & 0.87 & $16.95\pm0.09$ & $<1.2$ \\
00032613184\_2 & 1.07 & 2016-12-13.59 & 57735.59 & 1.27 & $17.25\pm0.09$ & $<1.0$ \\
00032613184\_3 & 1.13 & 2016-12-13.71 & 57735.72 & 1.40 & $17.37\pm0.09$ & $<1.1$ \\
00032613184\_4 & 0.99 & 2016-12-13.78 & 57735.79 & 1.47 & $17.36\pm0.10$ & $<1.6$ \\
00032613185\_1 & 1.35 & 2016-12-14.25 & 57736.26 & 1.94 & $17.78\pm0.10$ & $<0.8$ \\
00032613185\_2 & 0.21 & 2016-12-14.33 & 57736.33 & 2.01 & $17.79\pm0.16$ & $<5.3$ \\
00032613185\_3 & 1.26 & 2016-12-14.38 & 57736.39 & 2.07 & $17.88\pm0.10$ & $<0.9$ \\
00032613185\_4 & 0.90 & 2016-12-14.45 & 57736.45 & 2.13 & $17.89\pm0.11$ & $<1.3$ \\
00032613186\_1 & 1.31 & 2016-12-15.64 & 57737.65 & 3.33 & $18.61\pm0.12$ & $<1.0$ \\
00032613186\_2 & 0.58 & 2016-12-15.92 & 57737.92 & 3.60 & $18.94\pm0.19$ & $<1.9$ \\
00032613186\_3 & 1.35 & 2016-12-15.97 & 57737.97 & 3.65 & $18.37\pm0.11$ & $<0.8$ \\
00032613188\_1 & 1.10 & 2016-12-16.38 & 57738.38 & 4.06 & $18.65\pm0.13$ & $<1.0$ \\
00032613189\_1 & 1.71 & 2016-12-18.10 & 57740.10 & 5.78 & $19.27\pm0.15$ & $1.3\pm0.3$ \\
00032613189\_2 & 1.64 & 2016-12-18.16 & 57740.17 & 5.85 & $19.06\pm0.14$ & $<0.5$ \\
00032613189\_3 & 0.56 & 2016-12-18.23 & 57740.23 & 5.91 & $19.71\pm0.32$ & $<2.1$ \\
00032613190\_1 & 1.36 & 2016-12-19.49 & 57741.50 & 7.18 & $19.67\pm0.21$ & $<1.0$ \\
00032613190\_2 & 0.63 & 2016-12-19.56 & 57741.57 & 7.25 & $20.02\pm0.35$ & $<1.7$ \\
00032613190\_3 & 1.65 & 2016-12-19.62 & 57741.63 & 7.31 & $19.96\pm0.23$ & $0.5\pm0.2$ \\
00032613190\_4 & 0.42 & 2016-12-19.69 & 57741.69 & 7.37 & $>19.8$ & $1.7\pm0.8$ \\
00032613191\_1 & 1.65 & 2016-12-20.88 & 57742.89 & 8.57 & $20.49\pm0.33$ & $2.2\pm0.4$ \\
00032613191\_2 & 0.24 & 2016-12-20.97 & 57742.97 & 8.65 & - & $1.0\pm0.9$ \\
00032613191\_3 & 0.11 & 2016-12-20.97 & 57742.98 & 8.66 & $>19.0$ & $<9.9$ \\
00032613192\_1 & 1.67 & 2016-12-21.49 & 57743.49 & 9.17 & $>20.8$ & $1.9\pm0.4$ \\
00032613192\_2 & 1.61 & 2016-12-21.55 & 57743.56 & 9.24 & $20.62\pm0.35$ & $1.4\pm0.3$ \\
00032613192\_3 & 0.69 & 2016-12-21.62 & 57743.62 & 9.30 & $>20.2$ & $0.4\pm0.4$ \\
00032613193\_1 & 1.22 & 2016-12-22.68 & 57744.69 & 10.37 & $20.21\pm0.30$ & $2.5\pm0.5$ \\
00032613193\_2 & 1.32 & 2016-12-22.74 & 57744.75 & 10.43 & $20.33\pm0.32$ & $0.9\pm0.3$ \\
00032613194\_1 & 1.71 & 2016-12-23.67 & 57745.68 & 11.36 & $20.71\pm0.38$ & $1.8\pm0.4$ \\
00032613194\_2 & 1.24 & 2016-12-23.75 & 57745.75 & 11.43 & $>20.6$ & $0.9\pm0.3$ \\
00032613195\_1 & 0.92 & 2016-12-24.00 & 57746.01 & 11.69 & $19.87\pm0.29$ & $0.7\pm0.4$ \\
00032613195\_2 & 0.90 & 2016-12-24.07 & 57746.08 & 11.76 & $>20.4$ & $0.7\pm0.4$ \\
00032613195\_3 & 0.90 & 2016-12-24.20 & 57746.21 & 11.89 & $>20.4$ & $1.0\pm0.4$ \\
00032613195\_4 & 0.19 & 2016-12-24.34 & 57746.34 & 12.02 & $>19.2$ & $<6.0$ \\
00032613196\_1 & 1.60 & 2016-12-25.00 & 57747.01 & 12.69 & $>20.7$ & $0.4\pm0.2$ \\
00032613196\_2 & 0.99 & 2016-12-25.07 & 57747.07 & 12.75 & $>20.4$ & $1.0\pm0.4$ \\
00032613196\_3 & 0.21 & 2016-12-25.13 & 57747.14 & 12.82 & $>19.4$ & $<5.4$ \\
00032613197\_1 & 1.37 & 2016-12-26.20 & 57748.20 & 13.88 & $>20.6$ & $0.3\pm0.2$ \\
00032613197\_2 & 1.35 & 2016-12-26.26 & 57748.27 & 13.95 & $>20.6$ & $0.4\pm0.3$ \\
00032613198\_1 & 1.60 & 2016-12-27.72 & 57749.73 & 15.41 & $>20.7$ & $<0.7$ \\
00032613198\_2 & 1.25 & 2016-12-27.79 & 57749.79 & 15.47 & $>20.5$ & $<1.0$ \\
00032613199\_1 & 1.69 & 2016-12-28.19 & 57750.19 & 15.87 & $>20.7$ & $<0.5$ \\
00032613199\_2 & 1.56 & 2016-12-28.26 & 57750.26 & 15.94 & $>20.6$ & $<0.8$ \\
00032613200\_1 & 1.49 & 2016-12-29.45 & 57751.46 & 17.14 & $>20.7$ & $<0.7$ \\
00032613200\_2 & 0.88 & 2016-12-29.52 & 57751.53 & 17.21 & $>20.3$ & $<1.3$ \\
00032613200\_3 & 0.30 & 2016-12-29.65 & 57751.66 & 17.34 & $>19.6$ & $<3.4$ \\
00032613201\_1 & 0.79 & 2016-12-30.05 & 57752.05 & 17.73 & $>20.2$ & $<1.4$ \\
00032613201\_2 & 0.91 & 2016-12-30.12 & 57752.12 & 17.80 & - & $<1.2$ \\
00032613201\_3 & 0.69 & 2016-12-30.46 & 57752.46 & 18.14 & $>20.2$ & $<1.6$ \\
00032613201\_4 & 0.68 & 2016-12-30.65 & 57752.65 & 18.33 & $>20.1$ & $<1.7$ \\
00032613202\_1 & 1.65 & 2016-12-31.58 & 57753.58 & 19.26 & $>20.7$ & $<0.8$ \\
00032613202\_2 & 1.25 & 2016-12-31.65 & 57753.66 & 19.34 & $>20.6$ & $<0.9$ \\
\enddata
\tablenotetext{a}{Dead-time corrected exposure time}
\tablenotetext{b}{Start date of the snapshot}
\tablenotetext{c}{Time in days after the eruption of nova \nova in the optical on 2016-12-12.32 UT (MJD 57734.32; cf.\ Section~\ref{sec:time})}
\tablenotetext{d}{the \swift UVOT uvw2 filter has a central wavelength of 1930\,\AA; not all snapshots have UVOT aspect corrections}
\end{deluxetable}

\newpage

\section{Spectra of the 2016 eruption of \novak}

\begin{figure*}[h]
\begin{center}
\includegraphics[width=0.83\textwidth]{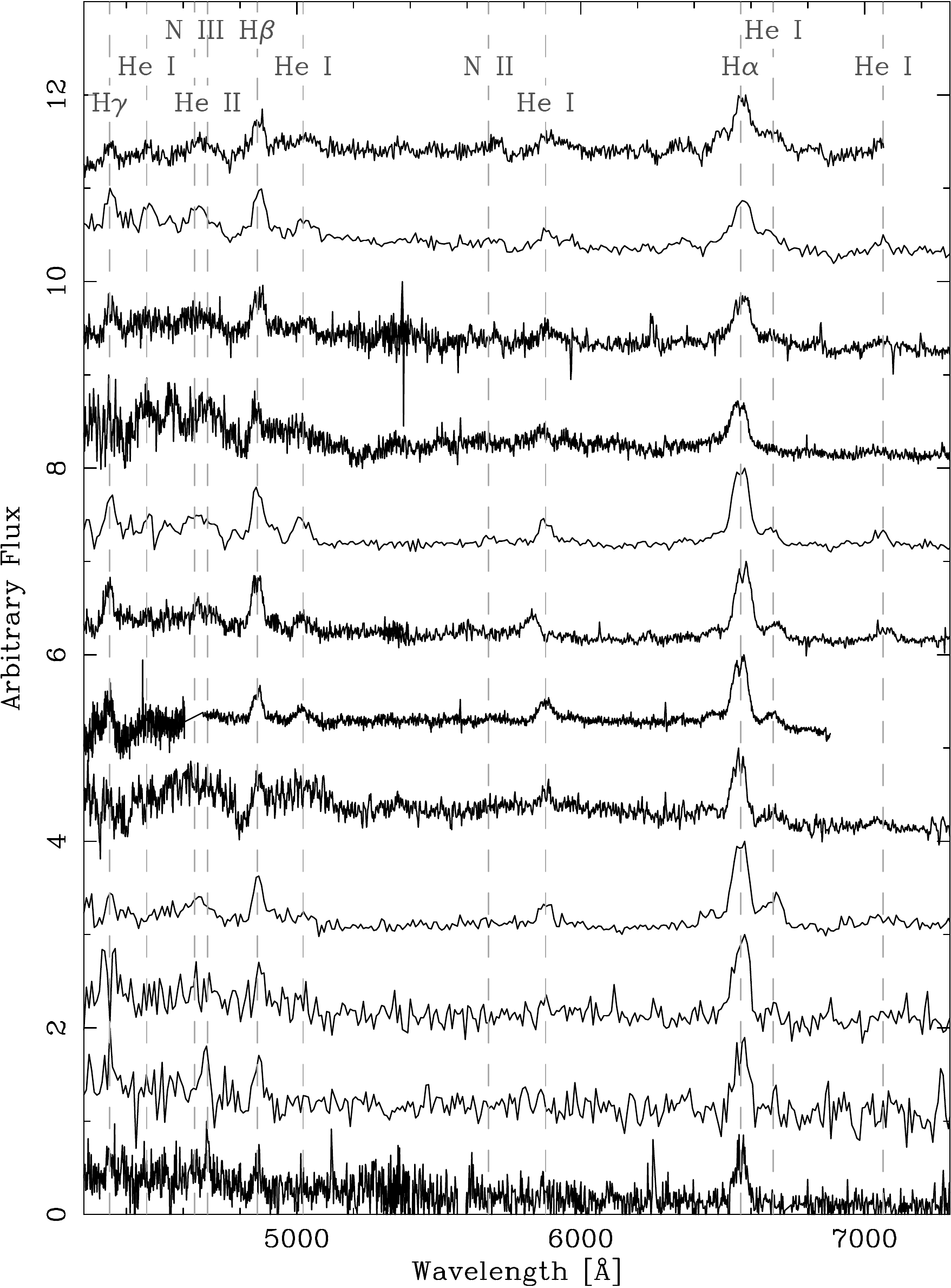}
\caption{All spectra of the 2016 eruption of \novak. The figure shows the spectra in date order (see Table~\ref{tab:spec}) from the 0.54\,d ALFOSC/NOT spectrum at the top to the 5.83\,d DIS/ARC spectrum at the bottom. The wavelengths of prominent lines are indicated.\label{specall}}
\end{center}
\end{figure*}

Figure~\ref{specall} presents all the spectra following the 2016 eruption of \novak, as recorded in Table~\ref{tab:spec}.  As it was not possible to obtain an absolute flux calibration of all the spectra, here they are presented with arbitrary flux.

\end{document}